\documentclass[10pt]{article}
\usepackage{latexsym,amsmath,amssymb,amsbsy,mathrsfs,graphics,amscd, empheq, mathtools, tikz-cd}
\usepackage{enumitem}
\mathtoolsset{showonlyrefs}
\usepackage{booktabs}
\usepackage{xcolor}
\usepackage{stackengine}
\usepackage[breaklinks,colorlinks,
  citecolor=blue,
  urlcolor=blue]{hyperref}
\usepackage{graphicx}
\usepackage{longtable}
\usepackage{subcaption}

\DeclareFontFamily{OT1}{pzc}{}
\DeclareFontShape{OT1}{pzc}{m}{it}{<-> s * [1.10] pzcmi7t}{}
\DeclareMathAlphabet{\mathpzc}{OT1}{pzc}{m}{it}

\addtocounter{footnote}{1}
\makeatletter
\@addtoreset{table}{bsection}
\def\thetable{\thesection.\@arabic\c@table}
\def\fps@table{h, t}
\@addtoreset{equation}{section}

\makeatother

\newtheorem{theorem}{Theorem}[section]
\newtheorem{definition}[theorem]{Definition}
\newtheorem{lemma}[theorem]{Lemma}
\newtheorem{remark}[theorem]{Remark}
\newtheorem{proposition}[theorem]{Proposition}

\newtheorem{corollary}[theorem]{Corollary}

\newcommand{\bfi}{\bfseries\itshape}
\newcommand{\vertiii}[1]{{\left\vert\kern-0.25ex\left\vert\kern-0.25ex\left\vert #1 
    \right\vert\kern-0.25ex\right\vert\kern-0.25ex\right\vert}}
\newsavebox{\savepar}

\makeatletter

\usepackage{scalerel,stackengine}
\stackMath
\newcommand\reallywidehat[1]{%
\savestack{\tmpbox}{\stretchto{%
  \scaleto{%
    \scalerel*[\widthof{\ensuremath{#1}}]{\kern-.6pt\bigwedge\kern-.6pt}%
    {\rule[-\textheight/2]{1ex}{\textheight}}%WIDTH-LIMITED BIG WEDGE
  }{\textheight}% 
}{0.5ex}}%
\stackon[1pt]{#1}{\tmpbox}%
}
\newcommand{\supp}{\operatorname{supp} }
\newcommand{\Basin}{\operatorname{{Basin}}}

\begin{document}

\title{\textbf{Learning the climate of dynamical systems with state-space systems}}
\author{James Louw$^{1}$, Juan-Pablo Ortega$^{1}$}
\date{}
\maketitle

\begin{abstract}
    State-space systems encompass a broad class of algorithms used for modeling and forecasting time series. For such systems to be effective, two objectives must be met: (i) accurate point forecasts of the time series must be produced, and (ii) the long-term statistical behaviour of the underlying data-generating process must be replicated. The latter objective, often referred to as learning the climate, is closely related to the task of producing accurate distribution forecasts. Empirical evidence shows that distribution forecasts are far more stable than point forecasts, which are sensitive to initial conditions. In this work, we rigorously study this phenomenon for state-space systems. The main result shows that, if the underlying data-generating process is structurally stable and possesses a mixing or an attracting measure, then a sufficiently regular initial probability distribution remains close to the true future distribution at arbitrarily long time horizons when forecasted by a $C^1-$close state-space proxy. Thus, under these conditions, learning the climate of a dynamic process with a universal family of state-space systems is feasible with arbitrarily high accuracy.
\end{abstract}

\bigskip


\makeatother
\makeatletter
\addtocounter{footnote}{1} \footnotetext{%
Division of Mathematical Sciences,
School of Physical and Mathematical Sciences,
Nanyang Technological University,
21 Nanyang Link,
Singapore 637371.
{\texttt{Louw0002@e.ntu.edu.sg, Juan-Pablo.Ortega@ntu.edu.sg}}}
\makeatother

\tableofcontents

\section{Introduction}

A challenge that, for millennia, has been a major driving force in humanity's quest for knowledge is predicting the future. While man has for centuries studied and experimented with different methods of prediction, ranging from the mystic to what today is called science, he is faced with the fact that there are inherent difficulties with this task: all his methods are fundamentally subject to certain limits which keep him from being able to know the future precisely. The advent of machine learning has marked a new phase in this quest. Nevertheless, researchers have long found that even these powerful algorithms are limited. Thus, the task has been to determine what these limits are: what knowledge we are able to obtain about the future, and with what confidence we can assert this knowledge coming from `blackbox' machine learning algorithms in forecasting.

A particular learning methodology that has gained prominence as a versatile and highly competitive predictive strategy is based on the use of {\bfi state-space systems} \cite{kalman1960new, Koopman:kalman, Sarkka2013, GuLi2023}. State-space systems extract underlying information from temporal data by mapping it to a higher-dimensional state space, where fundamental relationships among data points and their temporal structure are revealed. As a mathematical framework, state-space systems include theoretically derived methods, such as {\bfi Kalman filtering, forecasting, and smoothing}, as well as various machine learning prediction strategies which together form a concept class known as {\bfi recurrent neural networks}. In practice, these algorithms have been successfully applied across a broad range of applications, from robot control and aeronautics \cite{AstromMurray2008Feedback, Stengel2015FlightDynamics} to prediction in atmospheric dynamics \cite{LawStuartZygalakis2015DataAssimilation, Evensen2009EnKF} and finance \cite{tsay:book, Koopman:kalman}. It is worth noting a subfamily of these systems, called {\bfi reservoir computing (RC)}, in which a significant portion of the system architecture is randomly generated \cite{Jaeger04, lukosevicius, Rodan2011, rodanSimpleDeterministicallyConstructed2012}, leading to inexpensive training. The reservoir computing subfamily has gained prominence for its excellent performance in forecasting of time series. Notably, random feature maps have recently begun to emerge as a competitor to reservoir computing, having been shown to exceed the forecasting capabilities of reservoir computers on a number of classical benchmark systems \cite{rahimi2008random, gottwald_forecast_noisy_ds, nielsen_random_feature_model, gottwald2025_hitandrun, Gottwald_choice}. Random feature maps also fall in the state-space system family, but they are in fact much simpler as they bypass state space dynamics. The results of this paper also hold true for these models.

On the theoretical side, conditions on the underlying architecture that guarantee various mathematical properties have been established. Among these are {\bfi universal approximation properties} for various families of state-space systems \cite{schafer2007recurrent, RC6, RC7, RC8, RC20, RC12}, which describe the expressive power of the architecture, that is, its ability to approximate a given system to an arbitrary precision, {\bfi fading memory, state- and input-forgetting properties} \cite{RC30, RC31, RC32} which characterize the way in which memory is processed for sequential data, the {\bfi echo state property} \cite{gallicchio:esp, RC28, RC32}, which, when satisfied, means that the state-space system induces a functional relation between the inputs and the state space, as well as the existence of {\bfi generalized synchronizations} \cite{rulkov1995generalized, hunt:ott:1997, RC18, RC21}, which link the underlying system and the state space, so that the state space is in fact driven by dynamics in the underlying system, even though inputs may consist only of partial information. All these attributes are discussed in greater detail in Section \ref{sec:prelims} and provide a theoretical justification for the empirically observed ability of these algorithms to learn underlying dynamical patterns in data.

In forecasting deterministic time series, two objectives arise. (i) One might aim to produce accurate {\bfi point forecasts} of the value of the time series at some given future point in time. Alternatively, (ii) we might seek to produce a proxy system whose iterates possess similar {\bfi long-term statistical properties} to those of the given time series. Lorenz, in his paper \cite{lorenz1964problem}, referred to these long-term statistical properties of a dynamical system as its {\bfi climate}. By contrast, we may refer to point predictions as forecasts of the {\bfi weather} of the system. Empirically it has long been known that good weather predictions do not always lead to successful learning of the climate. For example, in \cite{pathak:chaos}, the authors produced two state-space systems, both exhibiting good short-term point forecasting abilities, though in the long term, due to the chaotic properties of the underlying system, the point forecasts failed. Nevertheless, the manner in which the two predictions failed was significantly different: one produced a time series entirely unrelated to the training data, while the other produced a trajectory which to the naked eye still looks as if it had come from the same generating system as the data on which it was trained -- it exhibited similar {statistical properties over time} to the training data. From this, the observation arises that, although weather forecasts are unstable due to sensitivity to initial conditions, learning the climate may prove stable.

In this paper, we study the accuracy of state-space systems in producing {\bfi density forecasts.} In density forecasting, rather than starting from a single initial condition of a time series, we forecast from an initial probability distribution on the space in which the time series lives, seeking to predict the evolution of this distribution over time. Density forecasts are not concerned so much with tracking individual trajectories of the time series as with producing a system with ergodic properties similar to those of the driving process. They rely on {\bfi replicating the climate} of the true system.

% In this work we study the notion of climate for time series data from the observations of dynamical systems more deeply. To this end, we present a first result for prediction accuracy of state-space systems for the {\bfi density forecasting task}. While point forecasts may be likened to predicting the {\bfi weather} of a system, density forecasts are closer to predicting the {\bfi climate}. When thinking of climate as opposed to weather, a number of thoughts immediately come to mind. While weather seems unpredictable even at short time horizons, perhaps a few weeks into the future, we expect to be able to be able to predict certain aspects of the climate much further into the future, perhaps a few years at least. Weather has to do with predicting atmospheric variables precisely, such as the temperature on a given day, or the amount of rainfall in a certain area; climate is global and has to do with long term aggregate statistical properties of the weather. Weather is volatile, while climate evolves over much larger time scales. From the outset, we stress that the application of the terms {\bfi climate} and {\bfi weather} to time series data merely form an analogy -- our work is applicable to datasets far beyond the reaches of atmospheric dynamics.

Work on establishing bounds for {\bfi error in the point-wise prediction task} using state-space systems began with the results in \cite{berry2022learning, RC22}. In these works, a horizon-dependent bound on the prediction errors from state-space systems was derived for time-series data arising from observations of a dynamical system. That bound demonstrates the link between the universal approximation properties of a state-space family and its point forecasting ability. According to this result, the prediction accuracy deteriorates exponentially over the forecasting horizon, at a rate given by the top Lyapunov exponent of the proxy system. The exponential character of the bound results from a theorem of Oseledets \cite{oseledets1968multiplicative} which illuminates the inherent mathematical property of chaotic systems that makes a time-uniform bound on the error of point predictions impossible in the presence of Lyapunov-type chaos.

On the surface, the density forecasting task may seem more complex because the prediction space is of much higher dimension. However, as was noted in \cite{pathak:chaos}, it is possible for a system to fail at the point forecasting task, yet still succeed in reproducing the climate. In our work, we give a rigorous mathematical proof to substantiate this observation. We present a first result for the {\bfi error of density forecasts} using state-space systems, showing that under appropriate conditions, the density forecasting task is stable at arbitrarily long time horizons even in the presence of Lyapunov-type chaos. Under these conditions, {\bfi the sensitivity to initial conditions inherent to chaos does not carry over to density forecasts.}

% A well-known work highlighting this dichotomy is \cite{Ott2018}, in which the authors present two state-space systems trained on a single dataset. Both systems are fairly accurate at short term point forecasts, though at long time horizons they fail. However, the manner in which the two predictions fail is significantly different: one eventually produces a time series completely unrelated to the training data, while the other, though it loses track of the true data, produces a trajectory which to the naked eye still looks as if it had come from the same generating system as the data on which it was trained -- it exhibits similar {\bfi statistical properties over time} as the system on which is was trained. To describe this phenomenon, the authors said that the latter state-space system replicated the {\bfi climate} of the attractor underlying the data, while the former had failed to do this.

We briefly discuss the theoretical notions that lead to this result. Under reasonable conditions on the space on which the dynamical system evolves (compactness and metrizability), the system is guaranteed to possess a stationary probability distribution -- indeed it may have many. In many cases, however, one arises as the `natural' distribution \cite{newman2025attractingmeasures}. In particular, in Theorem \ref{thm:transport_of_measures_with_state_space_system_bound}, our main result, we assume that the underlying system possesses a {\bfi physical} or {\bfi attracting} measures. These measures have different naturalness properties characteristic of SRB measures (discussed in greater detail later on) and characterize the long-term statistical behavior of most trajectories of the system in various ways. Accordingly, we will say that a state-space system can {\bfi learn the climate} of the underlying dynamical system if it is able to reproduce the natural distribution of the system.
Several stability notions exist in dynamical systems theory. We assume {\bfi structural stability} of the underlying system, which, in our context, implies stability of the natural measures described above under perturbations of the dynamical system map.

% In our work we show that, when the underlying system is structurally stable and possesses a physical or attracting measure, a sufficiently well-trained member of a universal class of state-space systems is indeed able to learn the climate of the system on which it is trained. Structurally stable means that the natural distribution of the system is stable under perturbations of the dynamical system.

Our main result shows that when the underlying system is structurally stable and possesses a measure that is mixing or attracting, density predictions from a sufficiently regular initial probability distribution by a (well-trained) proxy system will remain close to the ground-truth distribution: {\bfi density forecasts satisfy a time-uniform error bound.}  In the case of physical or ergodic measures, we derive a similar error bound, but in a Ces\`{a}ro sense.
%-- that is, while point predictions fail at an exponential rate, density forecasts {\bfi may be made arbitrarily precise, and will remain so across any time horizon}.
On the space of probability measures, a dynamical system acts as the restriction of its associated Perron-Frobenius operator to this space \cite{lasota2013chaos}. As we will see in the proofs, the time-uniform bound follows from the fact that mixing and attracting measures are asymptotically stable fixed points of the Perron-Frobenius operator acting on a subset of the space of measures.
By contrast, we show that omitting the hypothesis of such natural measures leaves us with a bound on prediction errors that grows exponentially, as in the point prediction task. The proofs for this bound make use of a theory of Lyapunov exponents on metric spaces applied to the weak topology on the space of probability measures \cite{Kifer_1983}. Thus, we see that what ultimately distinguishes point forecasts from density forecasts is the existence of natural measures that introduce stability in the way the dynamical system acts on the space of measures.

An additional observation that follows from the main results of this paper concerns the interplay between the universal approximation properties of state-space systems and their density-forecasting abilities. In particular, the main results rely on a {\bfi structural stability} assumption on the dynamics, thus requiring a sufficiently accurate approximation of the underlying dynamical system in the Whitney $C^1$ norm.

% \tod{could I take out the pragraph below?}

Learning dynamical systems is a multi-faceted task. Beyond approximating the transformation law of the dynamical system, various other aspects are integral to producing a reliable proxy to the dynamical system. This perspective was highlighted in the paper \cite{berry2025limits}, in which the authors discuss the interaction between strategies that target approximating the dynamical system map, finding its invariant sets, approximating the dynamical system as a Markov process, and learning its Koopman operator. Our work may be interpreted within this framework, too. The invariant measure of a dynamical system is closely related to its invariant sets, in addition to being associated with a view of the dynamical system as a Markov process. Moreover, the dual of the Perron-Frobenius operator of the Markov process is indeed the Koopman operator \cite{lasota2013chaos}. In our work, we study conditions under which an accurate approximation of the dynamical system map yields a faithful Markov representation and, hence, a replication of the system's natural measure.

 In machine learning, much concern is being raised as to the reliability of predictions made by these algorithms, which are typically seen as a `black box'. This study represents an important step toward understanding the capabilities of machine learning algorithms to replicate the climate of deterministic time-series data. Beyond this lies the realm of stochastic time series, in which, to our knowledge, no results have been established regarding the accuracy of predictions using state-space systems. Such understanding is crucial to the endeavor of providing theoretical guarantees for the accuracy of these algorithms and, hence, to the problem of determining their suitability for a given task.

\medskip

{\bf Outline of the paper.} The paper is organized as follows: Section \ref{sec:prelims} contains an introduction to the mathematical setting of dynamical systems and time series arising from their observations. The state-space learning methodology is also introduced here, along with a brief overview of Lyapunov exponents for metric spaces and selected concepts from ergodic theory. Section \ref{sec:stability_dyn_sys_probability} develops several results for the stability of trajectories of measures under the action of the Perron-Frobenius operator. These serve as intermediate mathematical results that build up to Section \ref{sec:main_result}, where the main theorems on the error of distribution forecasts using state-space systems are proved.
% For readers who are interested mainly in the results of this paper, \ref{sec:stability_dyn_sys_probability} may largely be skipped in deference to Section \ref{sec:main_result}.
Section \ref{sec:numerics} discusses the application of the theoretical results to the Lorenz system, illustrating them numerically. Finally, Section \ref{sec:conclusion} concludes the paper. A more detailed description of much of the theory used in the paper, as well as a few additional results, may be found in the Appendix \ref{sec:appendix}.

\medskip

{\bf Codes and data.} All code and data used in generating the numerical results presented in this paper may be found at this \href{https://github.com/Learning-of-Dynamic-Processes/Learning-the-climate-with-state-space-systems.git}{GitHub repository}.

% \begin{longtable*}[!t]
% \renewcommand{\arraystretch}{1.3}
% \caption{My data}
% \centering
% \begin{tabular}{|p{2.8cm}|p{0.6cm}|p{1.6cm}|p{2.2cm}|P{0.8cm}|p{2.6cm}|p{2.5cm}|}
% \hline
% \textbf{Name} & \textbf{Year} & \textbf{ID} & \textbf{Address} & \textbf{Salary} & \textbf{Skills} & \textbf{Qualifications} \\
% \hline
% Some text & some text & some text & 5 & some text & Som text
% \hline
% \end{tabular}
% \label{table3}
% \end{longtable*}

\section{Preliminaries}\label{sec:prelims}

This section reviews the theory used throughout the paper. Section \ref{subsec:learning with state-space systems} discusses state-space systems and how they are used in producing point and density forecasts of time series from the observations of dynamical systems. Section \ref{subsec:metric_le} introduces a metric Lyapunov exponent, and Section \ref{subsec:ergodicity} reviews some results from ergodic theory.

%(), In this section, we establish the necessary theory to prove the main results of the paper. In Section \ref{subsec:learning with state-space systems} we introduce state-space systems and show how they may be used in learning and predicting time series pointwise and in distribution. We focus on results in the context of learning time series from the observations of dynamical systems. In Section \ref{subsec:metric_le}, we discuss a formulation of Lyapunov exponents for metric spaces along with some basic results for these exponents. Section \ref{subsec:metrics on prob spaces} gives an overview of the weak topology on the space of probability measures and conditions under which this topology may be metrized. We name a number of possible metrics, of which the Wasserstein and MMD distances feature later in the paper. The main results of the paper, however, hold generally for any metric compatible with the weak topology. Section \ref{subsec:ds, erg, mixing} recalls the definitions of ergodicity and mixing in dynamical systems and some elementary results in connection with these notions. We also discuss physical and attracting measures. Finally, in \ref{subsec:stability} we consider stability of dynamical maps under $C^1$ perturbations and derive some implications of this stability.

\subsection{Learning time series with state-space systems.}\label{subsec:learning with state-space systems}

{\bf Observations from dynamical systems.} In this paper we present results for time series which consist of the observations from the trajectory of a dynamical system. In its most basic form, a dynamical system consists of a topological space $\mathcal{M}$ and a continuous map $\phi \colon \mathcal{M} \longrightarrow \mathcal{M}.$ In our analogy of climate and weather, we may think of the dynamical system as the entire system consisting of the atmosphere, hydrosphere, cryosphere, the sun, and other factors, all underlying and driving what we experience as the weather. Naturally, we do not expect to have access to complete information about the true state of the dynamical system. Rather, we gain information about a trajectory of the dynamical system, $(\dots, \underline{m}_{-1}, \underline{m}_0, \underline{m}_1, \underline{m}_2 \dots)$, where $\underline{m}_t = \phi(\underline{m}_{t-1})$ for all $t \in \mathbb{Z}$, through an observation function $\omega \colon\mathcal{M} \longrightarrow \mathcal{U},$ where $\mathcal{U}$ is called the {\bfi observation space}. This yields the parallel sequence $(\dots, \underline{u}_{-1}, \underline{u}_0, \underline{u}_1, \underline{u}_2, \dots)$ with $\underline{u}_t = \omega(\underline{m}_t)$ for all $t.$ For example we might consider the observation function $\omega$ which takes in the state of the entire dynamical system underlying the weather at a given time and then reads out from this the temperature at a certain location, at that time.

{\bf State-space systems for time series.} Given access to the left infinite time series of observations $\underline{{u}} = (\dots, \underline{u}_{-2}, \underline{u}_{-1}, \underline{u}_0) \in \mathcal{U}^{\mathbb{Z}_-}$, we seek to path-continue the time series using a state-space system. A state-space system consists of a {\bfi state map} $f \colon\mathcal{X} \times \mathcal{U} \longrightarrow\mathcal{X}$ and a {\bfi readout} $h \colon\mathcal{X} \longrightarrow \mathcal{U}$, where $\mathcal{X}$ is referred to as the {\bfi state space}, and $\mathcal{U}$ is called the {\bfi input space}. (In general, the readout may map into an {\bfi output space} distinct from the input space, but in the context of path-continuing the input time series, the input- and output spaces coincide.) Beyond the specific setting of observations from dynamical systems, state-space systems may be used in a wide class of learning tasks involving sequential data, though it is typically necessary to make the assumption that an underlying system drives the inputs to the state-space system. Here we discuss how in this general setting such sequential data may be processed using state-space systems. To the first-time reader it may be helpful nevertheless to keep the specific example of observations from a dynamical system in mind, and we also include a few remarks in this regard below.
% learn a map from a space of bi-infinite time series to another one by means of a filter; in our case, we focus on how they may be used to path-continue a given time series, which amounts to learning a causal filter that acts as the left-shift map on the time series. We denote the space in which the time series is defined by $\mathcal{U}.$ We are given a left-infinite time series $\underline{{u}} = (\dots, \underline{u}_{-2}, \underline{u}_{-1}, \underline{u}_0) \in \mathcal{U}^{\mathbb{Z}_-}$ In our work, the time series $\underline{u}$ is assumed to be driven by some deterministic underlying system.
%In our work we deal with time series arising from partial observations of a dynamical system (described below).
In our case, our task is to train the state-space system $(f,h)$ so as to make accurate predictions $(\hat{\underline{u}}_1, \hat{\underline{u}}_2, \hat{\underline{u}}_3, \dots)$ of the true future values $(\underline{u}_1, \underline{u}_2, \underline{u}_3, \dots)$ of the observations. In {\bfi reservoir computing} the state map $f$ is randomly generated, and subsequently the readout $h$ is trained. The results of this paper hold for the more general mathematical framework of state-space systems, while the numerics we implement in Section \ref{sec:numerics} utilize {\bfi echo state networks}, a particular reservoir computing architecture. A concept fundamental to the learning and prediction we aim to perform is the {\bfi echo state property}. The {\bfi state map} $f$ is said to have the echo state property when for any $\underline{{u}} \in \mathcal{U}^{\mathbb{Z}_-}$ there is a unique $\underline{{x}} \in\mathcal{X}^{\mathbb{Z}_-}$ such that the relation

\begin{equation}\label{eq:state_eq}
    \underline{x}_t = f(\underline{x}_{t-1}, \underline{u}_t)
\end{equation}

\noindent holds for all $t\leq 0.$ We refer to \eqref{eq:state_eq} as the {\bfi state equation.} When $f$ has the echo state property, we may associate to it
%a time-invariant {\bfi filter} $U^f \colon \mathcal{U}^{\mathbb{Z}_-} \longrightarrow \mathcal{X}^{\mathcal{Z}_-},$ $U^f(\underline{u}) = \underline{x}$ and 
a {\bfi functional} $H^f \colon \mathcal{U}^{\mathbb{Z}_-} \longrightarrow \mathcal{X},$ $\underline{u} \mapsto \underline{x}_0$ where $(\underline{u}, \underline{x})$ satisfy the state equation \eqref{eq:state_eq}. Thus the echo state property means that $f$ induces a mapping from an input sequence of observations to a corresponding sequence in the state space. Typically the dimension of the state space is much larger than that of the observations. Ideally, the state map $f$ will unfold the observations in the state space so that the dynamics of the underlying system are exposed. 
% State sequences $\underline{x}$ induced by inputs from $\underline{\mathcal{U}}^-$ through the state equation are typically found to come from a much lower dimensional, backward shift invariant, subspace $\underline{\mathcal{X}}^- \subseteq \mathcal{X}^{\mathcal{Z}_-}$.
The echo state property has been well studied in \cite{jaeger2001, Manjunath:Jaeger, RC7, manjunath:prsl, manjunath2022embedding} and can also be extended to stochastic settings \cite{RC27, RC28, RC31, RC32}. We briefly mention a few other properties which are dealt with in greater depth in \cite{RC26, RC32}: When the space ${\mathcal{X}}$ is compact, state-space systems that possess the ESP also possess the {\bfi fading memory property (FMP)}, a continuity condition on the filter induced by the state-space system. This property has at times been somewhat confused with the related {\bfi state forgetting property (SFP)} which allows us to approximate $\underline{x}_0$ from the sequence $\underline{{x}} = (\dots, \underline{x}_{-2}, \underline{x}_{-1}, \underline{x}_0)$ satisfying \eqref{eq:state_eq} given only a finite history of inputs $(\underline{{u}}_{-T}, \dots, \underline{u}_{-1}, \underline{u}_0).$ In particular, if the state-space system has the SFP, we may arbitrarily initialize our state sequence at some $\underline{x}_{-T}' \in \mathcal{X}$, and by iterating \eqref{eq:state_eq}, for large enough $T$ we will get a good approximation of $\underline{x}_0$.

Once we have $\underline{{x}}_0,$ we may make predictions by setting $\hat{\underline{x}}_0 = \underline{x}_0, \hat{\underline{u}}_0 = \underline{u}_0$ and iterating the state-space equations
\begin{empheq}[left = {\empheqlbrace}]{align}
    \hat{\underline{x}}_t &= f(\hat{\underline{x}}_{t-1}, \hat{\underline{u}}_t)\label{eq:SS_1}\\
    \hat{\underline{u}}_t &= h(\hat{\underline{x}}_{t-1})\label{eq:SS_2}
\end{empheq}
for $t =1,2,3,\dots$. Using the functional $H^f_h \colon \mathcal{U}^{\mathbb{Z}_-} \longrightarrow \mathcal{U},$ $H^f_h = h \circ H^f$ associated to the state-space system, we may write $\hat{\underline{u}}_t = H^f_h(\dots, \underline{u}_{-1}, \underline{u}_0, \hat{\underline{u}}_1, \dots, \hat{\underline{u}}_{t-1}).$
%The sequence $\hat{\underline{u}}_1, \hat{\underline{u}_2, \dots$ is generated by iteratively applying $H^f_h$ to the sequence and appending the output to the end of the sequence.
To ensure accuracy of the predictions generated by \eqref{eq:SS_1} and \eqref{eq:SS_2}, we require, besides the echo state property, that the family of state-space systems under consideration exhibit {\bfi universal approximation properties}. Consider the space of functionals $H \colon \mathcal{U}^{\mathbb{Z}_-} \longrightarrow \mathcal{U}$. A family of state-space systems is {\bfi universal} in this space if any functional within this family may be uniformly approximated to an arbitrary degree of accuracy by the functional $H^f_h$ induced by some member $(f,h)$ of the family of state-space systems. In our case, this allows us to approximate the one-step-ahead predictions of the time series to a desirably small degree of error using a state-space system from the family. In this paper, we examine the interplay between universality and the density-forecasting ability of state-space systems. Typically, the input and state spaces are Euclidean: $\mathcal{U} \subseteq \mathbb{R}^d$ and $\mathcal{X} \subseteq \mathbb{R}^L$ (with $d \ll L).$ When additionally the input time series is governed by a fading memory functional and comes from the uniformly bounded domain $K_M := \{ \underline{u} \in (\mathbb{R}^d)^{\mathbb{Z}_-} \colon \| \underline{u}_t \| \leq M, t \in \mathbb{Z}_- \}$ for some $M>0,$ it has been shown that several state-space families are universal \cite{RC6, RC7, RC12, sugiura2024nonessentiality, li2024simple}.

\medskip

{\bf Predicting dynamical systems with state-space systems.}  Having discussed how state-space systems may be used to process sequential data, we return to the specific setting of observations from a dynamical system. We require in our work that the dynamical system is {\bfi invertible}, that is, $\phi \in \operatorname{Hom}(\mathcal{M})$, the space of homemorphisms of ${\cal M} $. When the state map $f$ possesses the ESP, there exists a map $\zeta_{(\phi, \omega, f)} \colon\mathcal{M} \longrightarrow \mathcal{X}$ such that for any $\underline{m} \in\mathcal{M}$, whenever the sequences $\underline{x} = (\underline{x}_t)_{t \in \mathbb{Z}}$ and $\underline{u} = (\underline{u}_t)_{t \in \mathbb{Z}} = (\omega(\phi^t(\underline{m})))$ satisfy the state equation \eqref{eq:state_eq}, then for any $t \in \mathbb{Z}$
$$\underline{x}_t = \zeta_{(\phi, \omega, f)} (\phi^t(\underline{m})).$$
Thus, the dynamics of the sequence $\underline{x}$ in the state space are synchronized with the dynamical system $\phi$ underlying the observations $\underline{u}.$ Accordingly, we call the map $\zeta_{(\phi, \omega, f)}$ a {\bfi generalized synchronization} (GS) \cite{rulkov1995generalized}. The dynamical system, observation map and state map being clear from context, from hereon we abbreviate the GS map as $\zeta$ and write $\widetilde{\cal X} = \zeta({\cal M}).$ Sufficient conditions guaranteeing the existence, continuity, and differentiability of GSs have been established in \cite{RC18}. An important assumption used in many papers \cite{berry2022learning, RC26, berry2025limits} is that the GS is an embedding. This feature has been established only in very particular situations \cite{RC21} and its availability in general settings remains an open problem. Here we also make this assumption. The significance of this is two-fold. First of all, when the GS is an embedding, learnability of the system is guaranteed: the map $h \colon \widetilde{\cal X} \longrightarrow \mathcal{U}, h = \omega \circ \phi \circ \zeta^{-1}$ has the property that for any $m \in\mathcal{M}$, $\underline{u} = (\underline{u}_t)_{t \in \mathbb{Z}} = (\omega(\phi^t(\underline{m})))$, and $\underline{x}$ which together with  $\underline{u}$ satisfies the state equation \eqref{eq:state_eq}, we have $h(\underline{x}_{t-1}) = \omega(\phi^t(m)),$ for all $t \in \mathbb{Z}$ \cite[Proposition 2.1]{RC22}. Thus the learning task reduces to the problem of producing a sufficiently accurate approximation $\hat{h}$ of $h.$

\begin{minipage}[c]{0.7\textwidth}
    Secondly, when the GS is an embedding, the dynamical system $\Phi \colon \widetilde{\cal X} \longrightarrow \widetilde{\cal X},\,\, x \mapsto f({x}, h({x}))$ is conjugate to $\phi$ by $\zeta,$ that is, $\zeta \circ \phi = \Phi \circ \zeta,$ as depicted in the diagram on the right. This is important as it allows stability properties of the original dynamics $\phi$ to carry over to the state space dynamics $\Phi$. In particular, the original dynamics is structurally stable and possesses a mixing, attracting, ergodic or physical measure, if and only if the same is true for the state space dynamics. As the theoretical results proved below show, this gives stability in learning and predicting the dynamics in distribution.
\end{minipage}
\hfill
\begin{minipage}[c]{0.3\textwidth}
    \centering
    \begin{tikzcd}[column sep=large]
        \mathcal{M} \arrow[r, "\phi"] \arrow[d, "\zeta"'] & \mathcal{M} \arrow[d, "\zeta"] \\
        {\widetilde{\cal X}} \arrow[r, "\Phi"'] & {\widetilde{\cal X}}
    \end{tikzcd}
\end{minipage}

 Writing $\hat{\Phi}$ for the map ${x} \mapsto f({x}, \hat{h}({x})),$ the forecasting error in the state space $\mathcal{X}$, equipped with metric $d_{\mathcal{X}}$, is $d_{\mathcal{X}}(\underline{x}_t, \underline{\hat{x}}_t) = d_{\mathcal{X}}(\Phi^t(\underline{x}_0), \hat{\Phi}^t(\underline{x}_0)),$ while that in the observation space $\mathcal{U}$, with metric $d_\mathcal{U}$, is $d_\mathcal{U}(\underline{u}_t, \underline{\hat{u}}_t) = d_\mathcal{U}(h \circ \Phi^t(\underline{x}_0), \hat{h} \circ \hat{\Phi}^t(\underline{x}_0)).$ 

\medskip

{\bf Density forecasting.} Until this point we have considered only learning time series and predicting them {\bfi pointwise} through the equations \eqref{eq:SS_1} and \eqref{eq:SS_2}, something analogous to predicting the weather: we might train our system on time series data from temperature readings over many months and subsequently make a point prediction of the temperature at a certain time a few days in the future. Consider now the situation where instead of being given data in the form of a single trajectory of the time series, we are given an initial {\bfi probability distribution} on the space of the time series. Rather than making a point prediction for the time series, we now forecast the evolution of the probability distribution into the future, analogous to predicting a probability distribution for the temperature at a future time point.

The setup is as follows: equipping the space $\mathcal{M}$ with its Borel sigma algebra makes it a measurable space. Rather than starting at some initial condition $m_0 \in\mathcal{M}$, we now start with a measure $\mu_0 \in \mathcal{P}(\mathcal{M})$, where $\mathcal{P}(\mathcal{M})$ is the space of probability measures on $\mathcal{M}$. The dynamical system $\phi$ transports this measure through the Perron-Frobenius operator \cite{lasota2013chaos}, which in this case is the pushforward map $\phi_* \colon \mathcal{P}(\mathcal{M}) \longrightarrow \mathcal{P}(\mathcal{M})$. At time $t \in \mathbb{Z}$, the evolved measure is $\mu_t = \phi^t_* \mu_0$. Correspondingly in the observation space we start with the measure $\nu_0 = \omega_* \mu_0 \in \mathcal{P}(\mathcal{U}),$ and in the state space we start with the probability distribution $\xi_0 = (\zeta_{\phi, \omega, f})_* \mu_0 \in \mathcal{P}(\mathcal{X}).$ In the state space our predicted probability distribution is $\hat{\xi}_t = (\hat{\Phi}^t)_* (\zeta_{\phi, \omega, f})_* \mu_0$ and our forecast of the observation distribution is $\hat{\nu}_t = \hat{h}_* (\hat{\Phi}^{t-1})_* (\zeta_{\phi, \omega, f})_* \mu_0.$ Using a metric $d_{\mathcal{P}, \mathcal{U}}$ on $\mathcal{P}(\mathcal{U})$ (see the Appendix \ref{subsec:metrics on prob spaces} for details about metrics on probability spaces), we consider the prediction error $d_{\mathcal{P}, \mathcal{U}}(\nu_t, \hat{\nu}_t).$

{\it The main result of this paper shows that while the point predictions $\underline{\hat{u}}_t$ fail at an exponential rate, when a mixing or attracting measure exists for the system, predictions $\hat{\nu}_t = \hat{h}_* \hat{\Phi}^{t-1}_* (\zeta_{\phi, \omega, F})_* \mu_0$ of a sufficiently regular initial probability distribution $\mu_0$ remain accurate: even if predicting the weather is unstable, predicting the climate may still be stable.}

\subsection{A metric Lyapunov exponent.}\label{subsec:metric_le}

Density forecasts rely on the fact that the pushforward of the underlying map acts as a dynamical system on the space $\mathcal{P}(\mathcal{X})$ of probability measures. With point forecasts, a theorem of Oseledets shows that the exponential rate at which errors in prediction grow is related to the top Lyapunov exponent of the proxy system \cite{berry2022learning, RC22}. We require a Lyapunov exponent theory for metric spaces to transfer these results to the dynamical system $(\mathcal{P}(\mathcal{X}), \Phi_*).$ Here we introduce the metric Lyapunov exponent proposed by Kifer \cite{Kifer_1983}. A fuller exposition of the properties of this exponent may be found in the Appendix \ref{subsec:metric_le_appendix}.

Let $(X,d)$ be a separable metric space without isolated points and let $\varphi \colon X \longrightarrow X$ be a continuous dynamical system on $X$. For a point $x \in X$ and a nonnegative integer $t$, let $B_x( \delta, t)$ denote the set of points $y \neq x$ such that $\max\{d(\varphi^i(x), \varphi^i(y)): i = 0, \dots, t\} < \delta.$ Now, let $$A_\delta(x,t) = \sup_{y \in B_x(\delta, t)} \frac{d(\varphi^t(x), \varphi^t(y)}{d(x,y)}.$$

% Finally, let $\mu \in \mathcal{P}(X)$ be a $\varphi$-invariant Borel measure and suppose there exists $\delta_0>0$ such that for all $\delta  \in (0, \delta_0)$

% \begin{equation}\label{ineq:mu_integrability_cond}
%     \sup_{t \in \mathbb{N}} \frac{1}{t} \int_X |\log A_\delta(x, t)| d\mu(x) < \infty.     
% \end{equation}

% This condition is fulfilled, for example, when $\varphi$ is Lipschitz.
We define the {\bfi Kifer metric Lyapunov exponent} as

\begin{equation}\label{eq:kifer_exponent}
    \Lambda_{\varphi}(x) = \lim_{\delta \to 0^+} \limsup_{t \to \infty} \frac{1}{t} \log A_{\delta}(x,t).    
\end{equation}

Conditions under which this exponent exists and satisfies specific desirable properties may be found in the Appendix \ref{subsec:metric_le_appendix}.

\subsection{Ergodic theory.}\label{subsec:ergodicity}

The difference in stability between point and density forecasts results from the existence of invariant measures for the underlying dynamical system, which may possess various properties that we now recall.
%, thus introducing stability into the dynamical system $(\mathcal{P}(\mathcal{X}), \Phi_*).$

\medskip

{\bf Ergodic measures.} Let $(X,d)$ be a separable metric space and let $\varphi\colon X \longrightarrow X$ be a continuous dynamical system. A Borel probability measure $\mu \in \mathcal{P}(X)$ is said to be {\bfi invariant} if $\varphi_* \mu = \mu.$ Thus $\mu$ is a fixed point of the dynamical system $(\mathcal{P}(X), \varphi_*).$
%A set $A \subseteq X$ is said to be {\bfi invariant} if $\varphi^{-1}(A) = A$. An invariant probability measure $\mu$ is said to be {\bfi ergodic} with respect to the dynamical system $(X,\varphi)$ if every invariant Borel set $A$ either has full or zero measure.
An invariant probability measure $\mu$ is {\bfi ergodic} with respect to $(X, \varphi)$ if for any Borel sets $A, B \subseteq X,$
    
    \begin{equation}
        \lim_{T \to \infty} \frac{1}{T} \sum_{t=0}^{T-1} \mu( \varphi^{-t} (A) \cap B) = \mu(A) \mu(B).
    \end{equation}
    
If $X$ is compact, the existence of an ergodic measure is guaranteed \cite[Theorem 2.1]{Viana_Oliveira_2016_foundations_of_ergodic_theory}. In Appendix \ref{subsec:ergodic_measures_appendix}, we provide additional information on ergodic measures. In particular, Corollary \ref{cor:ergodicity_convergence} shows that measures which are absolutely continuous with respect to an ergodic measure are attracted to it in a Ces\`{a}ro sense.

\medskip

{\bf Mixing measures.} A strengthening of ergodicity is mixing. An invariant probability measure $\mu$ is said to be {\bfi mixing} with respect to $(X, \varphi)$ if for any Borel sets $A,B \subseteq X,$

\begin{equation}
    \lim_{t \to \infty} \mu( \varphi^{-t} (A) \cap B) = \mu(A) \mu(B).
\end{equation}

% Mixing measures exhibit a stronger attraction property than ergodic measures with respect to the dynamics of the pushforward map.
If the probability measure $\nu$ is absolutely continuous with respect to the mixing measure $\mu$, then $ \varphi^t_*\nu \longrightarrow \mu$ in the weak topology. See Proposition \ref{prop:strong_mixing_convergence}.

\medskip

{\bf Physical and attracting measures.} Under the dynamics of the pushforward map, mixing (respectively ergodic) measures attract measures that are absolutely continuous with respect to the mixing (respectively ergodic) measure. For deterministic dynamical systems, mixing and ergodic measures are often singular with respect to Lebesgue. Thus for experiments it is more meaningful if this attraction occurs for measures absolutely continuous with respect to some reference measure $\ell$ on $X$. For example, we might take $\ell$ to be the Lebesgue measure if $X$ is Euclidean. We require $\ell$ to be locally finite and Borel. Furthermore, we require that the dynamical system $\varphi$ be {\bfi nonsingular} with respect to the reference measure, that is, $\varphi_*\ell \ll \ell.$ We now introduce physical and attracting measures, which exhibit attraction properties similar to ergodic and mixing measures, respectively, when the initial distribution is absolutely continuous with respect to the reference measure $\ell.$

A compactly supported measure $\mu \in \mathcal{P}(X)$ is called a {\bfi physical measure} if its support has a neighborhood $U$ such that $\ell-$almost every point of $U$ is in $\Basin (\mu).$  Here the basin of the measure $\mu$ refers to the set of those points for which the empirical measure calculated by averaging along a trajectory started from the point converges to $\mu$. Analogous to Corollary \ref{cor:ergodicity_convergence}, Lemma \ref{lmm:physical_convergence} shows that if the probability measure $\nu$ is absolutely continuous with respect to the reference measure $\ell,$ and $\nu(U) = 1,$ then its iterates $\varphi^t_* \nu$ are attracted to $\mu$ in a Ces\`{a}ro sense.

A compactly supported measure $\mu \in\mathcal{P}(X)$ is called an {\bfi attracting measure} if there exists an open neighborhood $U$ of its support such that for every $\ell-$absolutely continuous $\nu \in \mathcal{P}(X)$ with $\nu(U) = 1$, we have $\varphi^t_*\nu \longrightarrow \mu$ as $t \longrightarrow \infty.$ 

Note that physical and attracting measures may still be singular with respect to the reference measure $\ell$. However, they exhibit attraction properties for a set of points of positive $\ell-$measure. More details on physical and attracting measures can be found in Appendix \ref{subsec:physical_attracting_measures_appendix}.

{\bf Verifying the existence of ergodic and related measures.} The existence of the measures discussed in this section is an integral assumption in our main result Theorem \ref{thm:transport_of_measures_with_state_space_system_bound}. While compactness and continuity of the dynamics map is enough to guarantee the existence of an ergodic measure, verifying the existence of physical and attracting measures is more difficult. Empirically, by checking whether ergodic averages converge for a set of positive $\ell-$measure we can confirm the existence of a physical measure. Likewise, checking decay of correlations for a set of positive $\ell-$measure confirms the existence of an attracting measure.

\section{Density forecasting and dynamical systems on $\mathcal{P}(X)$.}\label{sec:stability_dyn_sys_probability}

Before formulating our main results, we establish bounds on trajectories of the pushforward map. In Section \ref{sec:main_result} we proceed to apply these in the context of learning with state-space systems. Recall the density forecasting scheme presented at the end of Section \ref{subsec:learning with state-space systems}. There, we observed that probability distributions on the space $\mathcal{X}$ are transported by the action of the map $\Phi_*$, the pushforward of the state-space dynamical system conjugate to the original by a generalized synchronization. Density forecast errors arise due to the use of an approximation $\hat{\Phi}$ of $\Phi$. When the family of state-space systems we consider is universal, and additionally we select a state map $f$ from this family such that the generalized synchronization is an embedding, the error $d_\infty(\Phi, \hat{\Phi})$ may be made arbitrarily small by obtaining a sufficiently accurate approximation $\hat{h}$ of the readout $h$. More strongly, a result by Hornik et al. \cite{hornik:derivatives} allows us to make the error $d_1(\Phi, \hat{\Phi})$ arbitrarily small ($d_1 $ is a metric for the Whitney $C ^1$ topology) using readouts with a single hidden layer (see Remark \ref{rem:final_thm}).

\medskip

To avoid clashing with the notation used in the state-space learning setup, in this section, we consider a separable metric space $(X,d)$ with a continuous dynamical system $\varphi.$

\subsection{A baseline result on density forecasting errors.}\label{subsec:baseline_metric} To draw a distinction between the density and the point forecasting tasks and to showcase the striking difference made by mixing and related assumptions, we begin by deriving a result about forecasting errors without any of the mixing assumptions. We will see how the existence of an ergodic or physical measure, and finally of a mixing or attracting measure, improve from these baseline results. Note that, since $(X,d)$ is separable, and $\varphi$ is continuous, $\varphi_*$ acts as a continuous map on the metric space $\mathcal{P}(X)$ (see Appendix \ref{subsec:metrics on prob spaces}). Therefore, we begin by extending the result in \cite[Theorem 8]{berry2022learning} to metric spaces, relating this to the Lyapunov exponent theory for metric spaces introduced by Kifer (see Section \ref{subsec:metric_le}). Thereafter, we show how this may be applied to the dynamical system $(\mathcal{P}(X), \varphi_*).$

\vspace{1em}

{\bf A metric space bound on prediction errors.} Suppose that $\varphi$ is Lipschitz. We define the pointwise Lipschitz constant

\begin{equation}\label{eq:pw_lip_const}
    L_\varphi(x) = \lim_{ \delta \to 0^+} \sup_{y \in B_x(\delta, 0)} \frac{d(\varphi(x), \varphi(y))}{d(x,y)}.
\end{equation}

Furthermore, we define the quantity

\begin{equation}\label{eq:metric_spatial_lyap_exp}
    \lambda_{\varphi}(\mu) = \int_X \log L_\varphi(x) d\mu(x).
\end{equation}

We can think of the Kifer exponent $\Lambda_{\varphi}(\cdot)$ as a Lyapunov exponent defined along a trajectory, while $\lambda_{\varphi}(\cdot)$ is a Lyapunov exponent defined in terms of a space average. Therefore, we will refer to the latter quantity as the {\bfi metric spatial Lyapunov exponent.} Traditionally Lyapunov exponents are formulated in terms of a limiting separation of trajectories for nearby initial conditions. The spatial metric Lyapunov exponent instead looks at a spatial average of the one step separations induced by the map. Thus this exponent is not directly related to sensitivity to initial conditions, but still functions as a quantification of the expansive properties of the map. In the setting of \cite{abarbanel_local_lyap_exps}, the metric spatial Lyapunov exponent corresponds to the average of the top local Lyapunov exponent for one iteration of the map. By Theorem \ref{thm:bound1_kiferexp} and the remark that follows, $\lambda_{\varphi}(\mu)$ dominates the Kifer exponent $\Lambda_{\varphi}(\cdot)$ as well as other previously defined metric Lyapunov exponents, $\mu-$a.e. In the case of differentiable maps on a manifold, it will also dominate the traditional top Lyapunov exponent. These bounds result from the subadditivity relation of Lemma \ref{lmm:subadditivity_Adelta and f'}. In particular we note that the spatial Lyapunov exponent and trajectory-wise (Kifer) Lyapunov exponent need not coincide.

Theorem \ref{thm:bd_ffhat_t_x} below shows that in metric spaces the trajectory of a perturbed dynamical system separates from the true trajectory approximately at an exponential rate, governed by the spatial metric Lyapunov exponent of the perturbed system.
% The proof makes use of the quantity

% \begin{equation}
%     \lambda'_{\varphi}(\gamma, \mu) = \int_X \log( L_\varphi(x) + \gamma) d\mu(x),
% \end{equation}

% where $\gamma>0.$ When all the necessary integrals exist, an argument using the Dominated Convergence Theorem shows that $\lambda'_{\varphi}(\gamma, \mu)$ is a continuous perturbation of $\lambda_{\varphi}(\mu),$ that is, $\lim_{\gamma \to 0^+} \lambda'_{\varphi}(\gamma, \mu) = \lambda_{\varphi}(\mu).$

\begin{theorem}\label{thm:bd_ffhat_t_x}
    Let $(X,d)$ be a separable metric space without isolated points and consider two Lipschitz dynamical systems $\varphi, \hat{\varphi}$ on $X$. Suppose that $\hat{\mu}$ is an ergodic probability measure for $(X, {\hat{\varphi}})$ and that $\log L_{\varphi}(\cdot)$ and $\log (L_{\varphi}(\cdot) + \gamma_0)$ (for some $\gamma_0>0$) are $\hat{\mu}-$integrable. Then for any $\delta>0$ and any time horizon $T \in \mathbb{N}$ and $\hat{\mu}-$a.e. $x \in X$, there is some $\varepsilon(\delta, T, x)>0$, and some $R(\delta, x) \geq 1$ such that if $d_\infty(\varphi, \hat{\varphi}) < \varepsilon$ then
    \begin{equation}\label{eq:ffhat_error_bound}
        d(\varphi^t(x), \hat{\varphi}^t(x)) < d_\infty(\varphi,\hat{\varphi}) \left(1 + R e ^{t \left(\lambda_{{\hat{\varphi}}}^+(\hat{\mu})+ \delta\right)}\right) \text{ for all } t=0,\dots, T,    
    \end{equation}
    where $\lambda^+_{\hat{\varphi}}(\hat{\mu}) = \max \{ \lambda_{\hat{\varphi}}(\hat{\mu}), 0\}.$
\end{theorem}

Before proving Theorem \ref{thm:bd_ffhat_t_x}, we consider the following lemma.

\begin{lemma}\label{lmm:bd_sum_prod}
    Let $(X,d)$ be a separable metric space without isolated points and suppose $\varphi$ is a Lipschitz dynamical system on $X$ with ergodic measure $\mu.$  Suppose that $\log L_{\varphi}(\cdot)$ and $\log(L_{\varphi}(\cdot) + \gamma_0)$ (for some $\gamma_0>0$) are ${\mu}-$integrable. Then for any $\delta>0$ and $\mu-$a.e. $x \in X$ there exists $R(\delta, x)\geq 1$ and $0<\gamma<\gamma_0$ such that

    \begin{equation}
        \sum_{j=0}^{t-1} \prod_{i=j}^{t-1} (L_\varphi(\varphi^i(x)) + \gamma) < R e^{t (\lambda^+_{\varphi}(\mu) + \delta)} \quad \text{ for all } t \in \mathbb{N},
    \end{equation}
    where $\lambda^+_{\varphi}(\mu) = \max \{ \lambda_{\varphi}(\mu), 0\}.$
\end{lemma}

\noindent {\bf Proof.} For $0<\gamma<\gamma_0$ we define the quantity

\begin{equation}
    \lambda'_{\varphi}(\gamma, \mu) = \int_X \log( L_\varphi(x) + \gamma) d\mu(x).
\end{equation}

Fix $\delta>0$. By Birkhoff's Ergodic Theorem \ref{thm:birkhoff}, for $\mu-$a.e. $x \in X$ we have
\begin{equation}
    \lim_{t \to \infty} \frac{1}{t} \log \prod_{i=0}^{t-1} (L_\varphi(\varphi^i(x)) + \gamma) = \int_X \log (L_\varphi(u) + \gamma) d\mu(u) = \lambda'_{\varphi}(\gamma, \mu).
\end{equation}

Thus there exists $T(\delta,\gamma, x) \in \mathbb{N}$ such that 

\begin{equation}
    \left| \frac{1}{t} \log \prod_{i=0}^{t-1} (L_\varphi(\varphi^i(x)) + \gamma) - \lambda'_{\varphi}(\gamma, \mu) \right| < \delta/4 \quad \text{ for all } t \geq T.
\end{equation}

We may then pick $C(\delta, \gamma, x)\geq 1$ large enough such that

\begin{equation}
    \frac{1}{C}e^{t(\lambda'_{\varphi}(\gamma, \mu) - \delta/4)} < \prod_{i=0}^{t-1} (L_\varphi(\varphi^i(x)) + \gamma) < Ce^{t(\lambda'_{\varphi}(\gamma, \mu) + \delta/4)} \quad \text{ for all } t \in \mathbb{N}.
\end{equation}

From this we may bound $$\prod_{i=j}^{t-1} (L_\varphi(\varphi^i(x)) + \gamma) < C^2 e^{(t-j)\lambda'_{\varphi}(\gamma, \mu) + (t+j) \delta/4} \quad \text{ for } j=0,\dots,t-1.$$ If $\lambda'_{\varphi}(\gamma, \mu) > 0,$ then

\begin{equation}
    \sum_{j=0}^{t-1} \prod_{i=j}^{t-1} (L_\varphi(\varphi^i(x)) + \gamma) < C^2 e^{t (\lambda'_{\varphi}(\gamma, \mu) + \delta/2)} \sum_{j=0}^{t-1} e^{-j \lambda'_{\varphi}(\gamma, \mu)} < \frac{C^2}{1-e^{-\lambda'_{\varphi}(\gamma, \mu)}} e^{t (\lambda'_{\varphi}(\gamma, \mu) + \delta/2)}.
\end{equation}

On the other hand, if $\lambda'_{\varphi}(\gamma, \mu) \leq 0,$ then

\begin{equation}
    \sum_{j=0}^{t-1} \prod_{i=j}^{t-1} (L_\varphi(\varphi^i(x)) + \gamma) < C^2 e^{t \delta/4} \sum_{j=0}^{t-1} e^{j \delta/4} < C^2 e^{t \delta/4} \frac{e^{t \delta/4}}{e^{\delta/4}-1}   = \frac{C^2}{e^{\delta/4}-1} e^{t \delta/2}. 
\end{equation}

Now, by the dominated convergence theorem, $\lim_{\gamma \to 0^+} \lambda_{\varphi}'(\gamma, \mu) = \lambda_{\varphi}(\mu),$ so we can choose $0<\gamma<\gamma_0$ such that $\lambda_{\varphi}'(\gamma, \mu) < \lambda_{\varphi}(\mu) + \delta/2.$ The bound immediately follows. $\quad \blacksquare$

{\bf Proof of Theorem \ref{thm:bd_ffhat_t_x}.} Fix $\delta>0$ and $x \in X$. Take $0<\gamma<\gamma_0$ as supplied in Lemma \ref{lmm:bd_sum_prod}. For $t=0,\dots, T,$ let $\Delta x_t := d(\varphi^t(x), \hat{\varphi}^t(x)).$ Then, by definition of the pointwise Lipschitz constant $L_{\hat{\varphi}}(\cdot)$, there exist constants $\gamma_t(x, \gamma)>0$ such that if $\Delta x_t = d(\varphi^t(x), \hat{\varphi}^t(x))<\gamma_t$ then $d({\hat{\varphi}} \circ \varphi^t(x), {\hat{\varphi}} \circ \hat{\varphi}^t(x)) < \Delta x_t (L_{\hat{\varphi}}({\hat{\varphi}}^t(x)) + \gamma)$. Thus,

\begin{align}
    \Delta x_{t+1} &\leq d(\varphi \circ \varphi^t(x), {\hat{\varphi}} \circ {\varphi}^t(x)) + d({\hat{\varphi}} \circ {\varphi}^t (x), \hat{\varphi} \circ \hat{\varphi}^t(x))\\
    &\leq d_\infty(\varphi, \hat{\varphi}) + \Delta x_t(L_{\hat{\varphi}}({\hat{\varphi}}^t(x)) + \gamma) .
\end{align}

For any $t=0,\dots,T$, if $\Delta x_i < \gamma_i$ for $i=0,\dots,t-1$ then making use of Lemma \ref{lmm:bd_sum_prod} applied to ${\hat{\varphi}}$, we have

\begin{equation}
    d(\varphi^t(x), \hat{\varphi}^t(x)) = \Delta x_t \leq d_\infty(\varphi, \hat{\varphi}) \left( 1 + \sum_{j=0}^{t-1} \prod_{i=j}^{t-1} (L_{\hat{\varphi}}({\hat{\varphi}}^i(x)) + \gamma) \right)\\
    \leq d_\infty(\varphi, \hat{\varphi}) \left( 1 + R e ^{t \left(\lambda_{\hat{\varphi}}^+(\hat{\mu}) + \delta\right)}\right).
\end{equation}

Since $\varphi, \hat{\varphi}$ are Lipschitz, choosing some $\varepsilon(\delta, T, x)>0$ sufficiently small will ensure that if $d_\infty(\varphi, \hat{\varphi}) < \varepsilon$, then the inequalities $\Delta x_t < \gamma_t$ will hold for all $t=0,\dots, T-1,$ thus giving us the desired bound. $\quad \blacksquare$

\begin{remark}
    \normalfont
    \begin{enumerate}
        \item It is important to note that the bound in Theorem \ref{thm:bd_ffhat_t_x} is expressed in terms of the metric spatial Lyapunov {exponent of the {\bfi perturbed map} $\hat{\varphi}.$} Later we will take $\varphi$ to be an unknown function and $\hat{\varphi}$ a learnt approximation of it. Knowledge of $\hat{\varphi}$ yields knowledge of its Lyapunov exponent $\lambda_{\hat{\varphi}}( \hat{\mu}).$
        \item In Section \ref{subsec:metric_le_appendix} where the Kifer exponent is discussed in greater detail, the integrability condition \eqref{ineq:mu_integrability_cond} is required for the existence of the exponent. This condition also guarantees the integrability condition given in the statements of Theorem \ref{thm:bd_ffhat_t_x} and Lemma \ref{lmm:bd_sum_prod}. 
        \item Since $\hat{\varphi}$ is Lipschitz, say with Lipschitz constant $L$, we could easily derive the following bound using Lipschitz constants:
        \begin{equation}
            d(\varphi^t(x), \hat{\varphi}^t(x)) = \frac{L^t - 1}{L - 1} d_\infty(\varphi, \hat{\varphi}) \quad \text{ for all } t=0,1,2,\dots.
        \end{equation}
        While this bound is attractive in its simplicity and the fact that it holds for all $t \in \mathbb{N},$ its growth factor is $L$, which dominates the growth factor $e^{\lambda_{\hat{\varphi}}(\hat{\mu}) + \delta}$ of Theorem \ref{thm:bd_ffhat_t_x}. To be precise, for any $\gamma'>0$, there exists $\delta>0$ such that $e^{\lambda_{\hat{\varphi}}( \hat{\mu}) + \delta} < L+\gamma'.$ In practice, however we expect that $e^{\lambda_{\hat{\varphi}}( \hat{\mu}) +\delta}$ should be considerably smaller than $L$ since it relies on a pointwise Lipschitz constant which gives locally tight bounds on the expansion of the map.
        \item As has already been noted, $\lambda_{\hat{\varphi}}(\mu) \geq \Lambda_{\hat{\varphi}}(x)$, $\hat{\mu}-$a.e. Studying how sharp this bound is in practice is an open direction for further research.
    \end{enumerate}
\end{remark}

Appendix \ref{subsec:add_bd_without_mixing} contains an additional bound in the context of metric spaces with a structural stability assumption (see Theorem \ref{thm:metric_space_conjugate_maps}). That bound gives an exponential rate of separation governed by the Kifer's exponent $\Lambda_{\hat{\varphi}}$.

\begin{remark}\label{rem:usefulness of bound}
    \normalfont
    Some caution must be taken when interpreting the bounds in Theorems \ref{thm:bd_ffhat_t_x} and \ref{thm:metric_space_conjugate_maps}. The Lyapunov exponents $\Lambda_{\hat{\varphi}}$ and $\lambda_{\hat{\varphi}}$ are fundamentally related to the expansion properties of the ergodic map $\hat{\varphi}$ along trajectories as time tends to infinity. For $\Lambda_{\hat{\varphi}}$, this takes place through Kingman's subadditive theorem \cite{kingman1968ergodic} used in the proof of Theorem \ref{thm:kifer_exp_def}, while for $\lambda_{\hat{\varphi}}$ this appears in Lemma \ref{lmm:bd_sum_prod} through Birkhoff's ergodic theorem. The bounds we give translate these statements about {\it limits at infinity} to statements about expansion at a {\it finite time horizon}.

    The consequences of this are twofold. On the one hand, the ambiguity that results from converting a statement about a limit as time tends to infinity to a finite time statement is absorbed in the constants $R$ and $\delta$ in both bounds. The constant $\delta>0$ is equivalent to the `leakage rate' in \cite{berry2022learning}. It gives us `space' around the limiting value of the divergence rate, requiring that this divergence at a finite time be within a $\delta-$distance of this limiting value. The constant $R$ depends on the initial condition $x$ and the leakage rate $\delta$. It arises from the constant $C$ in the proof of Lemma \ref{lmm:bd_sum_prod}, which deals with the finite number of terms before the growth rate is sufficiently close to its limit. Thus it is independent of the time horizon $T$.
    
    On the other hand, the proofs of Lemmas \ref{lmm:bd_sum_prod} and \ref{lmm:metric_space_conjugate_maps} require that the separation between trajectories up to each time step be sufficiently small before the bound can be shown for the following time step. Thus the constant $\varepsilon$ depends not only on the initial condition and leakage rate, but also on the desired time horizon. This constant controls the neighborhood of $\varphi$ in which $\hat{\varphi}$ must be for the bound to hold. A larger time horizon may necessitate a smaller neighborhood so that the required separation bounds hold up to time $T-1$. Thus, the results should be interpreted as statements about the growth rate of the distance between trajectories {\it as long as they remain close}. Once the trajectories separate substantially, the proof no longer holds, though it is possible that in practice the bound might still be valid. This is a limitation of Theorem \ref{thm:bd_ffhat_t_x}, and indeed also of the similar bounds presented in \cite{berry2022learning, RC22}.
    
    We also remark that it remains to develop an algorithm to accurately approximate the two metric Lyapunov exponents used in our bounds, given the dynamical system map. This constitutes a direction for further investigation.

\end{remark}

The limitations noted in Remark \ref{rem:usefulness of bound} make way for the much better bounds in Section \ref{subsec:mixing_bounds}, which can be derived once mixing and related assumptions are added.

\vspace{1em}

{\bf Error bounds for density forecasting on $(\mathcal{P}(X), \varphi_*).$} We now take the bound on separation of trajectories for metric space dynamical systems and apply it to the dynamical system $(\mathcal{P}(X), \varphi_*).$ This brings us closer to the final application of these results for density forecasts that we will establish later on in Theorem \ref{thm:state_space_system_bd_metric_le_prob_space}. When $(X,d)$ is a separable metric space and $\varphi$ is a continuous dynamical system on $X$, $\varphi_*$ is a continuous dynamical system on the space $\mathcal{P}(X)$ endowed with the weak topology (see Appendix \ref{subsec:cts_maps_P(X)} for a self-contained presentation). As discussed in Appendix \ref{subsec:metrics on prob spaces}, various metrics may be used to metrize the weak topology on $\mathcal{P}(X).$ We denote the metric on this space by $d_{\mathcal{P}}$. Recall that the quantities $\Lambda_{\varphi}(\cdot)$ and $\lambda_{\varphi}(\cdot, \cdot)$ are related to an ergodic probability measure on the dynamical system. In this case we work with a probability measure $\Xi \in \mathcal{P}(\mathcal{P}(X))$ ergodic with respect to $\varphi_*.$ For $\Xi-$a.e. $\nu \in \mathcal{P}(X)$, the Kifer Lyapunov exponent is then

\begin{equation}\label{eq:kifer_exponent_measure_space}
    \Lambda_{\varphi_*}(\nu) = \lim_{\delta \to 0^+} \limsup_{t \to \infty} \frac{1}{t} \log \sup_{\hat{\nu} \in B_\nu (\delta, t)} \frac{d_{\mathcal{P}}(\varphi^t_*\nu, \varphi^t_*\hat{\nu})}{d_{\mathcal{P}}(\nu,\hat{\nu})}.    
\end{equation}

Similarly we define a pointwise Lipschitz constant for $\varphi_*:$

\begin{equation}
    L_{\varphi_*} (\nu) = \lim_{ \delta \to 0^+} \sup_{\hat{\nu} \in B_\nu(\delta, 0)} \frac{d_{\mathcal{P}} (\varphi_*\nu, \varphi_*\hat{\nu})}{d_{\mathcal{P}}(\nu, \hat{\nu})}.
\end{equation}

Finally, the spatial metric Lyapunov exponent for $(\mathcal{P}(X), \varphi_*)$ with ergodic probability measure $\Xi$ is

\begin{equation}
    \lambda_{\varphi_*}(\Xi) = \int_{\mathcal{P}(X)} \log L_{\varphi_*} (\nu)  d\Xi(\nu).
\end{equation}

\begin{theorem}\label{thm:bd_metric_le_prob_space}
    Let $(X,d)$ be a separable metric space without isolated points and consider two dynamical systems $\varphi, \hat{\varphi}$ on $X$. Equip the space $\mathcal{P}(X)$ with a metric $d_\mathcal{P}$ metrizing the weak topology and suppose that $\varphi_*, \hat{\varphi}_*$ are Lipschitz with respect to this metric. Let $\hat{\Xi}$ be an ergodic measure for $(\mathcal{P}(X), \hat{\varphi}_*)$ and suppose that $\log L_{\hat{\varphi}_*}(\cdot)$ and $\log (L_{\hat{\varphi}_*}(\cdot) + \gamma_0)$ (for some $\gamma_0>0$) are $\hat{\Xi}-$integrable. Then for any $ \delta>0$ and any time horizon $T \in \mathbb{N}$ and $\hat{\Xi}-$a.e. $\nu \in \mathcal{P}(X)$, there is some $\varepsilon(\delta, T, \nu)>0$, and some $R(\delta, \nu) \geq 1$ such that if $d_{\mathcal{P}, \infty}(\varphi_*, \hat{\varphi}_*) < \varepsilon$ then
    \begin{equation}
        d_{\mathcal{P}}(\varphi^t_*\nu, \hat{\varphi}^t_*\nu) < d_{\mathcal{P}, \infty}(\varphi_*, \hat{\varphi}_*) \left(1 + R e ^{t \left(\lambda_{\hat{\varphi}_*}^+( \hat{\Xi})+ \delta\right)}\right) \text{ for all } t=0,\dots, T,
    \end{equation}
    where $\lambda^+_{\hat{\varphi}_*}(\hat{\Xi}) = \max \{ \lambda_{\hat{\varphi}_*}( \hat{\Xi}), 0\}.$
\end{theorem}

\noindent {\bf Proof.} The dynamical systems $\varphi_*$ and $\hat{\varphi}_*$ on $\mathcal{P}(X)$ satisfy all the conditions of Theorem \ref{thm:bd_ffhat_t_x}. Thus, the result follows immediately. $\quad \blacksquare$

\medskip

At this point, we make two remarks. The first is on the relation between the Lipschitz condition on $\varphi$ and that on $\varphi_*$. Continuity of the map $\varphi$ guarantees continuity of $\varphi_*$ (see Lemma \ref{lmm:pushforward_cts_in_nu}). However, the Lipschitz property for $\varphi$ does not necessarily imply the same for $\varphi_*$.
%In a few cases, however, we can say more: For integral probability metrics such as the first two given in Section \ref{subsec:metrics on prob spaces} which metrize the strong topology on $\mathcal{P}(X),$ if $\varphi$ is merely continuous, $\varphi_*$ will be 1-Lipschitz. Thus for these metrics, $\varphi_*$ is nonexpanding. We, however, are concerned with the weak topology on $\mathcal{P}(X).$ 
In the case of the Wasserstein metric, Lemma \ref{lmm:wass_mumutilde_Lipschitz} shows that any Lipschitz constant for $\varphi$ is also a Lipschitz constant for $\varphi_*$. However, for the MMD metric which we use later in Section \ref{sec:numerics}, we cannot as easily make statements about the Lipschitz property for $\varphi_*$. See \cite{sriperumbudur_hilbert_space_embedding}[Theorem 21] for a comparison of MMD to Wasserstein distance, the Dudley metric and total variation distance. A similar question has to do with relating $d_\infty(\varphi, \hat{\varphi})$ to $d_{\mathcal{P}, \infty}(\varphi_*, \hat{\varphi}_*)$, the latter quantity being the uniform metric distance induced by $d_{\mathcal{P}}$ on maps in $C(\mathcal{P}(X), \mathcal{P}(X)).$ For the Wasserstein 1-distance, Corollary \ref{corr:wass_fftilde} tells us that $d_{\mathcal{P}, \infty}(\varphi_*, \hat{\varphi}_*) \leq d_\infty(\varphi, \hat{\varphi}).$ Beyond the Wasserstein metric, it is hard to make general comments.
Applying this to Theorem \ref{thm:bd_metric_le_prob_space}, if we choose $d_\mathcal{P}$ as the Wasserstein-1 metric on a bounded space $X$, then $d_{\mathcal{P}, \infty}(\varphi_*, \hat{\varphi}_*)$ may be replaced with $d_\infty(\varphi, \hat{\varphi})$.

The second remark concerns the ergodic measure $\Xi$ for the dynamical system $(\mathcal{P}(X), \varphi_*).$ When $X$ is compact and $\varphi$ is continuous, $\mathcal{P}(X)$ will be compact and $\varphi_*$ will be continuous. Thus, the Krylov-Bogolyubov Theorem (see, for example, \cite{Viana_Oliveira_2016_foundations_of_ergodic_theory}, Theorem 2.1) guarantees the existence of an ergodic measure $\Xi.$ Now, any invariant probability measure $\mu \in \mathcal{P}(X)$ is a fixed point of $\varphi_*,$ so the Dirac $\delta_\mu$ is ergodic for $\varphi_*$. We could go slightly further by noting that the inclusion $\iota \colon X \longrightarrow \mathcal{P}(X)$, $x \mapsto \delta_x$ provides a conjugacy between $(X, \varphi)$ and $(\iota(X), \varphi_*)$. For any $\varphi-$ergodic measure $\mu \in \mathcal{P}(X)$, its pushforward $\iota_* \mu$ is ergodic for $\varphi_*,$ supported on $\iota(X).$ If $\mu$ is nontrivial, so will $\iota_*\mu$ be. Another construction is as follows: suppose $X$ is compact and $\varphi$ is a homeomorphism of $X$. Take any non-$\varphi$-invariant measure $\mu$ (for example, if $X$ were Euclidean, we could consider taking a non-$\varphi$-invariant measure, absolutely continuous with respect to the Lebesgue measure) and consider its bi-infinite orbit $O = \{\varphi^t_* \mu \colon t \in \mathbb{Z}\}.$ 
The closure $\overline{O}$ is compact and $\varphi_*-$invariant. Hence there is an ergodic measure $\Xi \in \mathcal{P}(\mathcal{P}(X))$ supported on $\overline{O}.$ If $\varphi$ does not have a mixing or attracting measure, we expect that $\overline{O}$ will not contain an invariant measure of $\varphi$. Hence, an ergodic measure supported on this set is expected to be nontrivial. Investigating more generally what kind of ergodic measures exist for $(\mathcal{P}(X), \varphi_*)$ is an interesting research question by itself.

\subsection{Density forecasting error bounds using mixing and related assumptions.}\label{subsec:mixing_bounds}

Having observed the inherent exponential growth in bounds that do not rely on mixing assumptions, we now examine the differences that arise from applying such assumptions. {\bfi These assumptions will eventually enable us to prove the stability of density forecasts and, by extension, to learn the climate of the underlying system.} As we will see, the change in error bounds results from the fact that mixing and attracting measures are asymptotically stable fixed points of the pushforward dynamical system, within a suitable set of measures, while ergodic and physical measures possess a similar asymptotic limiting property, only in a Ces\`{a}ro sense. All our theorems require a structural stability assumption on the dynamical system. Structural stability is further discussed in Section \ref{subsec:stability}. In short, a $C^1$ map $\varphi$ is said to be ($C^1$) {\bfi structurally stable} if it has a neighborhood $U$ in the $C^1$ topology such that all maps $\hat{\varphi} \in U$ are topologically conjugate to $\varphi$. If in addition the conjugating homeomorphism may be made arbitrarily close to the identity map by decreasing the distance $d_1(\varphi, \hat{\varphi}),$ we say that $\varphi$ is ($C^1$) {\bfi strongly structurally stable.} Several weaker stability notions exist; in particular, {\bfi statistical stability} is closely related to our results \cite{ALVES_VIANA_2002_statistical_stability}. Statistical stability concerns how invariant measures vary continuously under perturbations of the dynamical system map, a key concept in this paper. For statistically stable systems, however, it is unclear whether properties such as mixing would be preserved under these perturbations. An in-depth discussion of this and other notions of stability is beyond the scope of this paper. Nevertheless, we note in Section \ref{sec:numerics} that the Lorenz system, despite violating the structural stability assumption, empirically still satisfies the outcomes of the theorems. We now formulate four conditions which we subsequently use in our theorems.

\medskip

{\bf Ergodic and physical measures.} The tuple $(X, \varphi, g, \mu, \nu)$, consisting of a dynamical system $(X,\varphi)$ with invariant probability measure $\mu$, a homeomorphism $g \colon X \longrightarrow X$ and a probability measure $\nu \in \mathcal{P}(X),$ is said to satisfy the conditions {\bf A1} and {\bf A2} whenever:
\begin{enumerate}
    \item[{\bf A1.}] if $\mu$ is an {\bfi ergodic} measure for $\varphi$, $\nu$ is absolutely continuous with respect to $\mu$ and $g^{-1}$ is nonsingular with respect to $\mu,$ and
    \item[{\bf A2.}] if $\mu$ is a {\bfi mixing} measure for $\varphi$, $\nu$ is absolutely continuous with respect to $\mu$ and $g^{-1}$ is nonsingular with respect to $\mu.$
\end{enumerate}

{\bf Mixing and attracting measures.} The tuple $(X, \varphi, g, \ell, \mu, \nu)$ consisting of a dynamical system $(X,\varphi)$ with Borel reference measure $\ell$, invariant probability measure $\mu$, a homeomorphism $g \colon X \longrightarrow X$ and a probability measure $\nu \in \mathcal{P}(X),$ is said to satisfy the conditions {\bf {\rm {\bf B1}}} and {\bf B2} whenever:
\begin{enumerate}
    \item[{\bf B1.}] if $\mu$ is a {\bfi physical} measure for $\varphi$, with open neighborhood $U$ of its support as defined in Section \ref{subsec:ergodicity}, $\nu$ is absolutely continuous with respect to $\ell$ with $\nu(U \cap g(U)) = 1$, and $\varphi, g, g^{-1}$ are nonsingular with respect to $\ell,$ and
    \item[{\bf B2.}] if $\mu$ is an {\bfi attracting} measure for $\varphi$, with open neighborhood $U$ of its support  as defined in Section \ref{subsec:ergodicity}, $\nu$ is absolutely continuous with respect to $\ell$ with $\nu(U \cap g(U)) = 1$ and $\varphi, g, g^{-1}$ are nonsingular with respect to $\ell.$
\end{enumerate}

\begin{remark}\label{rem:assumptions_validity}
    \normalfont
    \begin{enumerate}
        \item The requirement that $\varphi, g, g^{-1}$ be nonsingular with respect to $\ell$ will be met if $\varphi^{-1}, g, g^{-1}$ are absolutely continuous functions \cite{Bog07}[Chapter 9.9]. For the definition of nonsingular maps, see Section \ref{subsec:ergodicity}.
        \item According to Proposition \ref{prop:phys_att_alternate_def}, if $\supp (\mu)$ is an attractor, then we may take $U = \Basin_\varphi(\supp(\mu))$ in conditions {\bf B1} and {\bf B2}. 
        \item An important question is whether the set $U \cap g(U)$ is empty, or possibly has measure zero under the reference measure $\ell $. In what follows, we take $g$ to be the conjugating homeomorphism for two nearby maps $\varphi$ and $\hat{\varphi}$ arising from a structural stability assumption on $\varphi$. Since the maps $\varphi$ and $\hat{\varphi}$ are close, it is reasonable to expect that $g$ will be close to the identity map. Indeed, if $\varphi$ is strongly structurally stable, this can be guaranteed when $\hat{\varphi}$ is a sufficiently accurate approximant of $\varphi$. Under these circumstances, since $U$ is open, it can be ensured that $U \cap g(U)$ is nonempty. Since the intersection is also open, if we take $\ell$ to be the Riemannian volume, this set will have positive $\ell-$measure.
    \end{enumerate}
\end{remark}

We now proceed by outlining some consequences of the existence of an ergodic or physical measure. As noted in the Appendix \ref{subsec:ds, erg, mixing}, any continuous dynamical system on a compact space possesses an ergodic measure. The existence of a physical measure is more subtle to prove. The original works \cite{Ruelle1976AMA, Bowen2004} proved its existence for Axiom A systems. Subsequently, the existence of physical measures has been demonstrated in more general settings \cite{Pesin_Sinai_1982, COWIESON_YOUNG_2005, alves2015srbmeasurespartiallyhyperbolic}. In Theorem \ref{thm:ergodicity_structurally_stable_bound_main} we see that for a sufficiently regular initial distribution, when transported under a perturbed dynamical system, its trajectory in the space of probability measures remains close to the true trajectory, in a Ces\`{a}ro sense.

\begin{theorem}[Convergence for ergodic and physical measures.]\label{thm:ergodicity_structurally_stable_bound_main}
    Let $(X,d)$ be a smooth compact manifold and let $\varphi \in \operatorname{Diff}^1(X)$ be a $C^1$ structurally stable map. Thus there exists $\delta>0$ such that for $\hat{\varphi} \in \operatorname{Diff}^1(X)$ if $d_1(\varphi, \hat{\varphi})<\delta$, then $\varphi$ is topologically conjugate to $\hat{\varphi}$ by a homeomorphism $g \colon X \longrightarrow X$, that is $g \circ \varphi = \hat{\varphi} \circ g$. Fix a reference Borel probability measure $\ell$ on $X$, and let $\mu \in \mathcal{P}(X)$ be an invariant measure for $\varphi$ and $\nu \in \mathcal{P}(X)$ any other measure such that condition {\rm {\bf A1}} or {\rm {\bf B1}} is fulfilled. Then
    
    \begin{equation}
        \lim_{T \to\infty} d_{\mathcal{P}}\left(\frac{1}{T} \sum_{t=0}^{T-1} \varphi^t_* \nu, \frac{1}{T} \sum_{t=0}^{T-1} \hat{\varphi}^t_* \nu \right) = d_{\mathcal{P}}(\mu, g_*\mu).    
    \end{equation}
    
    If $\varphi$ is {\bfi strongly structurally stable}, $d_{\mathcal{P}}(\mu, g_*\mu)$ can be made arbitrarily small by decreasing $\delta$.
\end{theorem}

\noindent {\bf Proof.} We deal first with the case {\rm {\bf A1}}. We note that, by Lemma \ref{lmm:struct_stab_pres_mixing_erg} (i), $(X, \hat{\varphi}, \hat{\mu})$ is ergodic where $\hat{\mu} = g_* \mu$. Now, since $g^{-1}$ is nonsingular with respect to $\mu$, and since $\mu = g^{-1}_* \hat{\mu},$ $\nu$ is absolutely continuous with respect to both $\mu$ and $\hat{\mu}.$ Thus by Corollary \ref{cor:ergodicity_convergence},

\begin{equation}
    \lim_{T \to \infty} \frac{1}{T} \sum_{t=0}^{T-1}  \varphi^t_*\nu = \mu \text{ and } 
    \lim_{T \to \infty} \frac{1}{T} \sum_{t=0}^{T-1}  \hat{\varphi}^t_*\nu = \hat{\mu}. \label{eq:cesaro_lim}
\end{equation}

On the other hand, if condition {\rm {\bf B1}} is fulfilled, then by Lemma \ref{lmm:struct_stab_pres_att_phys} (i), $\hat{\mu}$ is a physical measure for $\hat{\varphi}$ with neighborhood $g(U)$ of its support as defined in Section \ref{subsec:ergodicity}. Furthermore, $\nu(U) = 1$ and $\nu(g(U)) =1.$ Thus by Lemma \ref{lmm:physical_convergence}, \eqref{eq:cesaro_lim} also holds under condition {\rm {\bf B1}}. In particular, by continuity of the metric $d_{\mathcal{P}}$, in both cases {\rm {\bf A1}} and {\rm {\bf B1}},
\begin{equation}
    \lim_{T \to \infty} d_{\mathcal{P}} \left(\frac{1}{T} \sum_{t=0}^{T-1} \varphi^t_*\nu, \frac{1}{T} \sum_{t=0}^{T-1} \hat{\varphi}^t_*\nu \right) 
% \leq \limsup_{t \to \infty} d_{\mathcal{P}} \left(\frac{1}{T} \sum_{t=0}^{T-1} \varphi^t_*\nu, \mu \right) + d_{\mathcal{P}}(\mu, \hat{\mu}) + d_{\mathcal{P}} \left(\hat{\mu}, \frac{1}{T} \sum_{t=0}^{T-1} \hat{\varphi}^t_*\nu \right) 
= d_{\mathcal{P}} (\mu, \hat{\mu}).
\end{equation}
If $\varphi$ is strongly structurally stable, continuity of the map $ g \mapsto g_* \mu$ (Lemma \ref{lmm:pushforward_cts_in_f}) means that $d_{\mathcal{P}}(\mu, g_*\mu) \longrightarrow 0$ as $\delta \longrightarrow 0. \quad \blacksquare$

\begin{remark}
    \normalfont
    If $d_{\mathcal{P}}$ is the Wasserstein-1 metric, by Corollary \ref{corr:wass_fftilde} we furthermore have $d_{\mathcal{P}}(\mu, g_*\mu) \leq d_\infty(id, g).$
\end{remark}

Finally, we consider mixing and attracting measures. The intuition behind the proof is simple: a sufficiently regular initial distribution will approach the mixing (respectively attracting) measure of the true system when transported by it, and likewise, when transported by the perturbed dynamical system, it approaches the mixing (respectively attracting) measure of the perturbed system. Since these measures are close, the limiting distributions are also close.

\begin{theorem}[Convergence for mixing and attracting measures.]\label{thm:strong_mixing_structurally_stable_bound_main}
    Let $(X,d)$ be a smooth compact manifold and let $\varphi \in \operatorname{Diff}^1(X)$ be a $C^1$ structurally stable map. Thus there exists $\delta>0$ such that for $\hat{\varphi} \in \operatorname{Diff}^1(X)$ if $d_1(\varphi, \hat{\varphi})<\delta$, then $\varphi$ is topologically conjugate to $\hat{\varphi}$ by a homeomorphism $g \colon X \longrightarrow X$, that is $g \circ \varphi = \hat{\varphi} \circ g$. Fix a reference Borel probability measure $\ell$ on $X$, and let $\mu \in \mathcal{P}(X)$ be an invariant measure for $\varphi$ and $\nu \in \mathcal{P}(X)$ any other measure such that condition {\rm {\bf A2}} or {\rm {\bf B2}} is fulfilled. Then
    
    \begin{equation}
        \lim_{t \to\infty} d_{\mathcal{P}}(\varphi^t_* \nu, \hat{\varphi}^t_* \nu) = d_{\mathcal{P}} (\mu, g_* \mu).
    \end{equation}

    If $\varphi$ is {\bfi strongly structurally stable}, $d_{\mathcal{P}} (\mu, g_* \mu)$ may be made arbitrarily small by decreasing $\delta$.
\end{theorem}

\noindent {\bf Proof.} We deal first with the case when {\rm {\bf A2}} is fulfilled. By Lemma \ref{lmm:struct_stab_pres_mixing_erg} (ii), $(X, \hat{\varphi}, \hat{\mu})$ is mixing, where $\hat{\mu} = g_* \mu.$ As in the proof of Theorem \ref{thm:ergodicity_structurally_stable_bound_main}, $\nu$ is absolutely continuous with respect to both $\mu$ and $\hat{\mu}.$  In particular, by Proposition \ref{prop:strong_mixing_convergence},

\begin{equation}
    \varphi^t_* \nu \longrightarrow \mu \text{ and } \hat{\varphi}^t_* \nu \longrightarrow \hat{\mu} \text{ as } t \longrightarrow \infty.\label{eq::conv_lim}
\end{equation}

On the other hand, if condition {\rm {\bf B2}} is fulfilled, $\mu$ is an attracting measure for $\varphi$ and by Lemma \ref{lmm:struct_stab_pres_att_phys} (ii), so is $\hat{\mu}$ for $\hat{\varphi}$. Since $\nu(U \cap g(U)) = 1$, \eqref{eq::conv_lim} again holds. Thus in both cases

\begin{equation}
    \lim_{t \to\infty} d_{\mathcal{P}}(\varphi^t_* \nu, \hat{\varphi}^t_* \nu)
    % \leq \limsup_{t \to\infty} d_{\mathcal{P}}(\varphi^t_* \nu, \mu) + d_{\mathcal{P}} (\mu, \hat{\mu}) + d_{\mathcal{P}}(\hat{\mu}, \hat{\varphi}^t_* \nu) 
    = d_{\mathcal{P}}(\mu, \hat{\mu}).
\end{equation}

If $\varphi$ is strongly structurally stable, the continuity of the map $ g \mapsto g_* \mu$ (Lemma \ref{lmm:pushforward_cts_in_f}) means that $d_{\mathcal{P}}(\mu, g_*\mu) \longrightarrow 0$ as $\delta \longrightarrow 0. \quad \blacksquare$

\medskip

We conclude this section by briefly discussing the genericity of the assumptions made in Theorems  \ref{thm:ergodicity_structurally_stable_bound_main} and \ref{thm:strong_mixing_structurally_stable_bound_main}. Structural stability is one of the stronger notions of stability; as noted earlier, many weaker notions exist. Nevertheless, structural stability holds for a well-studied class of dynamical systems. By the Smale-Palis Theorem for $C^1$ diffeomorphisms on a compact manifold, structural stability is equivalent to Axiom A plus the so-called strong transversality condition. (For a discussion of these notions, see \cite{anosov_DS_with_Hyp_behaviour}). In particular, every Anosov diffeomorphism is structurally stable. As discussed in Section \ref{subsec:ds, erg, mixing}, every continuous dynamical system on a compact manifold will have an ergodic measure. Mixing is a stronger condition, yet many well-known systems still satisfy it, including the Lorenz system.
%The set of Anosov diffeomorphisms that is mixing is open and dense in the set of all Anosov diffeomorphisms [cite Katok and Hasselblatt introduction to modern theory of DSs, check this reference].
The existence of physical and attracting measures is complex. Much study has been conducted into the existence of SRB measures, which combine both properties. For a discussion on this see \cite{young_whataresrbmeasures}.

\section{Climate forecasting with state-space systems.}\label{sec:main_result}

In this section, we formulate our main results which {\bfi provide a mathematical foundation for the stability of learning the climate, and more generally of making density forecasts, with state-space systems.} The setup is as in Section \ref{subsec:learning with state-space systems}. We work with a dynamical system $\phi \colon \mathcal{M} \longrightarrow \mathcal{M}$. Data is supplied to us through the observation function $\omega \colon \mathcal{M} \longrightarrow \mathcal{U}$ and learning occurs by means of a state map $f \colon \mathcal{X} \times \mathcal{U} \longrightarrow \mathcal{X}$. Throughout this section, we assume that the state map is differentiable. We further assume the existence of an injective generalized synchronization $\zeta_{(\phi, \omega, f)} \colon \mathcal{M} \longrightarrow \mathcal{X}$. Here we abbreviate it simply as $\zeta$ and we write $\widetilde{\mathcal{X}} := \zeta(\mathcal{M}) \subseteq \mathcal{X}.$ This guarantees the existence of a readout $h = \omega \circ \phi \circ \zeta^{-1} \colon \widetilde{\mathcal{X}} \longrightarrow \mathcal{U}$ such that the dynamical system $\Phi \colon \widetilde{\mathcal{X}} \longrightarrow \widetilde{\mathcal{X}},\,\, x \mapsto f({x}, h({x}))$ is conjugate to $\phi$ by $\zeta,$ that is, $\Phi \circ \zeta = \zeta \circ \phi.$ We employ our trained state-space system with approximation $\hat{h}$ (respectively ${\hat{\Phi}}$) of $h$ (respectively $\Phi$) to forecast probability distributions in the observation space.

By definition of the readout $h$, $\Phi$ maps $\widetilde{\cal X}$ into $\widetilde{\cal X}.$ An important question is whether this still holds true when we use the approximation $\hat{h}$ of $h$, that is, do we have $f(x,\hat{h}(x)) \in \widetilde{\cal X}$ for all $x \in \widetilde{\cal X}$? In \cite{McGehee2014SomeMP}, McGehee introduces the quantity

\begin{equation}
    \beta_{\Phi}(B) = \sup \{ \varepsilon\colon N_\varepsilon(\Phi(B)) \subseteq B \},
\end{equation}

for a set $B \subseteq {\cal X}$, where $N_\varepsilon(\Phi(B))$ denotes the $\varepsilon-$neighborhood of $\Phi(B).$ The quantity $\beta_{\Phi}(B)$ measures how `far' $\Phi$ maps $B$ into itself. Now, in the terminology of McGehee, any attractor $A$ of $\Phi$ possesses an attractor block, that is,  a nonempty compact set $B$ such that $\Phi(B) \subseteq B^\circ$, where $B^\circ$ denotes the interior of $B$. Moreover, for any attractor block $B$, $\beta_{\Phi}(B)>0.$ Choose $\mathcal{M}$ to be an attractor block for $\phi$. Then $\widetilde{\cal X} = \zeta({\cal M})$ is an attractor block for $\Phi$.  As noted at the beginning of Section \ref{sec:stability_dyn_sys_probability}, universality of the state-space family being used allows us to make the uniform approximation errors $d_\infty(\Phi, \hat{\Phi})$ and $d_\infty(h, \hat{h})$ arbitrarily small. In particular, if $d_\infty(\Phi, \hat{\Phi})< \beta_{\Phi}(\widetilde{\cal X}),$ then $\hat{\Phi}(\widetilde{\cal X}) \subseteq \widetilde{\cal X}.$ A slight difficulty that remains is that \cite{McGehee2014SomeMP} works in the setting of metric spaces. It is not clear whether every attractor possesses an attractor block {\it which is a manifold}, as required in Theorem \ref{thm:transport_of_measures_with_state_space_system_bound}.
In Theorem \ref{thm:transport_of_measures_with_state_space_system_bound} we require $\hat{\Phi}$ to be close to $\Phi$ in a stronger $C^1$ sense. Remark \ref{rem:final_thm} shows that, for a readout with a single hidden layer, $d_1(h, \hat{h}),$ and hence $d_1(\Phi, \hat{\Phi}),$ may also be made arbitrarily small.

%Theoretically, forecasting takes place by means of the pushforward map $\hat{\Phi}_*$ of the learnt dynamical system. In practice, one method of implementing this is by approximating the distributions above with an ensemble of points and then applying the forecasting strategy \eqref{eq:SS_1} and \eqref{eq:SS_2} to each point, thus transporting the distribution forward.
In what follows, we see how the approximation errors in learning $h$ relate to the errors in density forecasts. To measure the distance between forecasted and true distributions, we select metrics $d_{\mathcal{P}, \mathcal{M}}, d_{\mathcal{P}, \mathcal{U}},$ and $d_{\mathcal{P}, \widetilde{\mathcal{X}}}$ compatible with the weak topology on the spaces $\mathcal{P}(\mathcal{M}), \mathcal{P}(\mathcal{U})$ and $\mathcal{P}(\widetilde{\mathcal{X}})$, respectively. An initial probability measure $\mu_0 \in \mathcal{P}(\mathcal{M})$ is transported to $\mu_t = \phi^t_* \mu_0$ under the action of the underlying dynamical system. Correspondingly in the state space $\xi_0 = \zeta_* \mu_0 \in \mathcal{P}(\widetilde{\mathcal{X}})$ is transported to $\xi_t = \Phi^t_*\zeta_* \mu_0$ and $\hat{\xi}_t = \hat{\Phi}^t_*\zeta_* \mu_0$ under the true and proxy systems, respectively. In the observation space, the proxy system predicts $\hat{\nu}_t = \hat{h}_* \hat{\Phi}^{t-1}_* \zeta_* \mu_0$ as opposed to the true distribution $\nu_t = h_* \Phi^{t-1}_* \zeta_* \mu_0 = \omega_* \phi^t_* \mu_0.$ Using the theory of Section \ref{sec:stability_dyn_sys_probability}, we now proceed to bound the error of such predictions.

\subsection{Predictions without mixing and related assumptions.}\label{subsec:main_result_exponential}

We begin by showing that when we do not make use of mixing and related assumptions, the exponential bound of Theorem \ref{thm:bd_metric_le_prob_space} carries over to the state-space system learning setup. (For an additional bound when a structural stability assumption is added, we refer the reader to Theorem \ref{thm:state_space_system_bd_metric_le_prob_space_conjugate_maps}.) For the following result we take the metrics on $\mathcal{P}(\widetilde{\mathcal{X}})$ and $\mathcal{P}(\mathcal{U})$ to be the Wasserstein-1 distance. Note the similarity between this bound and that given in \cite{berry2022learning}[Theorem 4.2] and \cite{RC22}[Theorem 3.11] for pointwise forecasts.

\begin{theorem}\label{thm:state_space_system_bd_metric_le_prob_space}
    In the notation introduced above, suppose that $(\widetilde{\mathcal{X}},d)$ is a separable metric space without isolated points and $\Phi, \hat{\Phi}$ are two Lipschitz dynamical systems on $\widetilde{\mathcal{X}}$. Assume also that the learnt readout $\hat{h}$ is Lipschitz. Let $L_z$ be a Lipschitz constant for the second coordinate of the state map $f$, and $L_{\hat{h}}$ a Lipschitz constant for $\hat{h}$. Take $d_{\mathcal{P}, \widetilde{\mathcal{X}}}$ and $d_{\mathcal{P}, \mathcal{U}}$ to be Wasserstein-1 distances on their respective spaces. Let $\hat{\Xi}$ be an ergodic measure for the dynamical system $(\mathcal{P}(\widetilde{\mathcal{X}}), \hat{\Phi}_*)$ and suppose that $\log L_{\hat{\Phi}_*}(\cdot)$ and $\log (L_{\hat{\Phi}_*}(\cdot) + \gamma_0)$ (for some $\gamma_0>0$) are $\hat{\Xi}-$integrable. Then for any $\delta>0$ and any time horizon $T \in \mathbb{N}$ and $\hat{\Xi}-$a.e. $\xi_0 = \zeta_* \mu_0 \in \mathcal{P}(\widetilde{\mathcal{X}})$, there is some $\varepsilon(\delta, T, \xi_0)>0$, and some $R(\delta, \xi_0) \geq 1$ such that if $d_{\infty}(h, \hat{h}) < \varepsilon$ then
    \begin{equation}
        d_{\mathcal{P}, \mathcal{U}}(\nu_t, \hat{\nu}_t) < d_\infty(h, \hat{h}) \left( 1 + L_{\hat{h}} L_z \left( 1 + R e ^{(t-1) \left(\lambda_{\Hat{\Phi}_*}^+(\hat{\Xi})+ \delta\right)}\right) \right) \text{ for all } t=1,\dots, T,
    \end{equation}
    where $\lambda^+_{\hat{\Phi}_*}(\hat{\Xi}) = \max \{ \lambda_{\hat{\Phi}_*}( \hat{\Xi}), 0\}.$
\end{theorem}

% \tod{As we discussed, when we are done with this project we have to investigate the question of how to satisfy the condition $d_{\infty}(h, \hat{h}) < \varepsilon$ with $\hat{h} $ linear in the ESN and SAS case. This is a very important issue. Please make a note.}

\noindent {\bf Proof.} Since $\Phi, \hat{\Phi}$ are Lipschitz and $d_{\mathcal{P}, \widetilde{\mathcal{X}}}$ is the Wasserstein-1 distance, $\Phi_*, \hat{\Phi}_*$ are also Lipschitz. By Theorem \ref{thm:bd_metric_le_prob_space}, for $\hat{\Xi}-$a.e. $\xi_0 = \zeta_* \mu_0 \in \mathcal{P}(\widetilde{\mathcal{X}})$, there is some $\varepsilon'(\delta,  T, \xi_0)>0$, and some $R(\delta, \xi_0) \geq 1$ such that if $d_{\mathcal{P}, \widetilde{\mathcal{X}}, \infty}(\Phi_*, \hat{\Phi}_*) < \varepsilon'$ then

\begin{equation}
    d_{\mathcal{P}, \widetilde{\mathcal{X}}}(\xi_t, \hat{\xi_t}) = d_{\mathcal{P},  \widetilde{\mathcal{X}}} ( \Phi^t_* \xi_0, \hat{\Phi}^t_* \xi_0) < d_{\mathcal{P}, \widetilde{\mathcal{X}}, \infty} ( \Phi_*, \hat{\Phi}_*) \left( 1 + R e ^{t\left(\lambda_{\Hat{\Phi}_*}^+( \hat{\Xi})+ \delta\right)}\right) \text{ for all } t=0,\dots, T.
\end{equation}

Here $ d_{\mathcal{P}, \widetilde{\mathcal{X}}, \infty} (\Phi_*, \hat{\Phi}_*)= \sup_{\xi \in \mathcal{P}(\widetilde{\cal X})} d_{{\cal P}, \widetilde{\cal X}} (\Phi_* \xi, \hat{\Phi}_* \xi).$ Now, since we use Wasserstein-1 distance, $d_{\mathcal{P},  \widetilde{\mathcal{X}}, \infty} (\Phi_*, \hat{\Phi}_*) \leq d_\infty (\Phi, \hat{\Phi}) \leq L_z d_\infty(h, \hat{h}).$ Thus taking $\varepsilon := \varepsilon'/L_z$ guarantees that if $d_\infty(h, \hat{h})< \varepsilon$ then $d_{\mathcal{P},  \widetilde{\mathcal{X}}} (\xi_t, \hat{\xi_t}) < L_z d_{\infty} ( h, \hat{h}) \left( 1 + R e ^{t\left(\lambda_{\Hat{\Phi}_*}^+( \Xi)+ \delta\right)}\right)$ for all $t=0,\dots, T.$ Next note that $d_{\mathcal{P}, \mathcal{U}} (\nu_t, \hat{\nu_t}) = d_{\mathcal{P}, \mathcal{U}} (h_* \xi_{t-1}, \hat{h}_* \hat{\xi}_{t-1}) \leq d_{\mathcal{P}, \mathcal{U}} (h_* \xi_{t-1}, \hat{h}_* \xi_{t-1}) + d_{\mathcal{P}, \mathcal{U}} (\hat{h}_* \xi_{t-1}, \hat{h}_* \hat{\xi}_{t-1}).$ Since $d_{\mathcal{P}, \mathcal{U}}$ is the Wasserstein-1 distance, $d_{\mathcal{P}, \mathcal{U}, \infty} (h_*, \hat{h}_*) \leq d_\infty(h, \hat{h})$ and a Lipschitz constant for $\hat{h}_*$ is $L_{\hat{h}}.$ This completes the proof. $\quad \blacksquare$

\begin{remark}
    \normalfont
    In general for any metrics $d_{\mathcal{P}, \widetilde{\mathcal{X}}}$ and $d_{\mathcal{P}, \mathcal{U}}$ metrizing the weak topologies on $\mathcal{P}(\widetilde{\mathcal{X}})$ and $\mathcal{P}(\mathcal{U}),$ by Lemmas \ref{lmm:pushforward_cts_in_f} and \ref{lmm:pushforward_cts_in_nu}, the maps $\xi \mapsto d_{\mathcal{P}}(h_* \xi, \hat{h}_* \xi)$ and $(\xi, \hat{\xi}) \mapsto d_{\mathcal{P}}(\hat{h}_* \xi, \hat{h}_* \hat{\xi})$ will be continuous. If we take $\widetilde{\mathcal{X}}$ compact, they are bounded and so $d_{\mathcal{P}}(\nu_t, \hat{\nu}_t)$ is also bounded. But this is not very useful. Using the Wasserstein-1 distance allows the exponential separation of Theorem \ref{thm:bd_metric_le_prob_space} to carry through to the observation space.
\end{remark}

\subsection{Predictions with mixing and related assumptions.}\label{subsec:main_result_mixing}

Using mixing and related assumptions, the results of Section \ref{subsec:mixing_bounds} allow us to derive {\bfi time-uniform bounds for density predictions using state-space systems.} In addition to the results presented here, we refer the reader to the Appendix \ref{subsec:stab_empirical_measure} in which we derive a result for the stability of empirical measures under perturbations of the dynamical system map. In particular, when the underlying system possesses an ergodic or physical measure whose support is an attractor, Theorem \ref{thm:trajectory_average_with_state_space_system_bound} and Remark \ref{rem:stab_emp_measure} show that for a large set of initial conditions the empirical measures calculated using the proxy system are accurate approximations of this measure. When training state-space systems, trajectory averages of observables are computed, and their distributions are compared with those of the true data to assess whether successful learning has occurred. In the terms Lorenz used in \cite{lorenz1964problem}, we seek to know whether our system has {\bfi accurately replicated the climate of the underlying system}. Theorem \ref{thm:trajectory_average_with_state_space_system_bound} provides a theoretical justification for this heuristic showing that, under these conditions, {\bfi the climate remains stable under approximation of the system map.} 

We now state our main result. Recall the conditions {\rm {\bf A1}} (ergodic measure) and {\rm {\bf B1}} (physical measure), {\rm {\bf A2}} (mixing measure) and {\rm {\bf B2}} (attracting measure). The latter pair of conditions yields bounds on the convergence of distributions transported under the true and proxy systems, while the former pair yields the same in a Ces\`{a}ro sense. Thus, under these conditions, {\bfi density forecasts remain stable at arbitrary time horizons}.

\begin{theorem}\label{thm:transport_of_measures_with_state_space_system_bound}
    Suppose that $(\widetilde{\mathcal{X}}, d)$ is a smooth compact manifold, and $\Phi \in \operatorname{Diff}^1(\widetilde{\mathcal{X}})$ is a $C^1$ structurally stable map. Thus there exists $\delta>0$ such that if $\hat{\Phi} \in \operatorname{Diff}^1(\widetilde{\mathcal{X}})$, and $d_1(\Phi, \hat{\Phi}) < \delta$, then $\Phi$ is topologically conjugate to $\hat{\Phi}$ by a homeomorphism $g \colon \widetilde{\mathcal{X}} \longrightarrow \widetilde{\mathcal{X}}$, that is $ g \circ \Phi = \hat{\Phi} \circ g.$ Fix a reference Borel probability measure $\ell$ on $\widetilde{\mathcal{X}}$, and let $ \xi \in \mathcal{P}(\widetilde{\mathcal{X}})$ be an invariant probability measure for $\Phi$. 

\medskip
    
    {\bf Ergodic and physical measures.} Suppose that $\xi_0 = \zeta_* \mu_0 \in \mathcal{P}(\widetilde{\mathcal{X}})$ is such that the tuple $(\widetilde{\mathcal{X}}, \Phi, g, \xi, \xi_0)$ satisfies condition {\rm {\bf A1}} or the tuple $(\widetilde{\mathcal{X}}, \Phi, g, \ell, \xi, \xi_0)$ satisfies condition {\rm {\bf B1}}. Then
     
    \begin{equation}
        \lim_{t \to\infty} d_{\mathcal{P}, \mathcal{U}}\left(\frac{1}{T} \sum_{t=0}^{T-1} \nu_t, \frac{1}{T} \sum_{t=0}^{T-1} \hat{\nu}_t \right) = d_{{\cal P}, {\cal U}} (h_* \xi, \hat{h}_* g_*{\xi}) \leq d_{\mathcal{P}, \mathcal{U}}(h_* \xi, \hat{h}_* \xi) + d_{\mathcal{P}, \mathcal{U}}(\hat{h}_* \xi, \hat{h}_* g_* \xi).
    \end{equation}
    
\medskip

    {\bf Mixing and attracting measures.} If the tuple $(\widetilde{\mathcal{X}}, \Phi, g, \xi, \xi_0)$ satisfies condition {\rm {\bf A2}} or the tuple $(\widetilde{\mathcal{X}}, \Phi, g, \ell, \xi, \xi_0)$ satisfies condition {\rm {\bf B2}}, then

    \begin{equation}
        \lim_{t \to \infty} d_{\mathcal{P}, \mathcal{U}}(\nu_t, \hat{\nu}_t) = d_{{\cal P}, {\cal U}} (h_* \xi, \hat{h}_* g_*{\xi}) \leq d_{\mathcal{P}, \mathcal{U}}(h_* \xi, \hat{h}_* \xi) + d_{\mathcal{P}, \mathcal{U}}(\hat{h}_* \xi, \hat{h}_* g_* \xi).
    \end{equation}

    If $\Phi$ is strongly structurally stable, then the bound may be made arbitrarily small as $d_1(h, \hat{h}) \longrightarrow 0.$
    
\end{theorem}

\noindent {\bf Proof.} We deal with the case where conditions {\rm {\bf A2}} or {\rm {\bf B2}} are fulfilled; the proof for the other case is similar, making use of Theorem \ref{thm:ergodicity_structurally_stable_bound_main}. Recall that the proof of Theorem  \ref{thm:strong_mixing_structurally_stable_bound_main} shows that when {\rm {\bf A2}} or {\rm {\bf B2}} are fulfilled $\Phi^t_* \xi_0 \longrightarrow \xi$ and $\hat{\Phi}^t_* \xi_0 \longrightarrow \hat{\xi},$ where $\hat{\xi} = g_* \xi,$ as $t \longrightarrow \infty.$ Thus by Lemma \ref{lmm:pushforward_cts_in_nu},

\begin{align}
    \lim_{t \to\infty} d_{\mathcal{P}, \mathcal{U}}(\nu_t, \hat{\nu}_t) 
    &= \lim_{t \to \infty} d_{\mathcal{P}, \mathcal{U}}(h_*\Phi^{t-1}_* \xi_0, \hat{h}_* \hat{\Phi}^{t-1}_* \xi_0) \\
    % &\leq \limsup_{t \to \infty} d_{\mathcal{P}, \mathcal{U}}(h_*\Phi^{t-1}_* \xi_0, h_* \xi) + d_{\mathcal{P}, \mathcal{U}}(h_* \xi, \hat{h}_* \hat{\xi}) + d_{\mathcal{P}, \mathcal{U}}(\hat{h}_* \hat{\xi}, \hat{h}_* \hat{\Phi}^{t-1}_* \xi_0) \\
    &= d_{{\cal P}, {\cal U}} (h_* \xi, \hat{h}_* \hat{\xi})
    %&\leq d_{\mathcal{P}, \mathcal{U}} (h_* \xi, \hat{h}_* g_* \xi) 
    \leq d_{\mathcal{P}, \mathcal{U}}(h_* \xi, \hat{h}_* \xi) + d_{\mathcal{P}, \mathcal{U}}(\hat{h}_* \xi, \hat{h}_* g_* \xi).
\end{align}

Note that $d_1(\Phi, \hat{\Phi}) \longrightarrow 0$ as $d_1(h, \hat{h}) \longrightarrow 0.$ Thus if $\Phi$ is strongly structurally stable, $d_\infty(\text{id}, g) \longrightarrow 0$ as $d_1(h, \hat{h}) \longrightarrow 0$ and so by Lemmas \ref{lmm:pushforward_cts_in_f} and \ref{lmm:pushforward_cts_in_nu}, $d_{\mathcal{P}, \mathcal{U}}(h_* \xi, \hat{h}_* \xi) + d_{\mathcal{P}, \mathcal{U}}(\hat{h}_* \xi, \hat{h}_* g_* \xi) \longrightarrow 0$ as $d_1(h, \hat{h}) \longrightarrow 0. \quad \blacksquare$

\begin{remark}\label{rem:final_thm}
    \normalfont
    \begin{enumerate}
        \item The bound in Theorem \ref{thm:transport_of_measures_with_state_space_system_bound} shows that the prediction error splits into two parts, the first having to do with the estimation error in choosing the readout $\hat{h}$ and the second also depending on this error, but now through the conjugacy $g$ arising from the structural stability assumption.
        \item If $\hat{h}$ is Lipschitz with constant $L_{\hat{h}}$ and we take $d_{\mathcal{P}, \mathcal{U}}$ and $d_{\mathcal{P}, \widetilde{\mathcal{X}}}$ to be Wasserstein-1 metrics, we can further write $d_{\mathcal{P}, \mathcal{U}}(h_* \xi, \hat{h}_* \xi) \leq d_\infty(h, \hat{h})$ and $d_{\mathcal{P}, \mathcal{U}}(\hat{h}_* \xi, \hat{h}_* g_* \xi) \leq L_{\hat{h}} d_\infty(\text{id}, g).$
        % Thus the error is composed of two terms, the first related to the approximation error in choosing $\hat{h}$ and the second related to the size of the conjugating homeomorphism $g$ and the expansive properties of the map $\hat{h}.$
        \item In practice state-space systems are typically trained to minimize $d_\infty(\Phi, \hat{\Phi})$ rather than $d_1(\Phi, \hat{\Phi})$. Theorem  \ref{thm:transport_of_measures_with_state_space_system_bound} suggests that it might be more advantageous to optimize the latter quantity. A result by Hornik et al. \cite{hornik:derivatives}, shows that if we take $\hat{h}$ to be a single-layer feedforward network, $d_1(h, \hat{h})$ may be made arbitrarily small by taking the hidden dimension large enough and selecting appropriate weights.
        \item Note that the assumptions of structural stability, mixing, ergodicity, etc. are made on the dynamical system $\Phi$ acting on the manifold $\widetilde{\mathcal{X}}$ in the state-space, rather than on the original dynamical system $\phi.$ Assume that the manifold $\mathcal{M}$ is compact. Recall that $\Phi$ is conjugate to $\phi$ by the GS $\zeta$. Thus by Lemmas \ref{lmm:struct_stab_pres_mixing_erg}, \ref{lmm:top_conj_basin_supp} and \ref{lmm:struct_stab_pres_att_phys}, if $\zeta$ is a homeomorphism, then $\Phi$ satisfies assumptions {\rm {\bf A1}}, {\rm {\bf A2}}, {\rm {\bf B1}} or {\rm {\bf B2}} on $\widetilde{\mathcal{X}}$ if and only if $\phi$ satisfies the corresponding assumption on $\mathcal{M}$. Similarly, if $\zeta \in \operatorname{Diff}^1(\mathcal{M}, \widetilde{\mathcal{X}})$ then $\Phi$ is (strongly) structurally stable if and only if $\phi$ is.
        \item We remind the reader of the discussion in Section \ref{subsec:ergodicity} on the ergodic, mixing, physical and attracting measures of assumptions {\rm {\bf A1}}, {\rm {\bf A2}}, {\rm {\bf B1}} and {\rm {\bf B2}}. While all of these measures may be singular with respect to a reference measure such as Lebesgue, assumptions {\rm {\bf B1}} and {\rm {\bf B2}} in particular allow us to assume that the initial measure $\xi_0$ is absolutely continuous with respect to the reference measure, rather than the invariant measure $\xi$.
    \end{enumerate}
\end{remark}
% \begin{remark}

    % \normalfont
%     \begin{enumerate}
%         \item[1.] In the case where we make the assumption that $\Phi$ is strongly structurally stable, we will need to ensure that $d_1(\Phi, \hat{\Phi})<\delta$ which can be ensured by taking $d_1(h, \hat{h})<\delta_0$ for sufficiently small $\delta_0.$ Reference a theorem of Hornik which says that this can be done, if we use a NN readout.
%     \end{enumerate}
% \end{remark}

% \begin{remark}

    % \normalfont
%     \begin{enumerate}
%         \item[1.] Theorem \ref{thm:transport_of_measures_with_state_space_system_bound} gives us a {\bfi time uniform bound} for the Wasserstein distance between the actual and predicted distributions: since the $\limsup$ is bounded, for any $\varepsilon>0$, for large enough $T$ we may bound 

%         \begin{equation}
%             W_1(\omega_* \phi^t_* \nu, \hat{h}_* \hat{\Phi}^t_* \varphi_* \nu) \leq \delta_0 + L_{\hat{h}} d_\infty(\operatorname{id}, g_{\hat{\Phi}}) + \varepsilon
%         \end{equation}
        
%         uniformly for all time $t \geq T.$

%         \item[2.] As noted in Remark \ref{}, we may exchange the topological stability assumptions for (strong) structural stability assumptions, in this case requiring instead that $d_1(h, \hat{h}) < \delta_0$ for sufficiently small $\delta_0.$

%     \end{enumerate}
% \end{remark}

In the introduction it was remarked that the results of this paper also apply to random feature models (RFMs) \cite{nielsen_random_feature_model}. Random feature models use a randomly generated feature map to embed the dynamical system states into the state space, after which a simple linear readout is trained to map the embedded state to the next system state. Thus all dynamics takes place on the original manifold of the dynamical system; state space dynamics do not play a role. In this case the proof of Theorem \ref{thm:transport_of_measures_with_state_space_system_bound} is simplified. An RFM, when run autonomously acts as a proxy dynamical system $\hat{\phi}$  to $\phi$ on $\mathcal{M}$. Thus, under the conditions stated in the theorem, namely that $\phi$ is structurally stable, and possesses an ergodic/physical or mixing/attracting measure, a similar bound may be derived. RFMs as presented in \cite{gottwald2025_hitandrun} require access to the full system states, although the authors mention that they may be expanded to the partially observed case via time delays and Takens' theorem. Thus in the partially observed case, state space dynamics occur via the time delay embedding, and the bounds of Theorem \ref{thm:transport_of_measures_with_state_space_system_bound} may be applied.

\section{Numerical illustration}\label{sec:numerics}

In this section we illustrate the main result Theorem \ref{thm:transport_of_measures_with_state_space_system_bound} using the Lorenz dynamical system.

\subsection{The setup.}\label{subsec:numerics_setup}

{\bf The Lorenz system.} The Lorenz system is a dynamical system on $\mathbb{R}^3$ defined by the equations

\begin{equation}\label{eq:lorenz}
    \left\{
    \begin{aligned}
        \dot{m^1} &= \sigma (m^2 - m^1), \\
        \dot{m^2} &= m^1(\rho - m^3) - m^2, \\
        \dot{m^3} &= m^1 m^2 - \beta m^3.
    \end{aligned}
    \right.
\end{equation}

We use the standard parameters $\sigma = 10, \rho = 28,$ and $\beta = 8/3.$ We discretize the flow of the Lorenz system to get a discrete-time dynamical system $\phi$ on $\mathbb{R}^3.$ Theorem \ref{thm:transport_of_measures_with_state_space_system_bound} requires compactness, which can be ensured by restricting to a compact forward invariant region $\mathcal{M}$ containing the attractor of the Lorenz system. In our case we take the observation $\omega$ as the identity, so that we have full access to the coordinates of the dynamical system. Subsequently this is useful as it allows us to measure convergence of distributions between the true and proxy systems along all three axes.

In 1999 Tucker famously solved Smale's 14$^{th}$ problem, proving that the Lorenz attractor is robust under small changes to the parameters \cite{TUCKER19991197_lorenz_exists}. In 2005 Luzzato et al. showed that the Lorenz attractor is mixing, indeed stably mixing, in that it maintains a mixing measure under sufficiently small $C^1$ perturbations \cite{luzzatto2005lorenz_mixing}. While uniformly hyperbolic systems are structurally stable, the Lorenz attractor has an equilibrium at the origin and is therefore only {\it singularly hyperbolic} \cite{morales_singular_hyp_systems}. The Lorenz attractor is not structurally stable, but nevertheless in our numerics we see that it satisfies the outcomes of Theorem \ref{thm:transport_of_measures_with_state_space_system_bound}. Numerical work has shown that the Lorenz system satisfies linear response \cite{Christian_linear_response_lorenz}. This together with the stable mixing of Luzzato et al. may be what guarantees that the results of Theorem \ref{thm:transport_of_measures_with_state_space_system_bound} still hold. An interesting question is whether the invariant measures continue to be attracting in the sense of \cite{newman2025attractingmeasures} under $C^1$ perturbations of the dynamics map. In any case, our numerics point to the possibility that Theorem \ref{thm:transport_of_measures_with_state_space_system_bound} might be proved under more general conditions.

\medskip

{\bf The Echo State Network (ESN).} To learn the Lorenz system, we make use of an ESN \cite{Matthews:thesis, Jaeger04, jaeger2001}. ESNs are determined by a state map of the form
\begin{equation}
    f(x,m) = \operatorname{tanh}\left( A x + C m \right),
\end{equation}
and a linear readout $\hat{h}(x) = \hat{W} x.$
The state variable $x$ belongs to some (typically) high-dimensional Euclidean space.
 For our work, we selected a state space of dimension 1000. The elements of the connectivity matrix $A$ and input matrix $C$ are independently selected according to $\mathcal{U}([-0.5, 0.5))$. Subsequently, a factor of $13/3$ rescales the input matrix, and the connectivity matrix is scaled to have spectral radius $ 1.2$. Only the readout matrix $\hat{W}$ is trained.

\medskip

{\bf Training.} To train the ESN, we use data from 100 initial conditions $\underline{m}_0 \sim \mathcal{U}([-20,20) \times [-20,20) \times [7, 47)).$ We use $\tau$ to denote the intrinsic time of the system, while $t$ will denote the time steps taken by iterating our state-space system. Trajectories are calculated using the RK4 method over 60 intrinsic time units with a discretization $\delta \tau = 0.02,$ amounting to $3000$ data points on each trajectory. Trajectories are scaled by a factor of 0.01 for training in order to ensure they do not saturate the extremities of the $\operatorname{tanh}(\cdot)$ function. For each trajectory, the state coordinate is initialized at $\underline{x}_0 = 0 \in \mathbb{R}^{1000},$ and subsequently a trajectory calculated by iterating the state equation
\begin{equation}\label{eq:state_eq_2}
    {\underline{x}}_t = f({\underline{x}}_{t-1}, \underline{m}_t) \quad \text{for } t=1, \dots, 3000.
\end{equation}
Thereafter, the first 1000 data points are discarded and, using the remaining 2000 time points from all 100 initial conditions, the readout matrix $\hat{W}$ is estimated by means of a ridge regression using SVD with regularization constant $10^{-14.5}$. Using SVD in the ridge regression has been shown to be more numerically stable \cite{numerical_instabilities_bollt}.

{\bf Prediction.} We approximate a distribution $\mu_0$ on the manifold $\mathcal{M}$ with Diracs from a large sample over the distribution. A difficulty with starting prediction immediately from a preselected sample of points is that the ESN state coordinate must be properly initialised to reflect the initial condition. One way to do this is to learn a {\bfi cold-starting map}, as in \cite{RC26}. In our case, we employ the traditional warmup method: 2200 initial conditions were drawn according to $\underline{m}_0 \sim \mathcal{U}([-20,20) \times [-20,20) \times [7, 47))$ and their trajectories calculated over 80 intrinsic time units, giving $4000$ data points on each trajectory. For each trajectory, the state coordinate is initialized at $\underline{x}_0 = 0 \in \mathbb{R}^{1000},$ and the state equation \eqref{eq:state_eq_2} iterated for $t=1,\dots,1000.$ This is called the {\bfi warmup.} By this time, due to the state-forgetting property, $\underline{x}_{1000}$ is a good approximation of the true value given by the GS $\zeta(\underline{m}_{1000}).$ From $t=1000$, the trained ESN is run autonomously according to
\begin{empheq}[left = {\empheqlbrace}]{align}
    \hat{\underline{x}}_t &= f(\hat{\underline{x}}_{t-1}, \hat{\underline{m}}_t)\\
    \hat{\underline{m}}_t &= h(\hat{\underline{x}}_{t-1})
\end{empheq}
for $t=1001, \dots, 4000,$ producing the predicted trajectory $\hat{\underline{m}}_{1001}, \dots, \hat{\underline{m}}_{4000}.$
As we already noted, using the traditional warmup approach bars us from directly initializing a sample of points $\underline{x}_0$ in the state-space, representing a distribution $\xi_0 \in \mathcal{P}(\mathcal{X})$ corresponding to the initial distribution $\mu_0$ on the manifold $\mathcal{M}.$ Thus, rather than specifying our distribution at time $t=0$ ($\tau=0$), we select our distribution from the trajectories at $t=1000$ ($\tau = 20$), when the state-coordinates are already warmed up and prediction starts. This is done as follows. Write $\underline{m}_t = (\underline{m}_t^1,\underline{m}_t^2,\underline{m}_t^3) \in \mathbb{R}^3$. Based on the first coordinate at $t=1001$ we separate the trajectories into two sets
\begin{align}
    &A_1 := \{ \underline{m} \colon \underline{m}^1_{1000} \geq 0 \}, 
    &A_2 := \{ \underline{m} \colon \underline{m}^1_{1000} < 0 \}
\end{align}
These two sets are subsequently downsampled to ensure that they both contained 1000 trajectories. Now we write
\begin{align}
    &\mu^1_t := \{ \underline{m}_t \colon \underline{m} \in A_1 \}
    &\mu^2_t := \{ \underline{m}_t \colon \underline{m} \in A_2 \}\\
    &\hat{\mu}^1_t := \{ \hat{\underline{m}}_t \colon \underline{m} \in A_1 \}
    &\hat{\mu}^2_t := \{ \hat{\underline{m}}_t \colon \underline{m} \in A_2 \}
\end{align}
for $t=1000, \dots, 4000.$ Thus $\mu^1_{1000}$ is a sample representing a distribution on the right lobe of the Lorenz attractor, while $\mu^2_{1000}$ is a sample representing a distribution on the left lobe of the Lorenz attractor (see Figure \ref{fig:densities_truetruepred}). For $t=1001,\dots, 4000$, $\hat{\mu}^1_t$ is the sample $\mu^1_{1000}$ transported under the proxy system, and likewise ${\mu}^1_t$ is the sample $\mu^1_{1000}$ transported under the true Lorenz system. In what follows we consider the distance $d_{\mathcal{P}}(\mu^1_t, \hat{\mu}^1_t)$ between the true and predicted distributions of $\mu^1_{1000},$ using the Wasserstein-1 distance $W_1$ (see Appendix \ref{subsec:wass_dist}) and MMD metric (see Appendix \ref{subsec:mmd}), comparing this to the results in Theorem \ref{thm:transport_of_measures_with_state_space_system_bound}. Considering the distance $d_{\mathcal{P}}(\mu^1_{1000}, \mu^2_{1000})$ will also give us a reference as to the diameter of the Lorenz attractor in the Wasserstein-1 distance and MMD metric since these two distributions are far apart, being located on the two opposing lobes of the Lorenz attractor. By this scale of reference we can judge the closeness of the distributions $\mu^1_t$ and $\hat{\mu}^1_t.$

\subsection{Results.}\label{subsec:results_numerics}

\begin{figure}[!htb]
\centering
\begin{subfigure}[b]{\textwidth}
    \centering
    \includegraphics[scale=0.45]{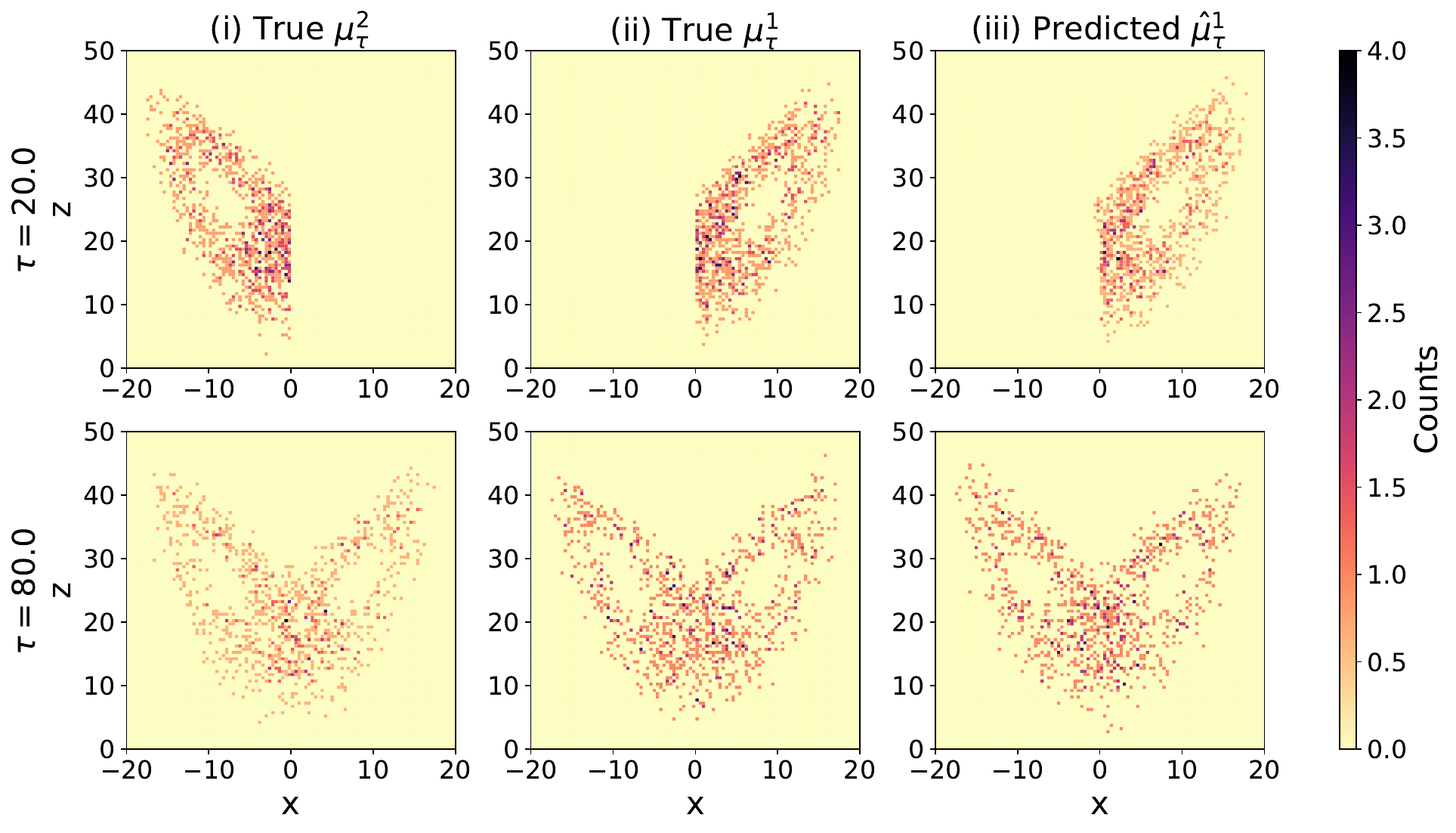}
    \caption{Projection onto the first and third coordinates of the samples (i) $\mu^2_{t}$, (ii) $\mu^1_{t}$, and (iii) $\hat{\mu}^1_t$.}
\end{subfigure}

\vspace{1em}

\begin{subfigure}[b]{\textwidth}
    \centering
    \includegraphics[scale=0.4]{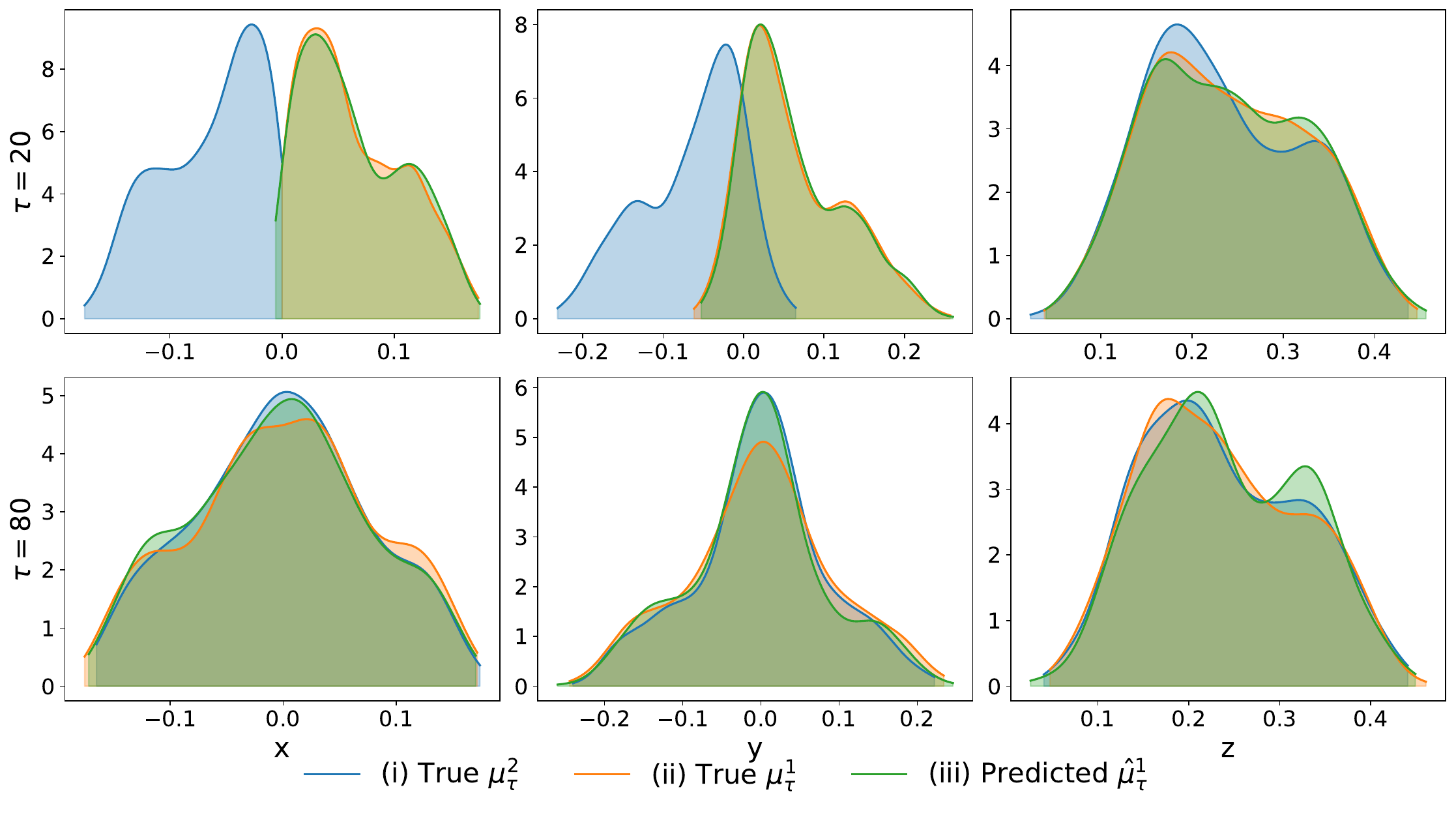}
    \caption{Marginal distributions of the samples (i) $\mu^2_{t}$, (ii) $\mu^1_{t}$, and (iii) $\hat{\mu}^1_t$. Upper row: distributions at $t=1000$ ($\tau = 20$), when prediction is started. Lower row: distributions at $t=4000$ ($\tau = 80$).}
\end{subfigure}
\caption{The mixing property of the Lorenz system causes all distributions to converge to the stationary distribution.}
\label{fig:densities_truetruepred}
\end{figure}

{\bf Mixing in the Lorenz system.} As we have described, at each time step $t=0,1,\dots,4000$ trajectories were separated into two samples $\mu^1_t = \{ \underline{m}_t \colon \underline{m} \in A_1 \},$ and $\mu^2_t = \{ \underline{m}_t \colon \underline{m} \in A_2 \}$ consisting of 1000 sampled points each. Prediction was started at $t=1000$ ($\tau = 20$), implemented according to the scheme described in Section \ref{subsec:numerics_setup}, producing the distributions $\{\hat{\mu}^1_t \colon t = 1001, 1002, \dots, 4000 \}.$ The upper rows of Figure \ref{fig:densities_truetruepred} (a) and (b) plot $\mu^1_{t}$, $\mu^2_{t}$ and $\hat{\mu}^1_{t}$ at $t=1000$ ($\tau = 20$) where prediction was started. The distributions $\mu^1_{t=1000}$ and $\mu^2_{t=1000}$ were selected so that they are on opposing lobes of the Lorenz attractor. Since prediction has not begun, $\hat{\mu}^1_{t=1000}$ coincides with $\mu^1_{t=1000}.$ The lower rows plot the same distributions at the terminal time $t=4000$ ($\tau = 80$). The marginal distributions in Figure \ref{fig:densities_truetruepred} (b) were smoothed using a Gaussian KDE method. The mixing property of the Lorenz system is clear as we see that despite being initialised on opposite lobes of the attractor, both $\mu^1_t$ and $\mu^2_t$ converge to the stationary distribution on the Lorenz system by the time $t=4000.$

The one drawback of the Lorenz system is that it is not structurally stable, a property required in Theorem \ref{thm:transport_of_measures_with_state_space_system_bound}. Nevertheless, the mixing property carries over to our ESN, as is seen in the lower right graph where $\hat{\mu}^1_{t=4000}$ has also converged to the stationary distribution after having been transported by the proxy system for $3000$ steps. As discussed in Section \ref{subsec:numerics_setup}, this may be due to the fact that the Lorenz system is stably mixing and satisfies linear response \cite{Christian_linear_response_lorenz, luzzatto2005lorenz_mixing}.

% Interestingly, the trained ESN is not conjugate to the Lorenz system. This is evident from Figure \ref{fig:invariant_measures} which plots the invariant measures of the true and proxy systems. The proxy ESN has an attracting fixed point off the Lorenz attractor. Nevertheless, as Figure \ref{fig:densities_truepred} shows, when trajectories are started sufficiently near to the Lorenz attractor, this attracting fixed point does not play a role.

% \begin{figure}[!htb]
% \centering
% \includegraphics[scale = 0.3]{Figures/invariant_measures.png}
% \caption{The first and third coordinates of the invariant measures of the true Lorenz and proxy ESN systems. The ESN has an attracting fixed point off the Lorenz attractor.}
% \label{fig:invariant_measures}
% \end{figure}

% \begin{figure}[!htb]
% \centering
% \includegraphics[scale = 0.3]{Figures/Prediction_one_trajectory.png}
% \caption{A single trajectory of the true Lorenz (blue) and proxy ESN (orange). Warmup ends at $t=20$, thereafter prediction continues till $t=80.$ Predictions fail by $t=30.$}
% \label{fig:prediction_one_trajectory}
% \end{figure}

\medskip

\begin{figure}[!htb]
\centering
\includegraphics[scale = 0.35]{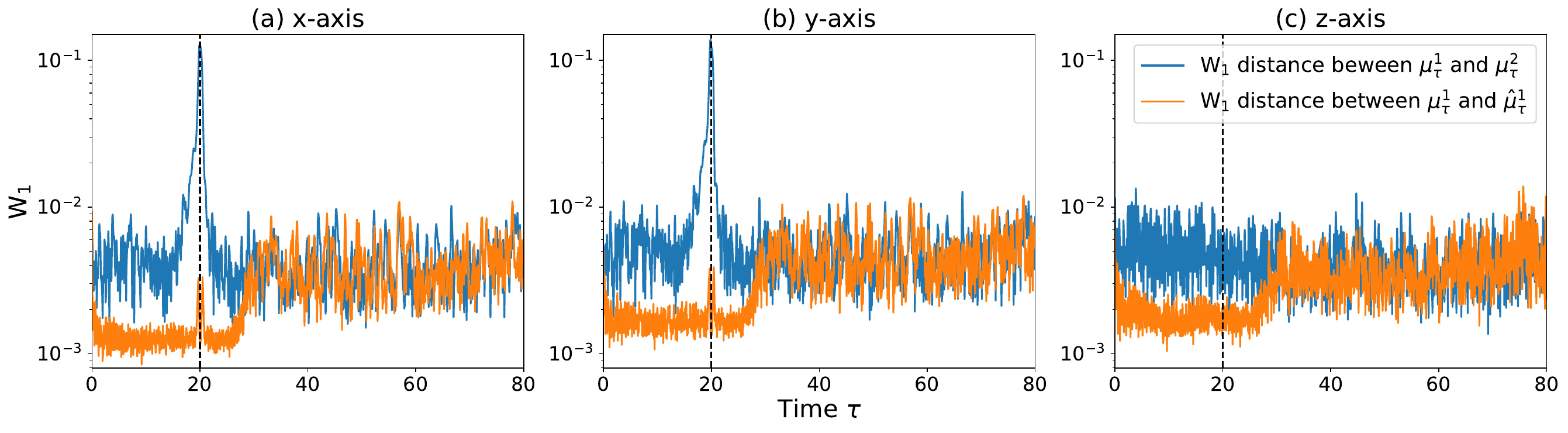}
\caption{Stability in distribution prediction error. Wasserstein-1 distance between $\mu^1_{t}$ and $\mu^2_{t}$ (blue) and between $\mu^1_{t}$ and $\hat{\mu}^1_{t}$ (orange) for each of the marginal distributions. The dashed line at $\tau=20$ indicates the start of prediction. The error in distribution predictions increases slightly with the prediction horizon, but after 9 Lyapunov times of prediction, it settles at a value one and a half orders of magnitude lower than the distance between the two lobes in the Wasserstein-1 metric (blue graph at $\tau = 20$), indicating that the distributions are nearly equal.}
\label{fig:Wass1_transport_figure_panel}
\end{figure}

\medskip

{\bf Time-uniform bound on distribution forecasts.} We now look at the errors in distribution predictions. To calculate the distance between distributions, we use the Wasserstein-1 and MMD metrics. For more details on Wasserstein distances, see Appendix \ref{subsec:wass_dist}; for details concerning the MMD metric and how it may be used to conduct two-sample hypothesis testing, see Appendix \ref{subsec:mmd}. Figure \ref{fig:Wass1_transport_figure_panel} depicts the estimated Wasserstein-1 distances between the marginal distributions of $\mu^1_{t}$ and $\mu^2_{t}$ in blue and between the samples $\mu^1_{t}$ and $\hat{\mu}^1_{t}$ in orange, for $t=0, 1, \dots, 4000$.

We briefly remark on the apparent noisiness in Figure \ref{fig:Wass1_transport_figure_panel}. Since we work with finite samples on the Lorenz attractor, the distances calculated, both with the Wasserstein-1 and MMD metrics, are approximations of the distance between the true distributions. These distributions are approximated by an ensemble of Diracs. This, paired with the oscillations and chaos inherent to the Lorenz attractor, are likely the cause of the observed noisiness. Nevertheless, some things may be clearly deduced from the graphs. Considering the Wasserstein-1 distance between $\mu^1_{t}$ and $\mu^2_{t}$ gives us a reference for the diameter of the attractor in the Wasserstein-1 metric. In particular, the first and second marginals of $\mu^1_{t=1000}$ and $\mu^2_{t=1000}$ are quite far apart since they are on opposite lobes of the attractor as depicted in the top rows of the graphs in Figure \ref{fig:densities_truetruepred}. In Figure \ref{fig:Wass1_transport_figure_panel} this is evidenced in the high value of the blue graph at $\tau=20$ ($t=1000$) for the first and second marginal distributions. Due to mixing, the blue graph soon drops to a significantly lower value, evidenced by the very similar distributions depicted in the bottom rows of the graphs in Figure \ref{fig:densities_truetruepred}. On the other hand, we consider the Wasserstein-1 distance between $\mu^1_t$ and $\hat{\mu}^1_t$ (orange in Figure \ref{fig:Wass1_transport_figure_panel}). When prediction is started at $\tau=20$ ($t=1000$), the distance is very small. (Note that it is nonzero since $\hat{\mu}^1_{1000}$ has been calculated from $\mu^1_{999}$ through the one-step-ahead predictions of the state-space system. There is a small error in one step-ahead predictions by the state-space system.) Once prediction commences, the distance between the true and predicted distributions rises slightly, but by around $\tau=30$ (9.06 Lyapunov times from when prediction is started) stabilizes around the same value as the blue graph when $\mu^1_t$ and $\mu^2_t$ are thoroughly mixed. The stabilized value is about one and a half orders of magnitude lower than the Wasserstein-1 distance between $\mu^1_{t=1000}$ and $\mu^2_{t=1000}$ on the opposing lobes of the attractor. This suggests that at this level the distributions can be regarded as equal. We make this more precise by conducting a two-sample hypothesis test on the distributions using the MMD metric.

\medskip

\begin{figure}[!htb]
\centering

\begin{subfigure}[t]{0.48\textwidth}
    \centering
    \includegraphics[width=\linewidth]{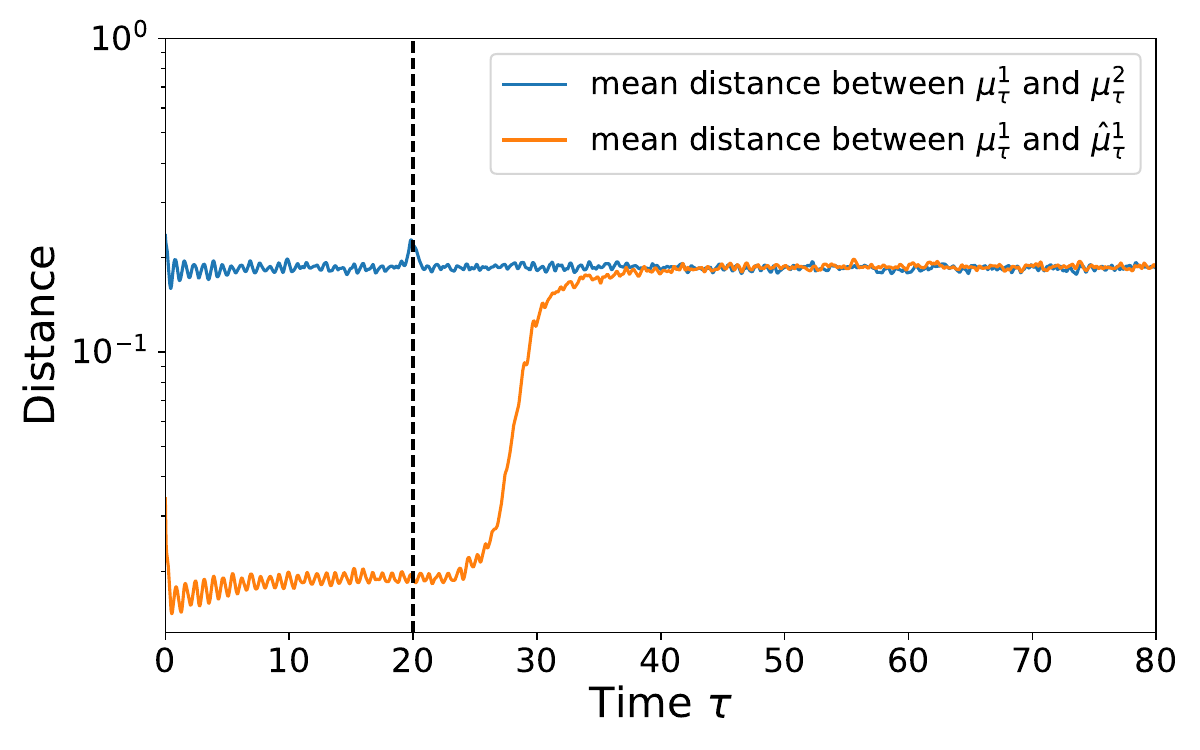}
    \caption{}
    \label{fig:trajectories_mean_distance}
\end{subfigure}
\hfill
\begin{subfigure}[t]{0.48\textwidth}
    \centering
    \includegraphics[width=\linewidth]{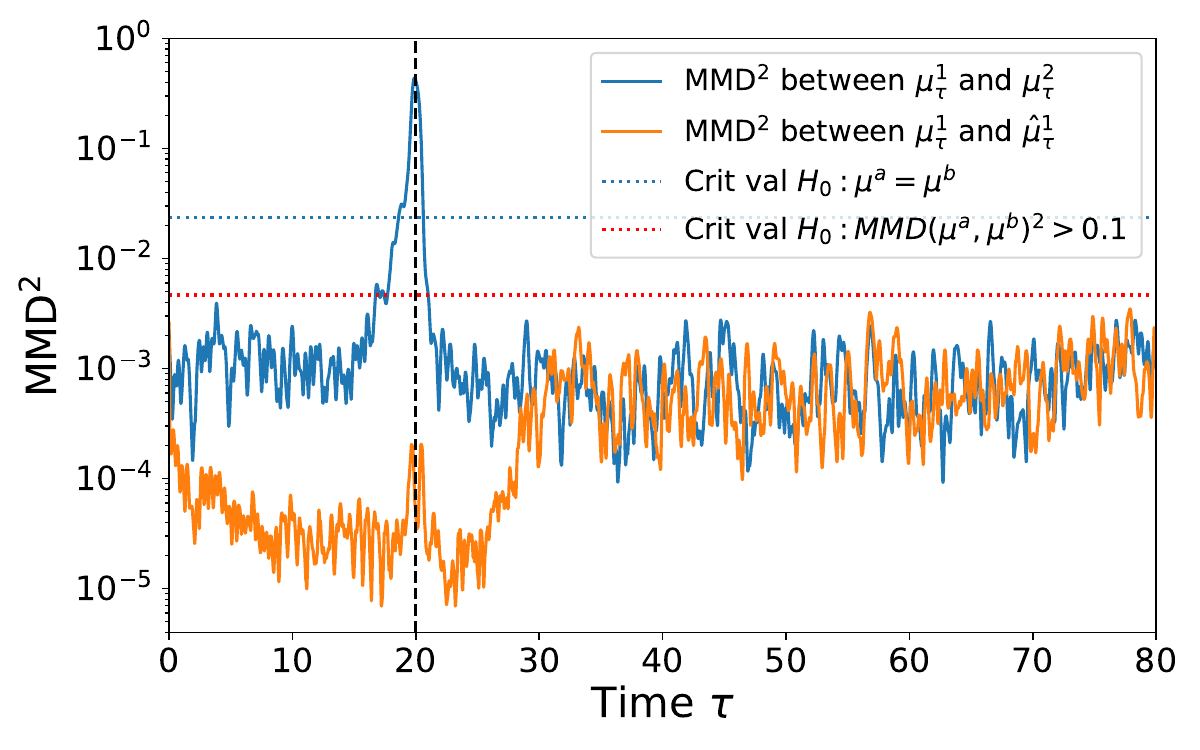}
    \caption{}
    \label{fig:mmd_transport_figure}
\end{subfigure}

\caption{Stability in distribution prediction error. In both graphs the dashed line at $\tau=20$ indicates the beginning of prediction. (a) Mean distance between arbitrarily paired points in $\mu^1_{t}$ and $\mu^2_{t}$ (blue) and mean prediction error for initial conditions in $\mu^1_{t}$ and $\hat{\mu}^1_{t}$ (orange). The error in point predictions (orange) settles at a value comparable to the distance between unrelated initial conditions (blue) after 18 Lyapunov times of prediction. (b) Squared distance in the MMD metric between $\mu^1_{t}$ and $\mu^2_{t}$ (blue) and between $\mu^1_{t}$ and $\hat{\mu}^1_{t}$ (orange). The error in distribution predictions increases slightly with the prediction horizon, but after 9 Lyapunov times of prediction, it settles at an acceptably low value, indicating that the distributions are nearly equal. The critical values for both statistical tests are calculated at level $\alpha = 0.05.$}
\label{fig:combined_figures}
\end{figure}

{\bf Two-sample hypothesis test.} An advantage of the MMD metric is that there are results which allow us to perform distribution-agnostic hypothesis tests \cite{gretton12a}. To judge the similarity between the distributions, we perform the two-sample statistical tests A and B described in Appendix \ref{subsec:mmd} using the MMD metric with a Gaussian RBF kernel
% Figure \ref{fig:combined_figures} (b) depicts the estimated squared distance between the samples $\mu^1_{t}$ and $\mu^2_{t}$ in blue and between the samples $\mu^1_{t}$ and $\hat{\mu}^1_{t}$ in orange, for $t=0, 1, \dots, 4000$, using the biased MMD estimator $d_{K,b}$ of \eqref{eq:mmd_biased_def}. In the construction of the MMD metric, we use a Gaussian RBF kernel
$K(x, y) = \exp \left(- \frac{\|x-y\|_2^2}{2 \sigma^2} \right),$
where $\| \cdot \|_2$ is the standard Euclidean norm on $\mathbb{R}^3$ and the kernel width $\sigma$ is set to the median distance between points from 100 trajectories chosen randomly from $\mu^1 \cup \mu^2$. This is a heuristic suggested in \cite{gretton12a}. In Test A, we take the null hypothesis to be that the two distributions are equal. By Corollary \ref{cor:mmd_testa}, at level $\alpha = 0.05$ with a sample size of 1000, we reject the null hypothesis if the squared biased estimator yields a distance between the samples greater than $0.02377$. This critical value is plotted in Figure \ref{fig:combined_figures} (b) as the dotted blue line. Since $d_{K, b}(\mu^1_{1000}, \mu^2_{1000})^2 > 0.02377$, the test clearly distinguishes that these two samples come from different distributions. On the other hand, in Test B, we take as the null hypothesis that the MMD between the two distributions exceeds $\varepsilon$. At level $\alpha = 0.05$ with sample size of 1000 and $\varepsilon^2 = 0.1$, according to Corollary \ref{cor:mmd_testb}, the test rejects the null hypothesis if the biased estimator yields a sample distance less than $0.00466.$ This critical value is plotted as the red dotted line in Figure \ref{fig:combined_figures} (b). Since for large $t$, $d_{K, b}(\mu^1_t, \mu^2_t)^2 < 0.00466$ and $d_{K, b}(\mu^1_t, \hat{\mu}^1_t)^2 < 0.00466$, we conclude that the true distributions from which they come are within $\sqrt{0.1}$ in the MMD metric.

This result is particularly important as it accords with the outcome of Theorem \ref{thm:transport_of_measures_with_state_space_system_bound}. Recall that $\mu^1_{1000}$ is taken to be a sample from some true distribution on the attractor. Let $\nu_t$ denote the true distribution from which $\mu^1_{t}$ is sampled for all $t$. Likewise, denote the actual distribution behind the samples $\hat{\mu}^1_t$ as they are transported under the proxy system by $\hat{\nu}_t.$ Figure \ref{fig:combined_figures} (b) shows that $\limsup_{t \to \infty} d_{K}(\nu_t, \hat{\nu}_t) < \sqrt{0.1}.$

\medskip

{\bf Exponential separation in point predictions.} In contrast to the distribution predictions, point predictions by the trained ESN soon fail. % When comparing the true and predicted trajectories with initial conditions in $\mu^1_{t=1000}$ it is evident that point predictions by the trained ESN soon fail.
%Figure \ref{fig:prediction_one_trajectory} plots the true and predicted trajectories for the first initial condition in the sample $\mu^1_{1000}.$ Prediction starts at time step $t=1000$ and loses track of the true trajectory by $t=1500.$
The orange graph of Figure \ref{fig:combined_figures} (a) shows that the average error in point predictions of trajectories starting in sample $\mu^1_{t=1000}$ increases approximately exponentially from the time when the prediction is started. This is in accord with the bounds derived in \cite{berry2022learning, RC22}. Since the Lorenz attractor is bounded, the mean error eventually begins to saturate at around $\tau=30$ and stabilizes at a constant value by $\tau=40$. Since the Lorenz system has a top Lyapunov exponent of $0.906$ \cite{viswanath2000}, these correspond to approximately $9.06$ and $18.1$ Lyapunov times from the time the prediction was started, respectively. In the blue graph of Figure \ref{fig:combined_figures} (a), initial conditions in $\mu^1_{t=1000}$ and $\mu^2_{t=1000}$ are arbitrarily paired up, the distances between them are calculated and averaged across the 1000 pairs. This is repeated at each time step $t=0,1,\dots,4000.$ There is no relation between the initial conditions in these two samples. Thus, Figure \ref{fig:combined_figures} (a) shows that eventually the error between the true and predicted trajectories started from the same initial condition is the same as if the true system had been started from two unrelated initial conditions.

\medskip

{\bf Stability of empirical measures.} Theorem \ref{thm:trajectory_average_with_state_space_system_bound} gives a bound on the error between the empirical measure associated to a given initial condition as calculated using the true and proxy systems. In Section \ref{sec:numerics_empirical_measure} this is validated by considering the convergence in Wasserstein-1 distance for the empirical measure calculated over the trajectory of a single initial condition. This showcases the stability of ergodic averages taken over trajectories.

{\bf A remark on numerical stability.} We remark briefly on the stability of trajectories of the proxy system. One of the main points of this paper is that while point predictions eventually fail, under the conditions presented in this paper, density predictions will remain accurate. This requires that the trajectories of the proxy system remain on or near the attractor of the true system. However, it has been observed that sometimes point predictions fall off the attractor and may even escape in unbounded trajectories \cite{pathak:chaos, numerical_instabilities_bollt}. Indeed, in our numerical work we observed that the attractor of the proxy state space system, mapped back into the observation space through the readout $h$, sometimes possessed a limiting cycle or attractive fixed point off the Lorenz attractor (see Figure \ref{fig:invariant_measures} in Section \ref{sec:invariant_distribution}). This shows that for some initial conditions trajectories will not stay on the Lorenz attractor, but might be attracted to these spurious attractors. Nevertheless, once we had trained our system well, trajectories that had been sufficiently warmed up so that they were on the Lorenz attractor did not later fall off. The theoretical results show that when the underlying system is structurally stable and the proxy is a sufficiently accurate $C^1-$approximation, we should not encounter these problems. As observed before, the Lorenz system is not structurally stable, and thus these problems can be expected when we attempt to learn it.

{\bf Discussion.} We briefly summarize the results shown in this section. In our numerical exercise, we demonstrated how the Lorenz system satisfies the results of Theorem \ref{thm:transport_of_measures_with_state_space_system_bound}. We took a sample $\mu^1_{1000}$ of points on the right lobe of the Lorenz attractor. On the one hand Figure \ref{fig:combined_figures} (a) shows that point predictions for these initial conditions soon fail. On the other hand, Figures \ref{fig:densities_truetruepred}, \ref{fig:Wass1_transport_figure_panel} and \ref{fig:combined_figures} (b) show that density predictions converge to the true stationary density of the Lorenz system, thus remaining accurate at arbitrarily long time horizons. While the Lorenz system is stably mixing \cite{luzzatto2005lorenz_mixing} with linear response \cite{Christian_linear_response_lorenz}, it does not satisfy the structural stability hypothesis of Theorem \ref{thm:transport_of_measures_with_state_space_system_bound}.
%Indeed, the proxy system is not topologically conjugate to the true Lorenz system.
Nevertheless, the theorem's conclusion remains valid. This raises the possibility that the theorem might be proved under more general conditions.

\section{Conclusions}\label{sec:conclusion}

Forecasting of deterministic time series is conducted under two paradigms: point and distribution forecasts. These involve learning the weather and climate of the underlying system, respectively. This paper gives results for the latter forecasting scheme. In particular, the accuracy of density forecasts from state-space systems for time-series data, which arise as observations of a deterministic dynamical system, is studied in this paper.

The main result, Theorem \ref{thm:transport_of_measures_with_state_space_system_bound}, shows that, when the underlying dynamical system is structurally stable and possesses a mixing or attracting measure, sufficiently regular initial distributions forecasted through a $C^1$ close proxy system will {\bfi remain close to the true future distribution at arbitrarily long time horizons.} When the underlying system is strongly structurally stable, this error may be made arbitrarily small by decreasing the distance between the true and proxy systems in the space of $C^1$ diffeomorphisms on the embedded manifold in the state space. When the underlying system possesses an ergodic or physical measure, the same results hold true, in a Ces\`{a}ro sense. {\bfi This provides a theoretical foundation for the empirically observed stability in learning the climate of dynamical systems} with state-space systems. This stands in contrast to point predictions, in which sensitivity to initial conditions leads to forecast instability.
% the presence of a positive Lyapunov exponent leads to an exponential decay in accuracy over time.

Beyond the main result, it has been shown that, when the mixing and related assumptions are dropped, an error bound with exponential growth can be derived, as in the point-prediction case. This makes use of a theory of Lyapunov exponents for metric spaces first introduced by Kifer \cite{Kifer_1983}, thus providing a use case for his Lyapunov exponent and also extending the previous results for point forecasts in \cite{berry2022learning, RC22} to metric spaces. In addition, the metric spatial Lyapunov exponent was introduced and related to Kifer's exponent.

Several insights can be drawn from the results presented in this paper. While ergodic, mixing, and physical measures are well-studied, attracting measures were first so named in \cite{newman2025attractingmeasures}. Attracting measures represent a second, less studied, `naturalness' property inherent to SRB measures. The results of this paper demonstrate the striking effect the existence of such measures has on the dynamics of the Perron-Frobenius operator acting on the space of probability measures. While this space is much richer than that of the original dynamical system, on certain regular subsets of the space of probability measures, the dynamics turns out to be much simpler. 

As in the point forecasting case, universality of the class of state-space systems employed in learning plays a role in the accuracy that can be achieved in prediction. The results presented here rely on a structural stability assumption that requires the proxy system to be close in the $C^1$ sense. When the system is strongly structurally stable, $C^1$ closeness directly affects the degree of accuracy that can be attained. This suggests that uniform approximation alone may be insufficient to obtain a system that reliably reproduces the underlying system's climate. Results by Hornik et al. \cite{hornik:derivatives} give classes of feedforward neural networks that are $C^1$ universal. A direction for future research is to investigate how $C^1$ approximations can be achieved in implementations of standard state-space systems and reservoir computing-based learning paradigms, such as echo state networks.

A further question arises regarding the assumptions underlying the main results of this paper. In particular, how essential is the existence of mixing and attracting measures? Is it possible to improve on the results in Theorems \ref{thm:state_space_system_bd_metric_le_prob_space} and \ref{thm:state_space_system_bd_metric_le_prob_space_conjugate_maps}, where the existence of such measures is not assumed? On the flip side, could adding an assumption on the rate of mixing provide a rate of convergence for the density forecasts? About the structural stability assumption, Section \ref{sec:numerics} shows that a well-selected ESN produces reliable density forecasts of the Lorenz system, despite this system not being structurally stable. This raises the possibility that weaker notions of stability could be used to derive similar results to those presented in this paper.

Finally, this paper addresses only deterministic dynamical systems. The question of deriving error bounds for predictions using state-space systems in the stochastic case remains open.

A common complaint in the implementation of reservoir computing systems is that the training process is highly volatile: some systems with highly accurate one-step ahead point predictions fail to produce sensible predictions in the future -- not only do they lose track of the time series, which is inevitable in the presence of chaos, but they often produce spurious attractors and dynamics unrelated to the true system. They {\bfi fail to learn the climate of the underlying system.} This paper lends insight into possible reasons for these observations: the underlying system may not be structurally stable, the proxy system, though accurate in one-step ahead point predictions, may fail to be $C^1$ close to the true system, or the underlying system may fail to possess a mixing or attracting measure.

In these ways, this work contributes to the theory of state-space systems and to their ability to reliably learn the climate of deterministic time series. Given the wide range of applications of state-space systems for predicting time series in today's world, such understanding is invaluable.

\medskip

{\bf Acknowledgments}
The authors thank Lyudmila Grigoryeva, G Manjunath, and Florian Rossmannek for many insightful discussions and remarks. J L is funded by a Graduate Scholarship from Nanyang Technological University. J-P O acknowledges partial financial support from the School of Physical and Mathematical Sciences of the Nanyang Technological University. 

\section*{Glossary of symbols}

{\bf State-space systems learning for observations from a dynamical system}

\begin{longtable}{p{9cm}lc}
    \toprule
    \textbf{Functions} & \textbf{Symbol} & \textbf{Domain and codomain} \\
    \midrule
    \toprule
    \endfirsthead
    \endhead
    \bottomrule
    \endfoot

    Dynamical system map & $\phi$ & $\mathcal{M} \longrightarrow \mathcal{M}$ \\
    Observation map & $\omega$ & $\mathcal{M} \longrightarrow \mathcal{U}$ \\
    State map & $f$ & $\mathcal{X} \times \mathcal{U} \longrightarrow \mathcal{X}$ \\
    Readout & $h$ & $\mathcal{X} \longrightarrow \mathcal{U}$ \\
    Approximated readout & $\hat{h}$ & $\mathcal{X} \longrightarrow \mathcal{U}$ \\
    Generalized synchronization & $\zeta$ & $\mathcal{M} \longrightarrow \mathcal{X}$ \\
    State-space dynamical system & $\Phi$ & $\tilde{\mathcal{X}} \longrightarrow \tilde{\mathcal{X}}$ \\
    Proxy state-space dynamical system & $\hat{\Phi}$ & $\tilde{\mathcal{X}} \longrightarrow \tilde{\mathcal{X}}$ \\
    \toprule
    \textbf{Spaces} & \textbf{Symbol} & \textbf{Trajectory} \\
    \midrule
    \toprule

    Dynamical system domain & $\mathcal{M}$ & $\underline{m} = (\underline{m}_t)_{t \in \mathbb{Z}}$ \\
    Observation space & $\mathcal{U}$ & $\underline{u} = (\underline{u}_t)_{t \in \mathbb{Z}}, \hat{\underline{u}} = (\hat{\underline{u}}_t)_{t \in \mathbb{Z}}$ \\
    State space & $\mathcal{X}$ & $\underline{x} = (\underline{x}_t)_{t \in \mathbb{Z}}, \hat{\underline{x}} = (\hat{\underline{x}}_t)_{t \in \mathbb{Z}}$ \\
    Embedded dynamical system domain & $\tilde{\mathcal{X}} = \zeta(\mathcal{M})$ & $\underline{x} = (\underline{x}_t)_{t \in \mathbb{Z}}, \hat{\underline{x}} = (\hat{\underline{x}}_t)_{t \in \mathbb{Z}}$ \\
    Probability measures on $\mathcal{M}$ & $\mathcal{P}(\mathcal{M})$ & $(\mu_t)_{t \in \mathbb{Z}}$ \\
    Probability measures on $\mathcal{U}$ & $\mathcal{P}(\mathcal{U})$ & $(\nu_t)_{t \in \mathbb{Z}}, (\hat{\nu}_t)_{t \in \mathbb{Z}}$ \\
    Probability measures on $\tilde{\mathcal{X}}$ & $\mathcal{P}(\tilde{\mathcal{X}})$ & $(\xi_t)_{t \in \mathbb{Z}}, (\hat{\xi}_t)_{t \in \mathbb{Z}}$ \\
\end{longtable}

{\bf Stability of dynamical systems transport}

\begin{longtable}{p{3cm}lc}
    \toprule
    & \textbf{Description} & \textbf{Symbol} \\
    \midrule
    \toprule
    \endfirsthead
    \endhead
    \bottomrule
    \endfoot
    {\bf Metrics} & Metric space & $(X, d)$ \\
    & Uniform metric & $(C(X,Y), d_\infty)$ \\
    & Wasserstein-$p$ distance & $(\mathcal{P}_p(X), W_p)$ \\
    & MMD metric & $(\mathcal{P}(X), d_K)$ \\
    & Whitney $C^1$ metric & $(\operatorname{Diff}^1(X), d_1)$ \\
    & Metric for weak topology & $(\mathcal{P}(X), d_{\mathcal{P}})$ \\
    & Uniform metric for maps between measures & $(C(\mathcal{P}(X), \mathcal{P}(Y)), d_{\mathcal{P}, \infty})$\\
    \midrule
    {\bf Metric LEs} & Dynamical system & $\varphi \colon X \longrightarrow X$ \\
    & Kifer metric Lyapunov exponent & $\Lambda_{\varphi}(x)$ \\
    & Spatial metric Lyapunov exponent & $\lambda_{\varphi}(\mu)$ \\
    & Pointwise Lipschitz constant & $L_{\varphi}(x)$ \\
    \midrule
    {\bf Section \ref{sec:stability_dyn_sys_probability}} & True and proxy dynamical systems & $\varphi, \hat{\varphi} \colon X \longrightarrow X$ \\
    & Probability measure to be predicted & $\nu \in \mathcal{P}(X)$\\
    & Ergodic measure for $(\mathcal{P}(X), \hat{\varphi}_*)$ & $\hat{\Xi} \in \mathcal{P}(\mathcal{P}(X))$ \\
    & Conjugating homeomorphism for $\varphi, \hat{\varphi}$ & $g: X \longrightarrow X$ \\
    & Ergodic, mixing, physical, or attracting measure & $\mu \in \mathcal{P}(X)$ \\
    & Reference Borel probability measure & $\ell$ \\
    & neighborhood of $\supp(\mu)$ & $U \subseteq X$ \\
    & Basin of attraction for measure $\mu$ & $\Basin_{\varphi}(\mu)$ \\
    & Basin of attraction for $\supp(\mu)$ & $\Basin_{\varphi}(\supp(\mu))$ \\
    \midrule
    {\bf Conditions} & Ergodic ({\rm {\bf A1}}) and mixing ({\rm {\bf A2}}) measures & {\rm {\bf A1}}, {\rm {\bf A2}} \\
    & Physical ({\rm {\bf B1}}) and attracting ({\rm {\bf B2}}) measures & {\rm {\bf B1}}, {\rm {\bf B2}} \\
\end{longtable}

{\bf Numerics}

\begin{longtable}{p{3cm}lc}
    \toprule
    & \textbf{Description} & \textbf{Symbol} \\
    \midrule
    \toprule
    \endfirsthead
    \endhead
    \bottomrule
    \endfoot

    {\bf ESN} & Lorenz system parameters & $\sigma, \rho, \beta$ \\
    & ESN connectivity and input matrices & $A, C$ \\
    & Trained readout matrix & $\hat{W}$ \\
    & Intrinsic time of Lorenz system & $\tau$ \\
    & Discrete time step of ESN & $t$ \\
    \midrule
    {\bf Predictions} & Point in $\mathbb{R}^3$ & $(m^1, m^2, m^3)$ \\
    & True trajectory & $\underline{m} = (\underline{m}_t)_{t \in \mathbb{Z}}$ \\
    & Predicted trajectory & $(\hat{\underline{m}}_t)_{t \geq 1000}$ \\
    & Sample distributions from right/left Lorenz lobes & $\mu_t^1, \mu_t^2$ \\
    & Predicted distribution transported by proxy system & $\hat{\mu}_t^1$ \\
    \midrule
    {\bf Statistical tests} & Biased MMD estimator & $d_{K,b}$ \\
    & Kernel function & $K: X \times X \longrightarrow \mathbb{R}$ \\
    & Test A & $H_0: \mu = \hat{\mu}$ \\
    & Test B & $H_0: d_K(\mu, \hat{\mu}) \geq \varepsilon$ \\
    & Hypothesis test significance level & $\alpha$ \\
\end{longtable}

\appendix

\section{Appendices}\label{sec:appendix}

These appendices provide more detailed descriptions of the theory used in the paper and additional results beyond those derived therein. Appendix \ref{subsec:metric_le_appendix} deals with the metric space Lyapunov exponent introduced in Section \ref{subsec:metric_le}. Appendix \ref{subsec:metrics on prob spaces} deals with metrizing the weak topology on $\mathcal{P}(X)$, and Appendix \ref{subsec:ds, erg, mixing} gives more details on ergodic, mixing, physical, and attracting measures which were introduced in Section \ref{subsec:ergodicity}. Appendix \ref{subsec:stability} discusses various consequences of structural stability.

\medskip

In Appendix \ref{subsec:add_bd_without_mixing} an additional bound on the error of density forecasts without making mixing or related assumptions is derived. Appendix \ref{subsec:stab_empirical_measure} gives a result on the stability of the climate as calculated from trajectory averages under approximation of the dynamical systems map.

\subsection{A metric space Lyapunov exponent.}\label{subsec:metric_le_appendix}

In the density forecasting task the underlying dynamical system $\phi \colon \mathcal{M} \longrightarrow \mathcal{M}$ induces a second dynamical system $\phi_* \colon \mathcal{P}(\mathcal{M}) \longrightarrow \mathcal{P}(\mathcal{M}).$ With point forecasts, a theorem of Oseledets shows that the rate at which prediction errors grow is related to the top Lyapunov exponent of the proxy system \cite{berry2022learning, RC22}. An immediate obstacle to applying such a result to a dynamical system on $\mathcal{P}(\mathcal{M})$ is the lack of a differential structure through which to define a Lyapunov exponent. In Appendix \ref{subsec:metrics on prob spaces} below, we discuss conditions under which $\mathcal{P}(\mathcal{M})$ is metrizable. This allows us to use a theory of Lyapunov exponents for metric spaces on the induced dynamical system. We will find that the metric space Lyapunov theory, while less developed than that for differentiable maps, is sufficient to obtain a result similar to the point-forecasting case.

Let $(X,d)$ be a separable metric space without isolated points and let $\varphi \colon X \longrightarrow X$ be a continuous dynamical system on $X$. In this section, $\varphi$ may possibly be non-invertible. Different definitions for a metric Lyapunov exponent exist \cite{MarioBessa2012DiscreteandContinuousDynamicalSystems, Luis_Barreira_2005_Discrete_and_Continuous_Dynamical_Systems, MORALES_THIEULLEN_VILLAVICENCIO_2018}. We use the maximal Lyapunov exponent of Kifer \cite{Kifer_1983}. Though each of the different definitions have advantages, Kifer's exponent is most natural mathematically, bearing resemblance to the definition of entropy for dynamical systems.

For a point $x \in X$ and  a nonnegative integer $t$, let $B_x( \delta, t)$ denote the set of points $y \neq x$ such that $\max\{d(\varphi^i(x), \varphi^i(y)): i = 0, \dots, t\} < \delta.$ Now, let $$A_\delta(x,t) = \sup_{y \in B_x(\delta, t)} \frac{d(\varphi^t(x), \varphi^t(y)}{d(x,y)}.$$ Finally, let $\mu \in \mathcal{P}(X)$ be a $\varphi$-invariant Borel measure and suppose that
\begin{equation}\label{ineq:mu_integrability_cond}
    \sup_{t \in \mathbb{N}} \frac{1}{t} \int_X |\log A_\delta(x, t)| d\mu(x) < \infty.     
\end{equation}
This condition is fulfilled, for example, when $\varphi$ is Lipschitz. Now we can define $$\Lambda_\delta(x) = \limsup_{t \to \infty} \frac{1}{t} \log A_\delta(x,t).$$ We have the following result:

\begin{theorem}[\cite{Kifer_1983}, Theorem 1]\label{thm:kifer_exp_def}
    Let $\mu \in \mathcal{P}(X)$ be a $\varphi$-invariant Borel probability measure such that \eqref{ineq:mu_integrability_cond} is fulfilled for some $\delta>0$. Then
    \begin{equation}
        \Lambda_\delta(x) = \lim_{t \to \infty} \frac{1}{t} \log A_\delta(x, t)
    \end{equation}
$\mu$-a.e. Additionally, $\Lambda_\delta(x)$ is $\mu$-a.e. invariant under $\varphi$ and hence if $\varphi$ is ergodic with respect to $\mu$ (see Section \ref{subsec:ds, erg, mixing}), then it is a constant $\mu$-a.e.    
\end{theorem}

If $\mu$ satisfies \eqref{ineq:mu_integrability_cond} for all $\delta  \in (0, \delta_0)$ then we may define the {\bfi Kifer exponent}
\begin{equation}
    \Lambda_{\varphi}(x) = \lim_{\delta \to 0^+} \Lambda_\delta(x).
\end{equation}

In summary, for $\mu-$a.e. $x \in X$ we have

\begin{equation}
    \Lambda_{\varphi}(x) = \lim_{\delta \to 0^+} \limsup_{t \to \infty} \frac{1}{t} \log \sup_{y \in B_x(\delta, t)} \frac{d(\varphi^t(x), \varphi^t(y))}{d(x,y)}.    
\end{equation}

We name some basic properties which are evident from the definition of the Kifer exponent.

\begin{lemma}\label{lmm:subadditivity_Adelta and f'}
    For any $s, t \in \mathbb{N}, \delta>0, x \in X$ we have

    \begin{enumerate}
        \item[(i)] $A_\delta(x, s+t) \leq A_\delta(\varphi^s(x), t) \cdot A_\delta (x, s),$ and so also
        \item[(ii)] $\log A_\delta(x, s+t) \leq \log A_\delta(\varphi^s(x), t) + \log A_\delta(x, s)$,
        \item[(iii)] $A_{\delta_1}(x, t) \leq A_{\delta_2}(x,t)$ whenever $ 0< \delta_1 < \delta_2,$ and
        \item[(iv)] $\Lambda_{\delta_1}(x) \leq \Lambda_{\delta_2}(x)$ whenever $0 < \delta_1 < \delta_2.$
    \end{enumerate}
        
\end{lemma}

Assuming additionally that $\varphi$ is Lipschitz, we may define a pointwise Lipschitz constant
\begin{equation}
    L_\varphi(x) = \lim_{ \delta \to 0^+} \sup_{y \in B_x(\delta, 0)} \frac{d(\varphi(x), \varphi(y))}{d(x,y)}
\end{equation}

Thus for $t \in \mathbb{N}, L_{\varphi^t}(x) = \lim_{\delta \to 0^+} A_\delta (x, t)$ and so $L_{\varphi^t}(x) \leq A_{\delta}(x,t)$ for all $\delta>0.$ We now relate the Kifer exponent, which is calculated as a trajectory average, to the spatial average of the pointwise Lipschitz constant, that is, the {\bfi metric spatial Lyapunov exponent} \eqref{eq:metric_spatial_lyap_exp},

\begin{equation}
    \lambda_{\varphi}(\mu) = \int_X \log L_\varphi(y) d\mu(y).
\end{equation}

\begin{theorem}\label{thm:bound1_kiferexp}
    Suppose $\mu \in \mathcal{P}(X)$ is ergodic (see Section \ref{subsec:ds, erg, mixing}) for $\varphi$. Then we have the following upper bound for the exponent $\Lambda_{\varphi}$:

    \begin{equation}\label{ineq:bd1_kexp}
        \Lambda_{\varphi}(x) \leq  \lambda_{\varphi}(\mu) \quad \mu - \text{a.e.}
    \end{equation}

\end{theorem}

\noindent {\bf Proof.} By Lemma \ref{lmm:subadditivity_Adelta and f'} (iii), $A_\delta(x,1)$ is monotone in $\delta$ at each $x \in X$. Thus by \eqref{ineq:mu_integrability_cond}, $\log L_\varphi(\cdot)$ is $\mu-$integrable and by the dominated convergence theorem 

\begin{equation}
    \lim_{\delta \to 0^+} \int_X \log A_\delta(y, 1) d\mu(y) = \int_X \log L_\varphi(y) d\mu(y).
\end{equation}

By Lemma \ref{lmm:subadditivity_Adelta and f'} (ii) we have $\log A_\delta(x,t) \leq \sum_{i=0}^{t-1} \log A_\delta(\varphi^i(x), 1)$ for all $x \in X, t \in \mathbb{N}.$ Birkhoff's Ergodic theorem tells us then that
\begin{equation}
    \Lambda_\delta(x) = \lim_{t \to \infty} \frac{1}{t} \log A_\delta(x,t) \leq \lim_{t \to \infty} \frac{1}{t}\sum_{i=0}^{t-1} \log A_\delta(\varphi^i(x), 1) = \int_X \log A_\delta(y, 1)d\mu(y) \quad \mu-\text{a.e.}    
\end{equation}
Taking the limit as $\delta \longrightarrow 0^+$ yields \eqref{ineq:bd1_kexp}.$\quad \blacksquare$

% We also have the following bound for fixed points:

% \begin{theorem}\label{thm:fp_le_bound}
%     Suppose $x \in X$ is a fixed point of $\varphi$. Then 

%     \begin{equation}\label{ineq:fp_le_bd}
%         \Lambda_{\varphi}(x) \leq \log L_\varphi(x).
%     \end{equation}
    
% \end{theorem}

% \noindent {\bf Proof.} Since $x$ is a fixed point, Lemma \ref{lmm:subadditivity_Adelta and f'} (ii) gives us $\frac{1}{t} \log A_\delta(x, t) \leq \frac{1}{t}\sum_{i=0}^{t-1} \log A_\delta(\varphi^i(x), 1) = \log A_\delta (x,1)$ for all $t \in \mathbb{N}.$ Taking the limit $\delta \longrightarrow 0^+$ yields \eqref{ineq:fp_le_bd}. Alternatively, note that the Dirac measure $\delta_x(\cdot)$ is ergodic for $\varphi$, and so taking $\mu = \delta_x$ yields \eqref{ineq:fp_le_bd}.$\quad \blacksquare$

\begin{remark}
    \normalfont It may similarly be proved that $\lambda_{\varphi}(\mu)$ dominates the exponent defined in  \cite{MarioBessa2012DiscreteandContinuousDynamicalSystems, Luis_Barreira_2005_Discrete_and_Continuous_Dynamical_Systems, MORALES_THIEULLEN_VILLAVICENCIO_2018}, $\mu-$a.e.
\end{remark}

The big difficulty in defining a Lyapunov exponent for metric spaces is that it is not clear how to define a substitute for the derivative of a map which satisfies the cocycle property. The current definitions satisfy only a subcocycle property. The metric spatial Lyapunov exponent, on the other hand, is not defined in terms of the limiting separation of a trajectory for nearby initial conditions. Rather, it looks at a spacial average of the one step separation induced by the map. Due to the subcocycle property, it necessarily upper bounds the previously defined exponents.

In \cite{abarbanel_local_lyap_exps}, the authors define local Lyapunov exponents for finite iterations of the dynamical system map, and consider the averages of these exponents with respect to the invariant measure. In this context, the metric spatial Lyapunov exponent \eqref{eq:metric_spatial_lyap_exp}, corresponds to the spatial average of the local Lyapunov exponent for one iteration of the dynamical system map.

\subsection{Metrics on probability spaces.}\label{subsec:metrics on prob spaces}

To study the dynamical system $\phi_* \colon \mathcal{P}(\mathcal{M}) \longrightarrow \mathcal{P}(\mathcal{M})$, it is necessary to establish a topology on $\mathcal{P}(\mathcal{M}).$ In this section, we define the weak topology on the space of measures and consider various possible metrics for it.
% More information on the {\bfi Wasserstein metric} is included in the Appendix Section \ref{subsec:wass_dist}, while the {\bfi Maximum Mean Discrepancy (MMD)} metric is discussed in greater detail in Section \ref{subsec:mmd}.

\medskip

\subsubsection{ The weak topology on $\mathcal{P}(X)$.}\label{subsec:cts_maps_P(X)} Let $(X, d_X)$ and $(Y, d_Y)$ be separable metric spaces and let $C_b(X, Y)$ be the space of continuous bounded functions $g \colon X \longrightarrow Y.$ On this space we impose the uniform topology induced by the {\bfi uniform metric} 

$$d_\infty(g,\hat{g}):= \sup_{x \in X} d(g(x),\hat{g}(x)). $$

When $X$ is compact, $C_b(X, \mathbb{R})$ is a Banach space under the uniform norm $\| g \|_\infty = \sup_{\|x\|_\infty \leq 1} \left|g(x)\right|$ and, by the Riesz-Markov-Kakutani representation theorem \cite{Rudin:real:analysis}, its dual is the space of finite signed (Baire) measures $\mathcal{SM}(X)$ with the norm

$$\| \mu \| = \sup_{ \|g \|_\infty \leq 1} \left|\int_X g d\mu \right|.$$

The topology induced by this norm is called the {\bfi strong topology}. When restricted to $\mathcal{P}(X)$, we retain the same name. Under this norm, the Perron-Frobenius operator acts as an isometry, and so convergence is impossible in this topology. For this reason, we choose to use the {\bfi weak topology} on $\mathcal{P}(X)$. It is defined as the coarsest topology such that the map $\mu \mapsto \int_X g d\mu$ is continuous for all $g \in C_b(X, \mathbb{R}),$ the space of continuous bounded functions on $X$. Thus $\mu_n \longrightarrow \mu$ whenever $\int_X g d\mu_n \longrightarrow \int_X g d\mu$ for all $g \in C_b(X, \mathbb{R})$. Separability of $X$ guarantees that the weak topology on $\mathcal{P}(X)$ is metrizable and separable. If, in addition, $X$ is compact or complete, the space $\mathcal{P}(X)$ will also possess the corresponding property \cite{billing_convergence_of_prob}. Note that the strong and weak topologies can be defined without requiring $X$ to be compact.

% We have already spoken of the weak topology for the space of measures. Here we define this precisely and consider various properties of metrics on this space. Let $(X, d)$ be a Polish space, that is, $X$ is a separable complete metric space with metric $d$. We define the weak topology on the space $\mathcal{P}(X)$ of measures on $X$ as the coarsest topology such that the map $\mu \mapsto \int_X f d\mu$ is continuous for all $\varphi \in C_b(X, \mathbb{R}),$ the space of continuous bounded functions on $X$. Thus $\mu_n \longrightarrow \mu$ whenever $\int_X f d\mu_n \longrightarrow \int_X f d\mu$ for all $\varphi \in C_b(X, \mathbb{R})$. Since $X$ is taken to be Polish, the space $\mathcal{P}(X)$ with the weak topology will also be Polish, and in particular metrizable. (In fact to guarantee metrizability of $\mathcal{P}(X)$, it is enough to require that $X$ be metrizable and separable. Requiring completeness of $X$, which in turn gives us completeness of $\mathcal{P}(X)$ will allow us to accomplish much more, and so we make this assumption throughout.) If, in addition, $X$ is compact, so is the space $\mathcal{P}(X).$

\medskip

{\bf Continuous maps on $\mathcal{P}(X)$.} Let $d_{\mathcal{P}} \colon \mathcal{P}(X) \times \mathcal{P}(X) \longrightarrow \mathbb{R}^{\geq 0}$ be a metric for the weak topology on $\mathcal{P}(X).$ Then $d_{\mathcal{P}}$ is continuous, and so in particular, $\mu_n \longrightarrow \mu$ in the weak topology if and only if $d_{\mathcal{P}}(\mu_n, \mu) \longrightarrow 0.$ We mention two simple results in connection with the weak topology on the space of probability measures, which are used throughout the proofs of this paper. For the first result, the authors could not find a precise reference, though \cite{Bog07} Theorem 8.4.1 (iii) is close.

\begin{lemma}\label{lmm:pushforward_cts_in_f}
    Let $(X,d_X)$ and $(Y, d_Y)$ be separable metric spaces with $(Y, d_Y)$ bounded. For any $\mu \in \mathcal{P}(X),$ the map $\varphi \in C(X,Y) \mapsto \varphi_* \mu \in \mathcal{P}(Y)$ is continuous with respect to the uniform topology on $C(X,Y)$ and the weak topology on $\mathcal{P}(Y).$
\end{lemma}

\noindent {\bf Proof.} Since the uniform topology on $C(X,Y)$ is a metric space, it is enough to show sequential continuity of the map. Let $\varphi_n \longrightarrow f$ in the uniform topology and consider any $g \in C_b(Y, \mathbb{R}).$ Then $g \circ \varphi_n \longrightarrow g\circ \varphi$ pointwise on $X,$ and so by the Dominated Convergence Theorem,

\begin{align}
    \lim_{n \to \infty} \int_Y g d (\varphi_{n})_* \mu &= \lim_{n \to \infty} \int_X g \circ \varphi_n d \mu\\
    &= \int_X \lim_{n \to \infty} g \circ \varphi_n d\mu\\
    &= \int_X g \circ \varphi d\mu = \int_Y g d\varphi_* \mu,
\end{align}

so that indeed $(\varphi_{n})_* \mu \longrightarrow \varphi_*\mu$ in the weak topology. $\quad \blacksquare$

\begin{lemma}[\cite{Bog07} Theorem 8.4.1 (i)]\label{lmm:pushforward_cts_in_nu}
    Let $(X,d_X)$ and $(Y, d_Y)$ be separable metric spaces. For any $\varphi \in C(X,Y)$, the map $\mu \in \mathcal{P}(X) \longrightarrow \varphi_* \mu \in \mathcal{P}(Y)$ is continuous with respect to the weak topology on $\mathcal{P}(X)$ and $\mathcal{P}(Y).$
\end{lemma}

\noindent {\bf Proof.} Again it is enough to check sequential continuity. Let $\mu_n \longrightarrow \mu$ in the weak topology. Then for any $g \in C_b(Y, \mathbb{R})$, 

\begin{align}
    \lim_{n \to \infty} \int_Y g d \varphi_* \mu_n &= \lim_{n \to \infty} \int_X g \circ \varphi d\mu_n\\
    &= \int_X g \circ \varphi d\mu = \int_Y g d \varphi_*\mu. \quad \blacksquare
\end{align}

{\bf Metrics on probability spaces.} Many different metrics for the weak topology exist, among these the L\'{e}vy-Prokhorov distance, the bounded Lipschitz distance, the weak-\textasteriskcentered{} distance, and the Toscani distance \cite{Dudley_2002}. A particular class which encompasses many of these is that of {\bfi integral probability metrics (IPMs).} \cite{Muller_1997} Let $\mathcal{G}$ be a class of real-valued bounded measurable functions on $X$. Then we may define the pseudometric
% Here we consider the {\bfi Wasserstein distance} and the {\bfi MMD} (Maximum Mean Discrepancy) metric and discuss some of their properties.

\begin{equation}\label{eq:integral_prob_metric}
    d_{\mathcal{G}}(\mu, \hat{\mu}) := \sup_{g \in \mathcal{G}} \left| \int_X g d\mu - \int_X g d\hat{\mu} \right|.
\end{equation}

Depending on our choice of $\mathcal{G}$, different metrics may be obtained \cite{sriperumbudur_hilbert_space_embedding}:

\begin{enumerate}
    \item $\mathcal{G} = C_b(X, \mathbb{R})$ yields a metric metrizing the strong topology on $\mathcal{P}(X).$
    \item $\mathcal{G} = \{ g : \| g \|_\infty \leq 1\}$ where $\|g\|_\infty = \sup_{x \in X} |g(x)|$ yields the total variation metric, also metrizing the strong topology.
    \item $\mathcal{G} = \{ g : \| g \|_L \leq 1\}$ where $\|g \|_L = \sup\{ |g(x) - g(y)|/d(x,y) : x \neq y, x,y\in X \}$ yields the Kantorovich distance, which coincides with Wasserstein-1 distance, since $X$ is separable. (Note that this distance is only defined for probability measures with finite first moment.) When $X$ is bounded, the Kantorovich distance metrizes the weak topology.
    \item $\mathcal{G} = \{ g : \| g \|_{BL} \leq 1 \}$, where $\|g\|_{BL} = \|g\|_L + \|g\|_\infty$ yields the Dudley metric. When $X$ is bounded, the Dudley metric also metrizes the weak topology.
    \item $\mathcal{G} = \{ g : \| g \|_H \leq 1 \}$ where $H$ is the reproducing kernel Hilbert space associated to a bounded continuous kernel $K$ which is characteristic. This yields the Maximum Mean Discrepancy (MMD) metric, which metrizes the weak topology. See the Appendix \ref{subsec:mmd} later on for a detailed description.
\end{enumerate}

While the main theorems of our paper hold for any metric that defines the weak topology on the space of measures, two metrics will be particularly useful; we review them in the following two subsections. First, {\bfi Wasserstein metrics} are particularly compatible with the Perron-Frobenius operator, and allow us to formulate more explicit bounds in certain cases. Second, in Section \ref{sec:numerics}, we use the {\bfi Maximum Mean Discrepancy (MMD)} metric, which exhibits excellent computational efficiency and enables statistical hypothesis testing.
%Details on the Wasserstein distance can be found in Section \ref{subsec:wass_dist}, while the MMD metric and kernels are discussed in greater detail in the context of our numerics in Section \ref{subsec:mmd}.

\subsubsection{The Wasserstein distance.}\label{subsec:wass_dist} Wasserstein metrics, and in particular the Wasserstein-1 distance, have some properties which make them especially suitable for studying the action of the Perron-Frobenius operator on probability distributions. Let $(X, d_X)$ be a separable metric space. In this section we moreover require $X$ to be complete, making it a Polish space. For $p \geq 1$, the Wasserstein distance of order $p$ is given by \cite{villani2009:optimal1}
\begin{equation}
    W_p(\mu, \hat{\mu}) = \inf_{\pi \in \Pi(\mu, \hat{\mu})} \left\{ \left( \int_X d_X(x,y)^p d\pi(x,y)\right)^{\frac{1}{p}} \right \},
\end{equation}
where $\Pi(\mu, \hat{\mu})$ is the set of all couplings between the measures $\mu$ and $\hat{\mu}$ on $X$. The distance is defined for measures coming from the Wasserstein$-p$ space $\mathcal{P}_p(X)$  given by
\begin{equation}
    \mathcal{P}_p(X) = \left\{ \mu \in \mathcal{P}(X) \colon \int_X d_X(x_0, x)^p d\mu(x)<\infty \right\},
\end{equation}
which is independent of the choice of $x_0$. For $p=1$, we have a duality formula relating the Wasserstein-1 distance to the Kantorovich-Rubinstein distance:
\begin{equation}
    W_1(\mu, \hat{\mu}) = \sup_{\|g\|_{\text{Lip}}\leq 1} \left\{ \int_X g d\mu - \int_X g d\hat{\mu} \right\}.
\end{equation}

Thus, the Wasserstein-1 distance is an integral probability metric. When the metric $d$ on $X$ is bounded (which can be guaranteed either by requiring that $X$ be compact or by switching to the equivalent metric $\tilde{d} = d/(1+d)$) we have $\mathcal{P}_p(X) = \mathcal{P}(X)$ for all $p \in [1, \infty)$, and the Wasserstein distances are all equivalent, inducing the weak topology on $\mathcal{P}(X)$ \cite{villani2009:optimal1}[Corollary 6.13]. Additionally, we have the following properties, which follow readily from the results in \cite[Chapter 6]{villani2009:optimal1} and \cite[Chapter 7]{ambrosio2008gradient}.

\begin{lemma}\label{lmm:wass_fftilde}
    Consider two Polish spaces $(X,d_X)$ and $(Y,d_Y)$ and two measurable maps $\varphi, \hat{\varphi} \colon X \longrightarrow Y.$ Then for any probability measure $\mu$ on $X$,
    %such that $\varphi_* \mu$ and $\hat{\varphi}_* \mu$ have bounded support,

    \begin{equation}
        W_1(\varphi_*\mu, \hat{\varphi}_*\mu) \leq \int_X d_Y(\varphi(x), \hat{\varphi}(x)) d \mu(x).
    \end{equation}
\end{lemma}

\noindent {\bf Proof.} By Kantorovich-Rubinstein duality, we have

\begin{align}
    W_1(\varphi_*\mu, \hat{\varphi}_*\mu) &= \sup_{g \in C(Y, \mathbb{R}), \text{Lip}(g) \leq 1} \int_Y g(y) d (\varphi_* \mu(y) - \hat{\varphi}_* \mu(y))\\
    &=\sup_{g \in C(Y, \mathbb{R}), \text{Lip}(g) \leq 1} \int_X g\circ \varphi(x) - g \circ \hat{\varphi} (x) d\mu(x) \\
    &\leq \int_X d_Y(\varphi(x), \hat{\varphi}(x)) d\mu(x). \quad \blacksquare
\end{align}

% \begin{remark}
%     \normalfont
%     We require $\varphi_* \mu$ and $\hat{\varphi}_* \mu$ to have bounded support so that the supremum on the right is finite.
% \end{remark}

% We have the following corollary to Lemma \ref{lmm:wass_fftilde}.

\begin{corollary}\label{corr:wass_fftilde}
    Consider two Polish spaces $(X,d_X)$ and $(Y,d_Y)$ and two bounded measurable maps $\varphi, \hat{\varphi} \colon X \longrightarrow Y.$ Then for any probability measure $\mu$ on $X$,

    \begin{equation}
        W_1(\varphi_*\mu, \hat{\varphi}_*\mu) \leq d_\infty(\varphi, \hat{\varphi}).
    \end{equation}
    
\end{corollary}

\begin{lemma}\label{lmm:wass_mumutilde_Lipschitz}
    Consider two Polish spaces $(X,d_X)$ and $(Y, d_Y)$ and a map $\varphi \colon X \longrightarrow Y$. If $\varphi$ is Lipschitz, then so is $\varphi_*$ in the Wasserstein $p$-distance, for any $p \in [ 1, \infty ).$ Moreover, any Lipschitz constant $L$ for $\varphi$ is also a Lipschitz constant for $\varphi_*,$ that is, for any two probability measures $\mu, \hat{\mu}$ on $X$

    \begin{equation}
        W_p(\varphi_* \mu, \varphi_* \hat{\mu}) \leq L W_p(\mu, \hat{\mu}).
    \end{equation}
    
\end{lemma}

\noindent {\bf Proof.} Consider a coupling $\pi$ of $(\mu, \hat{\mu}).$ Defining $F \colon X \times X \longrightarrow Y \times Y$ by $F(x_1,x_2) = (\varphi(x_1), \varphi(x_2)),$ we have that $F_* \pi$ is a coupling of $(\varphi_*\mu, \varphi_*\hat{\mu})$. In general, let $\Pi(\mu_1, \mu_2)$ denote the set of all couplings of two measures $(\mu_1, \mu_2).$ Then $F_* \Pi(\mu, \hat{\mu}) \subseteq \Pi(\varphi_*\mu, \varphi_* \hat{\mu}).$ Thus

\begin{align}
    W_p(\varphi_*\mu, \varphi_*\hat{\mu}) &= \inf_{\pi \in \Pi(\varphi_*\mu, \varphi_* \hat{\mu})} \left(\int_{Y \times Y} d_Y(y_1,y_2)^p d\pi(y_1,y_2)\right)^{1/p}\\
    &\leq \inf_{\pi \in F_*\Pi(\mu, \hat{\mu})} \left(\int_{Y \times Y} d_Y(y_1,y_2)^p d\pi(y_1,y_2)\right)^{1/p}\\
    &= \inf_{\pi \in \Pi(\mu, \hat{\mu})} \left(\int_{Y \times Y} d_Y(y_1,y_2)^p d F_*\pi(y_1,y_2)\right)^{1/p}\\
    &= \inf_{\pi \in \Pi(\mu, \hat{\mu})} \left(\int_{X \times X} d_Y(\varphi(x_1),\varphi(x_2))^p d\pi(x_1,x_2)\right)^{1/p}\\
    &\leq \inf_{\pi \in \Pi(\mu, \hat{\mu})} L \left(\int_{X \times X} d_X(x_1,x_2)^p d\pi(x_1,x_2)\right)^{1/p}\\
    &= L W_p(\mu, \hat{\mu}). \quad\blacksquare
\end{align}

\subsubsection{The Maximum Mean Discrepancy (MMD) metric.}
\label{subsec:mmd} 
The MMD metric is used in our statistical analysis of the numerics in Section \ref{sec:numerics}. The reasons for our choice of the MMD metric are two-fold: (i) the MMD metric is computationally simple to implement and, additionally, fast linear time algorithms for approximating it exist \cite{gretton12a}, which make it feasible to use when dealing with distances between large samples which need to be calculated at every time step across a long trajectory; (ii) a theory has been developed which allows us to perform non-parametric hypothesis testing using the MMD metric. The second point is desirable in our setting, since our systems are highly nonlinear and thus destroy any initial distribution from which we might sample points. Furthermore, a nonparametric hypothesis-testing framework enables us to apply it easily across different dynamical systems.

Let $K \colon X \times X \longrightarrow \mathbb{R}$ be a positive definite kernel, that is, $K$ is symmetric and furthermore satisfies

\begin{equation}
    \sum_{i=1}^p \sum_{j=1}^p c_i c_j K(x_i, x_j) \geq 0 \quad \text{for all } x_1, \dots, x_p \in X, c_1, \dots, c_p \in \mathbb{R}, p \in \mathbb{N}.
\end{equation}

By the Moore-Aronszajn theorem, $K$ induces a unique Hilbert space $(H, \langle \cdot, \cdot \rangle)$, called the {\bfi reproducing kernel Hilbert space (RKHS) associated to $K$}. The RKHS $H$ is the closure of the span of the functions of the type $K_x = K(x, \cdot) \colon X \longrightarrow \mathbb{R}$ for $x \in X,$ and the inner product is defined such that for all $g \in H$, and all $x \in X$, $g(x) = \langle g, K_x \rangle$ for all $x \in X.$ Using the integral probability metric definition \eqref{eq:integral_prob_metric} with $\mathcal{G} = \{ g \in H : \|g\|_H \leq 1 \}$, we can define the following semidistance on $\mathcal{P}(X)$:

\begin{equation}\label{eq:mmd_def}
   d_K(\mu, \hat{\mu}) = \sup_{\|g\|_H \leq 1} \left\{ \left | \int_X g d\mu - \int_X g d\hat{\mu} \right|\right\}.
\end{equation}

Another way of viewing \eqref{eq:mmd_def} is by noting that the kernel embedding $x \mapsto K_x$, which embeds $X$ as a subspace of the RKHS $H$, also produces a map $\mu \in \mathcal{P}(X) \mapsto K_\mu \in H$ where $K_{\mu} := \int_X K_x d\mu(x).$ As long as $\int_X \sqrt{ K(x,x) } d\mu(x) < \infty$ (which will be satisfied, for example, if we work with bounded $K$), $K_\mu \in H$ (\cite{gretton12a}, Lemma 3). Thus the MMD distance between two measures $\mu$ and $\hat{\mu}$ becomes the distance $d_K(\mu, \hat{\mu}) = \| K_\mu - K_{\hat{\mu}} \|_H$ between their images under this map in the space $H$. When this latter map is injective, the semidistance becomes a metric, and we call the kernel $K$ {\bfi characteristic}.

For $X$ compact, we say that the kernel $K$ is {\bfi universal} if its RKHS $H$ is dense in $C(X, \mathbb{R}),$ the space of continuous functions on $X$, with respect to the uniform norm. If $K$ is a universal continuous kernel on the compact space $X$, it is characteristic and $d_K$ metrizes the weak topology on $\mathcal{P}(X)$ \cite{sriperumbudur_hilbert_space_embedding}. Many well-known kernels are universal, including the Gaussian (RBF) and Laplacian kernels on compact finite-dimensional Euclidean space.

Using the inner product of the RKHS $H$, \eqref{eq:mmd_def} may be rewritten in a form that is more convenient for calculation by noting that

\begin{align}
   d_K(\mu, \hat{\mu})^2 &= \left\| \int_X K_x d\mu(x) - \int_X K_x d\hat{\mu}(x) \right\|_H^2 \\
    &= \int_X\int_X K(x,y) d\mu(x) d\mu(y) - 2 \int_X \int_X K(x,y) d\mu(x) d\hat{\mu}(y) + \int_X \int_X K(x,y) d\hat{\mu}(x) d\hat{\mu}(y).
\end{align}

For i.i.d. samples $\underline{x} = (x_1, \dots, x_p)$ and $\underline{y} = (y_1, \dots, y_q)$, sampled from two probability distribution $\mu, \hat{\mu} \in \mathcal{P}(X)$ respectively, an unbiased estimator of $d_K(\mu, \hat{\mu})^2$ is given by

\begin{equation}\label{eq:mmd_unbiased_def}
    d_{K,u}(\underline{x}, \underline{y})^2 = \frac{1}{p(p-1)} \sum_{i\neq j} K(x_i, x_j) + \frac{1}{q(q-1)} \sum_{i \neq j} K(y_i, y_j) - \frac{2}{pq} \sum_{i=1}^p \sum_{j=1}^q K(x_i, y_j).    
\end{equation}

Alternatively, a biased estimator is given by

\begin{equation}\label{eq:mmd_biased_def}
    d_{K,b}(\underline{x}, \underline{y})^2 = \frac{1}{p^2} \sum_{i, j}^p K(x_i, x_j) + \frac{1}{q^2} \sum_{i, j}^q K(y_i, y_j) - \frac{2}{pq} \sum_{i=1}^p \sum_{j=1}^q K(x_i, y_j).    
\end{equation}

Since \eqref{eq:mmd_unbiased_def} is unbiased, it may happen that it is negative since the kernel is positive definite. On the other hand, \eqref{eq:mmd_biased_def} is always nonnegative. The computation time for both statistics is $O( (p+q)^2 )$. In the case of a large sample size, a statistic exists which is an unbiased estimator of $d_K$, and which may be computed in linear time \cite{gretton12a}. Despite the quadratic-time computational cost, in our numerical experiments we use the biased estimator \eqref{eq:mmd_biased_def}. The reasons for this are threefold: (i) it produces a statistic which is always nonnegative, which is more realistic given that we are estimating $d_K(\mu, \hat{\mu})^2,$ (ii) while a statistic which may be computed in linear time is advantageous for large samples, such large samples would first need to be generated and their trajectories calculated using the state-space system, and this would become the bottleneck in calculations, (iii) the biased statistic proves empirically much more accurate than the linear time unbiased statistic, and as we are about to see, the biased statistic converges at a faster rate to the true value than the unbiased statistics, and this is advantageous for small samples.

\medskip

\noindent {\bf Two-sample hypothesis testing with the MMD metric.} As was already mentioned, the estimates above may be used in hypothesis testing to distinguish between distributions. Here we formulate two possible hypothesis tests and give results from \cite{gretton12a} which allow us to derive acceptance criteria for both tests.
%\tod{I need better notation here, it clashes with other notation.}
We are given two i.i.d. samples $\underline{x} = (x_1, \dots, x_p), \underline{y} = (y_1, \dots, y_q)$ sampled from the probability distributions $\mu, \hat{\mu} \in \mathcal{P}(X)$ respectively. We wish to decide whether or not they are equal. The standard statistical approach is to begin by assuming that they are equal and to test whether we are able to reject this hypothesis. When we reject the null hypothesis we may be confident that with probability $1-\alpha$ the distributions are different, where $\alpha$ controls the confidence level of the test. Our probability of committing a Type I error is $\alpha$, which is typically taken very small. However, we have no measure of our confidence that we do not commit a Type II error of accepting the null hypothesis when it is in fact false. Under the conditions given there, Theorem \ref{thm:transport_of_measures_with_state_space_system_bound} shows that after a sufficiently long time, the true and predicted future distributions (or a Ces\`{a}ro average of them) remain within some small $\varepsilon$ distance of one another. To test for this, we propose as our null hypothesis that the distance between them in the MMD metric is greater than the threshold value $\varepsilon>0$. If our test statistic is very small, we may accept that this hypothesis is false and with high confidence we can assert that the true distributions are within $\varepsilon$ of one another in the MMD metric. In some sense we are saying that the two distributions are `the same', up to a $\varepsilon$ error. Note that Theorem \ref{thm:transport_of_measures_with_state_space_system_bound} only guarantees the {\bfi existence} of $\varepsilon$, and in the case of strong structural stability of the underlying system, that $\varepsilon$ may be made arbitrarily small if our proxy approximates the true system to a sufficient accuracy; it does not not provide a means for calculating the value of $\varepsilon$ for a given system and its proxy, unless the conjugating homeomorphism $g$ is known. Thus we have the two hypothesis tests:

\begin{align}
    &\text{\bf Test A } &H_0: \mu = \hat{\mu}, \, &H_a: \mu \neq \hat{\mu},\\
    &\text{\bf Test B } &H_0: d_K(\mu, \hat{\mu}) \geq \varepsilon, \, &H_a: d_K(\mu, \hat{\mu})<\varepsilon.
\end{align}

In their paper \cite{gretton12a}, Gretton et al. give distribution-free bounds on the behavior of the estimates \eqref{eq:mmd_unbiased_def} and \eqref{eq:mmd_biased_def}. From these we may derive acceptance criteria for the two proposed hypothesis tests. While this may be done for the unbiased quadratic time and unbiased linear time estimates, here we restrict ourselves to using the biased estimate, which we implemented in our numerics.

\begin{theorem}[Gretton et al. \cite{gretton12a}, Theorem 7]\label{thm:prob_mmd_bound1}
    Let $(X,d)$ be a compact metric space and $K\colon X \times X \longrightarrow \mathbb{R}$ a continuous universal kernel such that $0 \leq K(x,y) \leq \kappa$ for all $x, y \in X.$ Let $\mu, \hat{\mu} \in \mathcal{P}(X)$ be two probability measures on $X$ and suppose $\underline{x} = (x_1, \dots, x_p), \underline{y} = (y_1, \dots, y_q)$ are independently identically distributed samples from $\mu, \hat{\mu}$ respectively. Fix some $\delta>0.$ Then

    \begin{equation}
        \mathbb{P} \left\{ |d_{K}(\mu, \hat{\mu}) - d_{K,b}(\underline{x}, \underline{y})| > 2 \left( \sqrt{\kappa/p} + \sqrt{\kappa/q} \right) + \delta \right\} \leq 2 \exp \left(\frac{ - \delta^2 pq}{2\kappa(p+q)} \right).
    \end{equation}
\end{theorem}

\begin{theorem}[Gretton et al. \cite{gretton12a}, Theorem 8]\label{thm:prob_mmd_bound2}
    Under the conditions of Theorem \ref{thm:prob_mmd_bound1}, assuming moreover that $\mu = \hat{\mu}$ and $p=q$,

    \begin{equation}
        \mathbb{P} \left\{ d_{K,b}(\underline{x}, \underline{y}) \leq \sqrt{2\kappa/p} + \delta \right\} \geq 1 - \exp\left(\frac{- \delta^2p}{4\kappa} \right).
    \end{equation}
\end{theorem}

\begin{corollary}[Test A: Gretton et al. \cite{gretton12a}, Corollary 9]\label{cor:mmd_testa}
    Suppose we have samples of equal size, i.e. $p=q$. A hypothesis test of level $\alpha \in (0,1)$ for the null hypothesis $\mu = \hat{\mu}$ has acceptance region

    \begin{equation}
        d_{K,b}(\underline{x}, \underline{y}) < \sqrt{2\kappa/p} \left( 1 + \sqrt{2 \log \alpha^{-1}} \right).
    \end{equation}
\end{corollary}

\begin{corollary}[Test B]\label{cor:mmd_testb}
    Suppose we have samples of equal size, i.e. $p=q$. A hypothesis test of level $\alpha \in (0,1)$ for the null hypothesis $d_K(\mu, \hat{\mu}) \geq \varepsilon$ has acceptance region

    \begin{equation}
        d_{K,b}(\underline{x}, \underline{y}) > \varepsilon - 2 \sqrt{\kappa/p} \left( 2 + \sqrt{\log(2/\alpha)} \right).
    \end{equation}
\end{corollary}
    
A final remark which may be made is that the probability of making Type I and Type II errors using the biased estimate converge to zero at a rate of $O(p^{-1/2})$ as opposed $O(p^{-1/4})$ for tests using the unbiased estimate. This faster rate of convergence makes the biased estimate tests advantageous when dealing with smaller samples. For details on this see \cite{gretton12a}. 

\subsection{Ergodic, mixing, physical, and attracting measures of dynamical systems.}\label{subsec:ds, erg, mixing}

From a mathematical point of view, the difference in stability between point and density forecasts is the result of dealing with two different dynamical systems. The dynamical system $(\mathcal{M}, \phi)$ induces a corresponding dynamical system $(\mathcal{P}(\mathcal{M}), \phi_*)$, on the space of measures. While $\mathcal{M}$ is typically taken to be finite-dimensional, $\mathcal{P}(\mathcal{M})$ is contained in the infinite-dimensional vector space of signed measures. The induced dynamical system on the space of measures is semiconjugate to the original dynamical system by the map $x \mapsto \delta_x$. Thus, the original dynamics on $\mathcal{M}$ are contained in the dynamics on the much richer space $\mathcal{P}(\mathcal{M})$. Nevertheless, under mild conditions, the dynamical system $(\mathcal{P}(\mathcal{M}), \phi_*)$ exhibits striking stability on certain subsets of $\mathcal{P}(\mathcal{M})$. Essential to this phenomenon are the concepts of ergodic, physical, mixing, and attracting measures, each carrying different types of stability in the dynamics of probability measures, while still allowing chaos and, in particular, sensitivity to initial conditions at the trajectory level of the dynamical system.

{\bf Invariant measures.} Throughout this section, let $(X,d)$ be a separable metric space equipped with a continuous dynamical system $\varphi \colon X \longrightarrow X.$ As before, we impose the Borel sigma algebra on $X$. A Borel probability measure $\mu \in \mathcal{P}(X)$ is said to be {\bfi invariant} if $\varphi_* \mu = \mu.$ Thus $\mu$ is a fixed point of the dynamical system $(\mathcal{P}(X), \varphi_*).$ If $X$ is additionally compact, the existence of such an invariant measure is guaranteed by the Krylov-Bogliubov Theorem (\cite{Viana_Oliveira_2016_foundations_of_ergodic_theory}, Theorem 2.1). Invariant probability measures provide information about how the dynamics operate in a stationary or steady-state flow. Unfortunately, in general, a dynamical system may support uncountably many invariant probability measures. Hence, the question arises as to which measure is meaningful in the sense that experiments may actually observe it.

\subsubsection{Ergodic measures.}\label{subsec:ergodic_measures_appendix} An invariant Borel probability measure $\mu \in \mathcal{P}(X)$ is said to be {\bfi ergodic} with respect to the dynamical system $(X,\varphi)$ if every invariant Borel set $A$ either has full or zero measure, that is, for every set $A$ such that $\varphi^{-1}(A) = A,$ we have that $\mu(A) \in \{0,1\}.$ The set of invariant measures is convex, and the ergodic measures are precisely the extreme points of this set. Thus, when $X$ is compact, we are also guaranteed the existence of an ergodic measure. The study of ergodic dynamical systems is a vast subject; here, we limit ourselves to naming only a few well-known results.

For $g \in L^1(\mu),$ $x \in X,$ we write $$\tilde{g}(x) := \lim_{T \to \infty} \frac{1}{T} \sum_{t = 0}^{T-1} g( \varphi^t(x)), \,\text{if the limit exists.}$$ This is the {\bfi time average} of $g$ over a trajectory of $ \varphi.$

\begin{theorem}[Birkhoff Ergodic Theorem (\cite{Viana_Oliveira_2016_foundations_of_ergodic_theory} Theorem 3.2.3)]\label{thm:birkhoff}
    Let $(X, \varphi)$ be a dynamical system as above with invariant probability measure $\mu.$ Then for any $g \in L^1(\mu)$, the time average is well-defined, integrable, and $ \varphi-$invariant $\mu-$a.e., that is, $\tilde{g} \in L^1(\mu)$ and $\tilde{g} \circ  \varphi= \tilde{g},\,\mu-$a.e. Furthermore,

    \begin{equation}
        \int_X g d\mu = \int_X \tilde{g} d\mu.
    \end{equation}

    If $ \varphi$ is ergodic with respect to $\mu$, then $\tilde{g}$ is constant $\mu-$a.e., and equal to $\int_X g d\mu.$
\end{theorem}

% kingman subadditive theorem, other theorems?

Associated to a point $x \in X$ we may define the empirical measure
\begin{equation}
    \mu_x = \lim_{T \to \infty} \frac{1}{T} \sum_{t=0}^{T-1} \delta_{\varphi^t(x)},
\end{equation}
when the weak limit exists. Here $\delta_x$ is the Dirac measure at the point $x$. For $g \in C_b(X, \mathbb{R})$, $\int_X g d\mu_x = \lim_{T \to \infty} \frac{1}{T} \sum_{t=0}^{T-1} \delta_{\varphi^t(x)}.$ Thus, integrating with respect to the measure $\mu_x$ is the same as averaging along the trajectory started at the point $x$. For a Borel probability measure $\mu$, we define its basin as the set of those points whose empirical measure under $\varphi$ coincides with $\mu$, that is,
\begin{equation}
    \Basin_\varphi (\mu) = \left\{ x \in X : \lim_{T \to \infty} \frac{1}{T} \sum_{t=0}^{T-1} \delta_{\varphi^t(x)} = \mu \right\}. 
\end{equation}

When the dynamical system $\varphi$ is clear from context, we omit the subscript. It can be shown that a Borel probability measure $\mu$ is ergodic with respect to $\varphi$ if and only if its basin has full $\mu-$measure. So ergodic measures are attracting in the sense that, starting from $\ mu$-almost every point, the corresponding empirical measure converges to $\mu$. Moreover, as the following result shows, measures that are absolutely continuous with respect to an ergodic measure are attracted to it in a Ces\`{a}ro sense. Although this result can be derived as a corollary of the Birkhoff Ergodic Theorem \ref{thm:birkhoff}, the authors could not find it stated in this exact form in standard texts on ergodicity.

\begin{corollary}\label{cor:ergodicity_convergence}
    Let $(X,  \varphi)$ be a dynamical system, ergodic with respect to the probability measure $\mu.$ Suppose the probability measure $\nu$ is absolutely continuous with respect to $\mu.$
    %and that both measures are $\sigma$-finite. 
    Then $\lim_{T \to \infty} \frac{1}{T} \sum_{t=0}^{T-1}  \varphi^t_*\nu = \mu$ in the weak topology on $\mathcal{P}( X).$
\end{corollary}
    
\noindent {\bf Proof.} We begin by noting that for any bounded continuous $g \colon X \longrightarrow \mathbb{R},$ and $\mu-$a.e. $x \in X,$

\begin{equation}
    \lim_{T \to \infty} \frac{1}{T} \sum_{t=0}^{T-1} g \circ  \varphi^t(x) = \int_X g d\mu.
\end{equation}

Since $\nu\ll \mu,$ it has a density $\rho \in L^1(\mu)$. Let $R$ be an upper bound for $|g|$. Thus for all $T$, $\left| \rho \cdot \frac{1}{T} \sum_{t=0}^{T-1} g \circ  \varphi^t \right| \leq R\cdot |\rho| \in L^1(\mu).$ By the Dominated Convergence Theorem,

\begin{equation}
    \lim_{T \to \infty} \int_X g d \left(\frac{1}{T} \sum_{t=0}^{T-1}  \varphi^t_* \nu \right) = \lim_{T \to \infty} \int_X \rho \cdot \frac{1}{T} \sum_{t=0}^{T-1} g \circ  \varphi^t d\mu = \int_X \rho \cdot \int_X g d\mu d\mu = \int_X g d \mu.
\end{equation}

In particular, $\lim_{T \to \infty} \frac{1}{T} \sum_{t=0}^{T-1}  \varphi^t_* \nu = \mu. \quad \blacksquare$

\medskip
 
\subsubsection{Mixing measures.}\label{subsec:mixing_measures_appendix} Having defined ergodic measures, we now consider mixing measures. We begin by noting that by \cite{Viana_Oliveira_2016_foundations_of_ergodic_theory}[Proposition 4.1.4] ergodic measures may equivalently be characterized as any invariant measure $\mu$ of a dynamical system $(X,\varphi)$ such that for any Borel sets $A, B \subseteq X,$
    
    \begin{equation}
        \lim_{T \to \infty} \frac{1}{T} \sum_{t=0}^{T-1} \mu( \varphi^{-t} (A) \cap B) = \mu(A) \mu(B).
    \end{equation}

A strengthening of ergodicity is mixing. We call the measure preserving system $(X,  \varphi, \mu)$ {\bfi mixing} if for any Borel sets $A,B$

\begin{equation}
    \lim_{t \to \infty} \mu( \varphi^{-t} (A) \cap B) = \mu(A) \mu(B).
\end{equation}

Analogous to Corollary \ref{cor:ergodicity_convergence}, we have the following result.

\begin{proposition}[\cite{newman2025attractingmeasures}, Definition 10]\label{prop:strong_mixing_convergence}
    Let $(X,  \varphi, \mu)$ be a mixing dynamical system. Suppose the probability measure $\nu$ is absolutely continuous with respect to $\mu.$
    %and that both measures are $\sigma$-finite. 
    Then $ \varphi^t_*\nu \longrightarrow \mu$ in the weak topology.
\end{proposition}

Thus mixing measures are attracting in a stronger sense than that of Ces\`{a}ro. Mixing measures are stable fixed points of the dynamical system $(\mathcal{P}(X), \varphi_*),$ within the set of measures absolutely continuous with respect to the mixing measure.

\medskip

\subsubsection{Physical and attracting measures.}\label{subsec:physical_attracting_measures_appendix} In \cite{newman2025attractingmeasures} the authors discussed the important question in ergodic theory of determining, amongst the set of invariant measures supported on an attractor of a dynamical system, which one is `natural' in that it characterizes how trajectories near the set behave. This initially led to the discovery of what are now called {\bfi SRB measures} on Axiom A attractors. While a full length introduction to SRB measures is beyond the scope of this paper, we proceed to highlight two `naturalness' properties which characterize SRB measures: an SRB measure of an Axiom A attractor is both a {\bfi physical measure} and an {\bfi attracting measure.} See \cite{young_whataresrbmeasures} for a detailed discussion on SRB measures.

To define physical and attracting measures, we first fix a reference measure $\ell$ on $X$ which is locally finite and Borel. Furthermore, we require that the dynamical system $\varphi$ be {\bfi nonsingular} with respect to the reference measure, that is, for any Borel set $A$, if $\ell(A) = 0$, then $\ell(\varphi^{-1}(A)) = 0,$ i.e. $\varphi_*\ell$ is absolutely continuous with respect to $\ell.$ (See \cite{newman2025attractingmeasures} for a slight generalisation of this requirement). For Euclidean space, the natural choice of reference measure would be the Lebesgue measure. Another example is when $X$ is a finite-dimensional Riemannian manifold with $d$ the geodesic distance and $\ell$ the Riemannian volume. In this case, the condition will be fulfilled if $\varphi \in \operatorname{Diff}^1(X).$

\begin{definition}\label{def:physical_meas}
    A compactly supported measure $\mu \in \mathcal{P}(X)$ is called a {\bfi physical measure} if its support has a neighborhood $U$ such that $\ell-$almost every point of $U$ is in $\Basin (\mu).$    
\end{definition}

Analogous to Corollary \ref{cor:ergodicity_convergence}, we have the following result.

\begin{lemma}\label{lmm:physical_convergence}
    Let $(X, \varphi)$ be a dynamical system, with physical probability measure $\mu$ and neighborhood $U$ of its support as in Definition \ref{def:physical_meas}. Suppose the probability measure $\nu$ is absolutely continuous with respect to the reference measure $\ell,$ and $\nu(U) = 1.$
    %and that both measures are $\sigma$-finite. 
    Then $\lim_{T \to \infty} \frac{1}{T} \sum_{t=0}^{T-1}  \varphi^t_*\nu = \mu.$
\end{lemma}

\noindent {\bf Proof.} We define the probability measure $\ell_U$ by assigning $\ell_U(A) = \ell(A \cap U) / \ell(U)$ for every Borel set $A$. By definition $\ell_U(\Basin(\mu)) = 1.$ Thus for any bounded continuous $g \colon X \longrightarrow \mathbb{R},$ and $\ell_U-$a.e. $x \in X,$

\begin{equation}
    \lim_{T \to \infty} \frac{1}{T} \sum_{t=0}^{T-1} g \circ  \varphi^t(x) = \int_X g d\mu.
\end{equation}

Since $\nu \ll \ell$ and $\nu(U) = 1$, we have that $\nu \ll \ell_U.$ Thus it has a density $\rho \in L^1(\ell_U)$. Let $R$ be an upper bound on $|g|$. Thus for all $T$, $\left| \rho \cdot \frac{1}{T} \sum_{t=0}^{T-1} g \circ  \varphi^t \right|\leq R\cdot |\rho| \in L^1(\ell_U).$ By the Dominated Convergence Theorem,

\begin{equation}
    \lim_{T \to \infty} \int_X g d \left(\frac{1}{T} \sum_{t=0}^{T-1}  \varphi^t_* \nu \right) = \lim_{T \to \infty} \int_X \rho \cdot \frac{1}{T} \sum_{t=0}^{T-1} g \circ  \varphi^t d\ell_U = \int_X \rho \cdot \int_X g d\mu d\ell_U = \int_X g d \mu.
\end{equation}

In particular, $\lim_{T \to \infty} \frac{1}{T} \sum_{t=0}^{T-1}  \varphi^t_* \nu = \mu. \quad \blacksquare$

Physical measures are well-studied. On the other hand, attracting measures have only recently been so named in the papers \cite{AshwinNewman2021, newman_physical_measures, newman2025attractingmeasures}.
 
\begin{definition}\label{def:att_meas}
    A compactly supported measure $\mu \in\mathcal{P}(X)$ is called an {\bfi attracting measure} if there exists an open neighborhood $U$ of its support such that for every $\ell-$absolutely continuous $\nu \in \mathcal{P}(X)$ with $\nu(U) = 1$, we have $\varphi^t_*\nu \longrightarrow \mu$ as $t \longrightarrow \infty.$
\end{definition}

Let $A \subseteq X$ be nonempty and compact. $A$ is said to be (forward) {\bfi invariant} if $\varphi(A) = A.$ For such an invariant set we define its basin as $\Basin_\varphi (A) = \{ x \in X : d(\varphi^t(x), A) \longrightarrow 0 \text{ as } t \longrightarrow \infty\}.$ When the dynamical system $\varphi$ is clear from context, we omit the subscript. We will call the invariant set $A$ an {\bfi attractor} if $\Basin (A)$ is a neighborhood of $A$. We have the following alternative characterization of physical and attracting measures.

\begin{proposition}[\cite{newman2025attractingmeasures}, Proposition 20.]\label{prop:phys_att_alternate_def}
    Let $\mu \in \mathcal{P}(X)$ be such that $\supp (\mu)$ is an attractor. Then

    \begin{enumerate}
        \item $\mu$ is physical if and only if $\Basin (\mu)$ contains $\ell-$almost every point in $\Basin (\supp (\mu)),$ and
        \item $\mu$ is attracting if and only if for every $\nu \in \mathcal{P}(X)$, absolutely continuous with respect to $\ell$ with $\nu(\Basin (\supp (\mu)) ) = 1,$ we have $\varphi^t_*\nu \longrightarrow \mu$ as $t \longrightarrow \infty.$
    \end{enumerate}
\end{proposition}

A physical measure is reminiscent of an ergodic measure. Whereas for an ergodic measure $\mu$, the averaging over trajectories started from $\ mu$- almost every initial condition is, in the limit, the same as integrating with respect to $\mu$, for physical measures, the same property holds, except that it is for $\ell-$almost every initial condition in a neighborhood of the support of $\mu.$ Similarly, attracting measures bear resemblance to mixing measures. Whereas for a mixing measure $\mu$, an initial distribution $\nu$ that is absolutely continuous with respect to $\mu$ approaches $\mu$ under transport by the dynamical system in the limit as time tends to infinity, for attracting measures the same property holds as long as $\nu$ is absolutely continuous with respect to the reference measure $\ell,$ and it is supported on the closure of a neighborhood of the support of $\mu.$ Physical and ergodic measures have to do with almost sure guarantees on the empirical measure obtained from initial conditions; attracting and mixing measures have to do with the limiting distribution of random trajectories \cite{newman2025attractingmeasures}. Despite these similarities, we should be careful to note that in general ergodic measures need not be physical, and vice versa. The same goes for attracting and mixing measures.

The appeal of physical and attracting measures is that they are experimentally more meaningful. The limiting properties of an ergodic or mixing measure $\mu$ hold with respect to a full measure set or an absolutely continuous distribution with respect to the same measure $\mu$, but this may be somewhat useless, since $\mu$ might not be absolutely continuous with respect to the Lebesgue measure -- indeed $\mu$ could very well be a Dirac. On the other hand, the properties of physical and attracting measures are guaranteed with respect to the reference measure $\ell,$ in a neighborhood of the support of $\mu,$ thus giving us real information we could expect to verify in experiments.

\subsection{Structural stability.}\label{subsec:stability}

The final ingredient we need in our theorems is a stability in the topological behavior of dynamical systems under perturbations. While mixing assumptions guarantee stability in the limiting behavior of the corresponding dynamical system on the space of probability measures under {\bfi perturbations of the initial measure}, topological and structural stability assumptions will allow us to conclude that these limits do not change drastically when {\bfi the dynamical system map is perturbed}. 

Let $X, Y$ be  smooth compact manifolds. We call a map $\hat{\varphi} \colon Y \longrightarrow Y$ a {\bfi topological factor} of $\varphi \colon X \longrightarrow X$ if there is a surjective continuous map $g \colon X \longrightarrow Y$ such that $g \circ \varphi = \hat{\varphi} \circ g,$ in which case we call the map $g$ a {\bfi semiconjugacy.} We call two maps $\varphi\colon X \longrightarrow X$ and $\hat{\varphi} \colon Y \longrightarrow Y$ {\bfi topologically conjugate} if this map $g$ is a homeomorphism. Let the manifold $X$ be equipped with a metric $d$ compatible with its topology. We equip the space of $C^1$ diffeomorphisms on $X$ with the Whitney $C^1$ topology, which is metrized by the metric $ d_1$. On the other hand, we equip the space of homeomorphisms on $X$ with the uniform topology. It is metrized by the uniform metric $d_\infty(\varphi, \hat{\varphi}) = \sup_{x \in X} d(\varphi(x), \hat{\varphi}(x)).$

% Let $\varphi$ be a $C^1$ diffeomorphism. Suppose that there exists a $C^0$ neighborhood $U$ of $\varphi$ in the space of homeomorphisms on $X$ such that for every $\hat{\varphi} \in U$ there is a continuous surjective map $g_{\hat{\varphi}} \colon X \longrightarrow X$ such that

% \begin{enumerate}
%     \item[(i)] $g_{\hat{\varphi}} \circ \hat{\varphi} = \varphi \circ g_{\hat{\varphi}}$, that is, $\varphi$ is a topological factor of $\hat{\varphi}$ by $g_{\hat{\varphi}},$ and
%     \item[(ii)] $g_{\hat{\varphi}} \longrightarrow \text{id}$ as $\hat{\varphi} \longrightarrow f$ in the $C^0$ topology.
% \end{enumerate}

% In this case we call the diffeomorphism $\varphi$ {\bfi topologically stable.}

Let $\varphi$ be a $C^1$ map. We say that $\varphi$ is ($C^1$) {\bfi structurally stable} if it has a neighborhood $U$ in the $C^1$ topology such that all maps $\hat{\varphi} \in U$ are topologically conjugate to $\varphi$. Suppose in addition that for any $\hat{\varphi} \in U$ we can pick the conjugating homeomorphism $g_{\hat{\varphi}}$ such that the following property holds: $\max\{d_\infty(g_{\hat{\varphi}}, id),d_\infty(g_{\hat{\varphi}}^{-1}, id)\} \longrightarrow 0$ as $d_1(\varphi, \hat{\varphi}) \longrightarrow 0$. Then we say that $\varphi$ is ($C^1$) {\bfi strongly structurally stable.}

In the following lemma we see that topological conjugacy preserves ergodicity and mixing. A consequence of the lemma is that, when a $C^1$ map $\varphi$ is strongly structurally stable, taking a map $\hat{\varphi}$ sufficiently close to it in the $C^1$ topology will guarantee that the corresponding ergodic or mixing measures may be made arbitrarily close in the weak topology. This fact is used in Theorems \ref{thm:ergodicity_structurally_stable_bound_main} and \ref{thm:strong_mixing_structurally_stable_bound_main}.

\begin{lemma}\label{lmm:struct_stab_pres_mixing_erg}
    Let $\varphi\colon X \longrightarrow X$ and $\hat{\varphi} \colon Y \longrightarrow Y$ be $C^1$ maps on the smooth compact manifolds $X$ and $Y$ respectively, topologically conjugate by the homeomorphism $g \colon X \longrightarrow Y$. Suppose that $\mu \in \mathcal{P}(X)$ is $\varphi$-invariant. Then $\hat{\mu} := g_* \mu \in \mathcal{P}(Y)$ is $\hat{\varphi}$-invariant and moreover
    \begin{enumerate}
        \item[(i)] if $(X,\varphi,\mu)$ is ergodic, so is $(Y, \hat{\varphi}, \hat{\mu}),$ and
        \item[(ii)] if $(X,\varphi,\mu)$ is mixing, so is $(Y, \hat{\varphi}, \hat{\mu}).$
    \end{enumerate}
\end{lemma}

\noindent {\bf Proof.} Since $g \circ \varphi = \hat{\varphi} \circ g$, we have $\hat{\varphi}_* \hat{\mu} = (\hat{\varphi} \circ g)_* \mu = (g \circ \varphi)_* \mu = g_*\mu = \hat{\mu}.$ (i) Now suppose $(X, \varphi, \mu)$ is ergodic and $A \subseteq Y$ is invariant under $\hat{\varphi}$, that is $\hat{\varphi}^{-1}(A) = A.$ Then $\varphi^{-1}(g^{-1}(A)) = g^{-1}(\hat{\varphi}^{-1}(A))$, so $g^{-1}(A)$ is invariant under $\varphi$. In particular, $\hat{\mu}(A) = \mu(g^{-1}(A)) \in \{0,1\}.$ (ii) If $(X,\varphi, \mu)$ is mixing, then for measurable sets $A, B \subseteq Y$ we have

\begin{equation}
    \hat{\mu}(\hat{\varphi}^t(A) \cap B) = \mu ( g^{-1} (\hat{\varphi}^t(A)) \cap g^{-1}(B)) = \mu (\varphi^t(g^{-1}(A)) \cap g^{-1}(B)) \longrightarrow \hat{\mu}(A) \cdot \hat{\mu}(B) \quad \text{as } t \longrightarrow \infty. \quad \blacksquare
\end{equation}

Lemma \ref{lmm:struct_stab_pres_att_phys} gives the same result for physical and attracting measures. In this case we require an extra condition on the nonsingularity of the conjugating homeomorphism with respect to the reference measures. To simplify the proof, we first check the following result.

\begin{lemma}\label{lmm:top_conj_basin_supp}
    Let $\varphi\colon X \longrightarrow X$ and $\hat{\varphi} \colon Y \longrightarrow Y$ be $C^1$ maps on the smooth compact manifolds $X$ and $Y$ respectively, topologically conjugate by the homeomorphism $g \colon X \longrightarrow Y$. Let $\mu \in \mathcal{P}(X)$ and $\hat{\mu} := g_*\mu.$ Then
    \begin{enumerate}
        \item[(i)] $g(\supp (\mu)) = \supp (\hat{\mu})$,
        \item[(ii)] $g(\Basin_\varphi(\mu)) = \Basin_{\hat{\varphi}}(\hat{\mu})$,
        \item[(iii)] $\supp (\mu)$ is invariant under $\varphi$, if and only if $\supp (\hat{\mu})$ is invariant under $\hat{\varphi},$
        \item[(iv)] if $\supp (\mu)$ is invariant, then $g(\Basin_\varphi(\supp (\mu))) = \Basin_{\hat{\varphi}}(\supp (\hat{\mu})),$ and
        \item[(v)] $\supp (\mu)$ is an attractor for the dynamical system $\varphi$, if and only if $\supp (\hat{\mu})$ is an attractor for the dynamical system $\hat{\varphi}.$
    \end{enumerate}
\end{lemma}

\noindent {\bf Proof.} (i) Let $y =g(x) \in g(\supp (\mu))$ and consider an open neighborhood $U$ of $y.$ Then $g^{-1}(U)$ is an open neighborhood of $x \in \supp (\mu)$ and hence $\hat{\mu}(U) = \mu(g^{-1}(U)) >0.$ Thus $y \in \supp (\hat{\mu}).$ Since $g$ is a homeomorphism, we may write $\mu = g^{-1}_* \hat{\mu}$, giving the reverse inclusion.

(ii) Now consider $y = g(x) \in g(\Basin_\varphi(\mu)).$ Since $x \in \Basin _\varphi(\mu),$ we have that $\frac{1}{T} \sum_{t=0}^{T-1} \delta_{\varphi^t(x)} \longrightarrow \mu$ weakly as $T \longrightarrow \infty.$ Now, note that

\begin{equation}
    \frac{1}{T} \sum_{t=0}^{T-1} \delta_{\hat{\varphi}^t(g(x))} = \frac{1}{T} \sum_{t=0}^{T-1} \delta_{g(\varphi^t(x))} = g_* \left(\frac{1}{T} \sum_{t=0}^{T-1} \delta_{\varphi^t(x)}. \right)
\end{equation}

By Lemma \ref{lmm:pushforward_cts_in_nu}, $\frac{1}{T} \sum_{t=0}^{T-1} \delta_{\hat{\varphi}^t(g(x))} \longrightarrow g_*\mu = \hat{\mu}$ as $T \longrightarrow \infty.$ Thus $y = g(x) \in \Basin _{\hat{\varphi}}(\hat{\mu}).$ Once again, using the fact that $\mu = g^{-1}_* \hat{\mu}$ gives the reverse inclusion.

Claim (iii) follows directly from (i). As for claim (iv), taking $y=g(x) \in g(\Basin _\varphi(\supp (\mu))),$ we know that $d(\varphi^t(x), \supp (\mu)) \longrightarrow 0$ as $t \longrightarrow \infty.$ Thus by continuity of $g$, $d(\hat{\varphi}^t(y), \supp (\hat{\mu})) = d(g \circ \varphi^t(x), g(\supp (\mu))) \longrightarrow 0 $ as $t \longrightarrow \infty.$ So $y \in \Basin _{\hat{\varphi}} (\supp (\hat{\mu})).$ The reverse inclusion follows from continuity of $g^{-1}.$ Finally, claim (v) follows from (iii) and (iv). $\quad \blacksquare$

\begin{remark}
    \normalfont
    As is evident from the proof, Lemma \ref{lmm:top_conj_basin_supp} holds more generally when $\varphi, \hat{\varphi}$ are continuous and $(X, d_X)$, $(Y, d_Y)$ are separable metric spaces. In this case, we note that $\supp (\mu)$ is compact if and only if $\supp (\hat{\mu})$ is. 
\end{remark}

\begin{lemma}\label{lmm:struct_stab_pres_att_phys}
    Let $\varphi\colon X \longrightarrow X$ and $\hat{\varphi} \colon Y \longrightarrow Y$ be $C^1$ maps on the smooth compact manifolds $X$ amd $Y$ respectively, topologically conjugate by the homeomorphism $g \colon X \longrightarrow Y$. Fix locally finite Borel reference measures $\ell_X$ and $\ell_Y$ on $X$ and $Y$ respectively such that $\varphi$ is nonsingular with respect to $\ell_X.$ Furthermore suppose that $g_* \ell_X \ll \ell_Y$ and $g^{-1}_* \ell_Y \ll \ell_X.$ Then $\hat{\varphi}$ is nonsingular with respect to $\ell_Y.$ Suppose that $\mu \in \mathcal{P}(X)$ is $\varphi$-invariant and take $\hat{\mu} := g_* \mu \in \mathcal{P}(Y)$. Then
    \begin{enumerate}
        \item[(i)] if $\mu$ is a physical measure for $(X,\varphi,\ell_X)$ with neighborhood $U$ of its support as in Definition \ref{def:physical_meas}, so is $\hat{\mu}$ for $(Y, \hat{\varphi}, \ell_Y),$ with neighborhood $g(U)$ and
        \item[(ii)] if $\mu$ is an attracting measure for $(X,\varphi,\ell_X)$ with neighborhood $U$ of its support as in Definition \ref{def:att_meas}, so is $\hat{\mu}$ for $(Y, \hat{\varphi}, \ell_Y)$ with neighborhood $U$.
    \end{enumerate}
\end{lemma}

\noindent {\bf Proof.} By the nonsingularity requirements on $g, g^{-1}$ and $\varphi$, we have $\hat{\varphi}_* \ell_Y = (g \circ \varphi \circ g^{-1})_* \ell_Y \ll \ell_Y.$ Thus $\hat{\varphi}$ is nonsingular with respect to $\ell_Y.$
(i) Suppose that $\mu$ is a physical measure for $(X, \varphi, \ell_X).$ Let $U$ be a neighborhood of $\supp (\mu)$ such that $\ell_X(\Basin (\mu) \cap U) = \ell_X(U).$ Then by Lemma \ref{lmm:top_conj_basin_supp} (i), $g(U)$ is a neighborhood of $\supp (\hat{\mu}).$ Now $\ell_X( U \setminus \Basin (U)) = 0, $ so by Lemma \ref{lmm:top_conj_basin_supp} (ii), $\ell_Y(g(U) \setminus \Basin (\hat{\mu})) = g^{-1}_* \ell_Y ( U \setminus \Basin (\mu)) = 0.$ Thus $\hat{\mu}$ is physical.

(ii) Suppose that $\mu$ is an attracting measure for $(X, \varphi, \ell_X).$ Let $U$ be a neighborhood of $\operatorname{{supp}}(\mu)$ such that for any $\nu \in \mathcal{P}(X)$, absolutely continuous with respect to $\ell_X$ and satisfying $\nu(U) = 1$, we have $\varphi^t_* \nu \longrightarrow \mu$ as $t \longrightarrow \infty.$ Now, $g(U)$ is a neighborhood of $\supp (\hat{\mu})$. Consider any $\nu \in \mathcal{P}(Y)$, absolutely continuous with respect to $\ell_Y$ and with $\nu(g(U)) = 1.$ Then $g^{-1}_* \nu$ is absolutely continuous with respect to $\ell_X$ and $g^{-1}_*\nu(U) = 1.$ Thus, $\varphi^t_*g^{-1}_* \nu \longrightarrow \mu$ as $t \longrightarrow \infty.$ By Lemma \ref{lmm:pushforward_cts_in_nu}, $\hat{\varphi}^t_* \nu = g_* \varphi^t_* g^{-1}_* \nu \longrightarrow g_* \mu = \hat{\mu}$ as $ t \longrightarrow \infty$.
% \begin{equation}
%     \hat{\varphi}^t_* \nu = g_* \varphi^t_* g^{-1}_* \nu \longrightarrow g_* \mu = \hat{\mu} \text{ as } t \longrightarrow \infty.
% \end{equation}
So $\hat{\mu}$ is an attracting measure for $(Y, \hat{\varphi}, \ell_Y). \quad \blacksquare$

\subsection{An additional forecasting error bound without mixing and related assumptions.}\label{subsec:add_bd_without_mixing}

\subsubsection{A metric space bound on the prediction error.} In Section \ref{subsec:mixing_bounds} we discussed how mixing and related assumptions improve the bound given in Theorem \ref{thm:bd_ffhat_t_x}. Another assumption which was needed alongside mixing was that of structural stability. In Theorem \ref{thm:metric_space_conjugate_maps} we see that making the structural stability assumption for metric spaces, while retaining the exponential growth, gives an improved growth factor of $e^{\Lambda_{\hat{\varphi}}(x) + \delta}$ as opposed to $e ^{\lambda_{\hat{\varphi}}^+(\gamma, \hat{\mu})+ \delta}$ in Theorem \ref{thm:bd_ffhat_t_x}.

\begin{theorem}\label{thm:metric_space_conjugate_maps}
    Let $(X,d)$ be a separable metric space without isolated points and $\varphi$ a continuous dynamical system on $X$. Suppose the dynamical system $\hat{\varphi}$ is conjugate to $\varphi$ by the homeomorphsim $g$, that is, $g \circ {\varphi} = \hat{\varphi} \circ g.$ Suppose that $\hat{\mu}$ is an ergodic probability measure for $(X, \hat{\varphi})$. Then for any $\delta>0$ and any time horizon $T \in \mathbb{N}$ and $\hat{\mu}-$a.e. $x$, there is some $\varepsilon(\delta, T, x)>0$, and some $R(\delta, x) \geq 1$ such that if $d_\infty(g, \text{id}) < \varepsilon$ then
    \begin{equation}\label{ineq:metric_space_conjugate_maps_bound}
        d(\varphi^t(x), \hat{\varphi}^t(x)) < d_\infty(g^{-1}, \text{id}) + d_\infty(g, \text{id}) e^{t (\Lambda_{\hat{\varphi}}(x) + \delta)} \quad \text{ for all } t=0,\dots,T.
    \end{equation}
    In particular, if $X$ is a smooth compact manifold with metric $d$ and $\varphi \in \operatorname{Diff}^1(X)$ is strongly structurally stable, then for $\hat{\varphi} \in \operatorname{Diff}^1(X)$, if $d_1(\varphi, \hat{\varphi})$ is sufficiently small the bound \eqref{ineq:metric_space_conjugate_maps_bound} holds for all $x$ such that $d_\infty(g, \text{id}) < \varepsilon(\delta, T, x).$
\end{theorem}

We first prove the following lemma.

\begin{lemma}\label{lmm:metric_space_conjugate_maps}
    Let $(X,d)$ be a separable metric space without isolated points and $\varphi$ a continuous dynamical system on $X$. Suppose that $\mu$ is an ergodic measure for $(X, \varphi)$. Then for any $\delta>0$ and any time horizon $T \in \mathbb{N}$ and $\mu-$a.e. $x$, there is some $\varepsilon(\delta, T, x)>0$, and some $R(\delta, x) \geq 1$ such that for any $y \in X$, if $d(x, y) < \varepsilon$ then
    \begin{align}
        d(\varphi^t(x), \varphi^t(y)) < d(x,y) R e^{t(\Lambda_{\varphi}(x) + \delta)} \text{ for all } t=0,\dots, T.
    \end{align}
\end{lemma}

\noindent {\bf Proof.} By definition of $\Lambda_{\varphi}(x)$, for sufficiently small $\delta'(x, \delta)>0$, $\Lambda_{\delta'}(x) < \Lambda_{\varphi}(x) + \delta/2.$ Again, by definition of $\Lambda_{\delta'}(x)$, there exists $R(\delta, x)\geq 1$ such that 
\begin{equation}
    A_{\delta'}(x,t) < Re^{t (\Lambda_{\delta'}(x) + \delta/2)}, \quad \text{ for } t=0,1,2,\dots.
\end{equation}
Since $\varphi$ is continuous, there exists $\varepsilon(\delta, T, x)>0$ such that if $d(x,y)<\varepsilon,$ then $y$ is in the ball $B_x(\delta, T)$ as defined in Section \ref{subsec:metric_le}. So,
\begin{equation}
    d(\varphi^t(x), \varphi^t(y)) \leq d(x,y) A_{\delta'}(x,t) < d(x,y) Re^{t (\Lambda_{\varphi}(x) + \delta)} \text{ for all } t=0,\dots, T. \quad \blacksquare
\end{equation}

{\bf Proof of Theorem \ref{thm:metric_space_conjugate_maps}.} Given the constants $\delta$ and $T$, for $\hat{\mu}-$a.e. $x \in X,$ Lemma \ref{lmm:metric_space_conjugate_maps} applied to ${\hat{\varphi}}$ gives us the two values $\varepsilon(\delta, T, x)$ and $R(\delta, x).$ Now, if $d_\infty(g, \text{id}) < \varepsilon,$ then

\begin{align}
    d(\varphi^t(x), \hat{\varphi}^t(x)) &\leq d(g^{-1} \circ \hat{\varphi}^t \circ g(x), \hat{\varphi}^t \circ g(x)) + d(\hat{\varphi}^t \circ g(x), \hat{\varphi}^t(x))\\
    &< d_\infty(\text{id}, g^{-1}) + d_\infty(\text{id}, g) R e^{t(\Lambda_{\hat{\varphi}}(x) + \delta)}  \text{ for all } t=0,\dots, T.
\end{align}

When $\varphi$ is strongly structurally stable, taking $d_1(\varphi, \hat{\varphi})$ sufficiently small, will shrink $d_\infty(g, \text{id}).$ The bound holds for $\hat{\mu}-$a.e. $x$ for which $d_\infty(g, \text{id}) < \varepsilon(\delta, T, x). \quad \blacksquare$

\subsubsection{Error bounds for prediction on $(\mathcal{P}(X), \varphi_*).$}

We apply Theorem \ref{thm:metric_space_conjugate_maps} to the dynamical system $(\mathcal{P}(X), \varphi_*).$

\begin{theorem}\label{thm:bd_metric_le_prob_space_conjugate_maps}
    Let $(X,d)$ be a separable metric space without isolated points and $\varphi$ a continuous dynamical system on $X$. Suppose the dynamical system $\hat{\varphi}$ is conjugate to $\varphi$ by the homeomorphsim $g$, that is, $g \circ {\varphi} = \hat{\varphi} \circ g.$ Equip the space $\mathcal{P}(X)$ with a metric $d_\mathcal{P}$ metrizing the weak topology. Let $\hat{\Xi}$ be an ergodic measure for $(\mathcal{P}(X), \hat{\varphi}_*)$. Then for any $\delta>0$ and any time horizon $T \in \mathbb{N}$ and $\hat{\Xi}-$a.e. $\nu \in \mathcal{P}(X)$, there is some $\varepsilon(\delta, T, \nu)>0$, and some $R(\delta, \nu) \geq 1$ such that if $d_{\mathcal{P}, \infty}(g_*, \text{id}_{\mathcal{P}(X)}) < \varepsilon$ then
    \begin{equation}\label{ineq:metric_space_conjugate_maps_bound_pushforward}
        d_{\mathcal{P}}(\varphi^t_*\nu, \hat{\varphi}^t_*\nu) < d_{\mathcal{P}, \infty}(g^{-1}_*, \text{id}_{\mathcal{P}(X)}) + d_{\mathcal{P}, \infty}(g_*, \text{id}_{\mathcal{P}(X)}) e^{t (\Lambda_{\hat{\varphi}_*}(\nu) + \delta)} \quad \text{ for all } t=0,\dots,T.
    \end{equation}
    In particular, if $X$ is a smooth compact manifold with metric $d$ and $\varphi \in \operatorname{Diff}^1(X)$ is strongly structurally stable, then for $\hat{\varphi} \in \operatorname{Diff}^1(X)$, if $d_1(\varphi, \hat{\varphi})$ is sufficiently small the bound \eqref{ineq:metric_space_conjugate_maps_bound_pushforward} holds for all $\nu$ such that $d_\infty(g_*, \text{id}_{\mathcal{P}(X)}) < \varepsilon(\delta, T, \nu).$
\end{theorem}

\noindent {\bf Proof.} The conjugacy between $\varphi$ and $\hat{\varphi}$ carries over to their pushforwards: $\hat{\varphi}_* = g^{-1}_* \varphi_* g_*.$ Thus the result follows from Theorem \ref{thm:metric_space_conjugate_maps}. $\quad \blacksquare$

\begin{remark}
    \normalfont
    If we choose $d_\mathcal{P}$ as the Wasserstein-1 metric on a bounded space $X$, then $d_{\mathcal{P}, \infty}(g^{-1}_*, \text{id}_{\mathcal{P}(X)})$ and $d_{\mathcal{P}, \infty}(g_*, \text{id}_{\mathcal{P}(X)})$ may be replaced with $d_\infty(g^{-1}, \text{id})$ and $d_\infty(g, \text{id})$, respectively.
\end{remark}

\subsubsection{An additional result on climate forecasting with state-space systems.} 

Finally, we apply the result to the state-space learning setup. The notation is as in Section \ref{sec:main_result}.

\begin{theorem}\label{thm:state_space_system_bd_metric_le_prob_space_conjugate_maps}
    Suppose that $(\widetilde{\mathcal{X}},d)$ is a Polish space without isolated points and $\Phi$ a continuous dynamical system on $\widetilde{X}$. Suppose that the readout function $\hat{h}$ is Lipschitz with constant $L_{\hat{h}}$ and the state map $f$ is also Lipschitz. Suppose that the dynamical system $\hat{\Phi}$ is conjugate to $\Phi$ by the homeomorphsim $g$, that is, $g \circ {\Phi} =  \hat{\Phi} \circ g.$  Take $d_{\mathcal{P}, \widetilde{\mathcal{X}}}$ and $d_{\mathcal{P}, \mathcal{U}}$ to be Wasserstein-1 distances on their respective spaces. Let $\hat{\Xi}$ be an ergodic measure for the dynamical system $(\mathcal{P}(\widetilde{\mathcal{X}}), \hat{\Phi}_*)$. Then for any $\delta>0$ and any time horizon $T \in \mathbb{N}$ and $\hat{\Xi}-$a.e. $\xi_0 = \zeta_* \mu_0 \in \mathcal{P}(\widetilde{\mathcal{X}})$, there is some $\varepsilon(\delta, T, \xi_0)>0$, and some $R(\delta, \xi_0) \geq 1$ such that if $d_{\infty}(g, \text{id}_{\widetilde{\mathcal{X}}}) < \varepsilon$ then
    \begin{equation}\label{ineq:state_space_conjugate_maps_bound_pushforward}
        d_{\mathcal{P}, \mathcal{U}}(\nu_t, \hat{\nu}_t) < d_\infty(h, \hat{h}) + L_{\hat{h}} \left( d_{\infty}(g^{-1}, \text{id}_{\widetilde{\mathcal{X}}}) + d_{\infty}(g, \text{id}_{\widetilde{\mathcal{X}}}) e^{(t-1) (\Lambda_{\hat{\Phi}_*}(\xi_0) + \delta)} \right)  \text{ for all } t=1,\dots,T.
    \end{equation}
    In particular, if $\widetilde{\mathcal{X}}$ is a smooth compact manifold with metric $d$ and $\Phi \in \operatorname{Diff}^1(\widetilde{\mathcal{X}})$ is strongly structurally stable, then for $\hat{\Phi} \in \operatorname{Diff}^1(\widetilde{\mathcal{X}})$, if $d_1(\Phi, \hat{\Phi})$ is sufficiently small the bound \eqref{ineq:state_space_conjugate_maps_bound_pushforward} holds for all $\xi_0$ such that $d_\infty(g, \text{id}_{\widetilde{\mathcal{X}}}) < \varepsilon(\delta, T, \xi_0).$
\end{theorem}

\noindent {\bf Proof.} % Since $f$ and $\hat{h}$ are Lipschitz, so is $\hat{\Phi}$. Since $d_{\mathcal{P}, \widetilde{\mathcal{X}}}$ is the Wasserstein-1 distance, $\hat{\Phi}_*$ is also Lipschitz.
By Theorem \ref{thm:bd_metric_le_prob_space_conjugate_maps}, for $\hat{\Xi}-$a.e. $\xi_0 = \zeta_* \mu_0 \in \mathcal{P}(\widetilde{\mathcal{X}})$, there is some $\varepsilon(\delta, T, \xi_0)>0$, and some $R(\delta, \xi_0) \geq 1$ such that if $d_{\mathcal{P}, \widetilde{\mathcal{X}}, \infty}(g_*, \text{id}_{\mathcal{P}(\widetilde{\mathcal{X}})}) < \varepsilon$ then

\begin{equation}
    d_{\mathcal{P}, \widetilde{\mathcal{X}}}(\xi_t, \hat{\xi}_t) = d_{\mathcal{P}}(\Phi^t_* \xi_0, \hat{\Phi}^t \xi_0) < d_{\mathcal{P}, \widetilde{\mathcal{X}}, \infty}(g_*, \text{id}_{\mathcal{P}(\widetilde{\mathcal{X}})}) + d_{\mathcal{P}, \widetilde{\mathcal{X}}, \infty}(g^{-1}_*, \text{id}_{\mathcal{P}(\widetilde{\mathcal{X}})}) e^{t (\Lambda_{\hat{\Phi}}(\xi_0) + \delta)}  \quad \text{ for all } t=0,\dots,T.
\end{equation}

Since $d_{\mathcal{P}, \widetilde{\mathcal{X}}}$ is the Wasserstein-1 distance, by Corollary \ref{corr:wass_fftilde}, 
\begin{equation*}
d_{\mathcal{P}, \widetilde{\mathcal{X}}, \infty}(g_*, \text{id}_{\mathcal{P}(\widetilde{\mathcal{X}})}) \leq d_\infty(g, \text{id}),
    d_{\mathcal{P}, \widetilde{\mathcal{X}}, \infty}(g^{-1}_*, \text{id}_{\mathcal{P}(\widetilde{\mathcal{X}})}) \leq d_\infty(g^{-1}, \text{id}),
\end{equation*}
and $d_{\mathcal{P}, \widetilde{\mathcal{X}}, \infty} (h_*, \hat{h}_*) \leq d_\infty(h, \hat{h})$.
By Lemma \ref{lmm:wass_mumutilde_Lipschitz}, a Lipschitz constant for $\hat{h}_*$ is $L_{\hat{h}}.$  Finally, $d_\mathcal{P}(\nu_t, \hat{\nu_t}) = d_{\mathcal{P}} (h_* \xi_{t-1}, \hat{h}_* \hat{\xi}_{t-1}) \leq d_{\mathcal{P}} (h_* \xi_{t-1}, \hat{h}_* \xi_{t-1}) + d_{\mathcal{P}} (\hat{h}_* \xi_{t-1}, \hat{h}_* \hat{\xi}_{t-1}).$ This completes the proof. $ \quad \blacksquare$

\subsection{Stability of climate forecasting using state-space learning.}\label{subsec:stab_empirical_measure}

In his paper \cite{lorenz1964problem}, Lorenz refers to the climate of a dynamical system as `the set of long-term statistical properties' of the system, in particular identifying it with averages taken along trajectories of the system. Ergodic and physical measures thus describe the climate of the dynamical system as calculated from a large set of points. In this section, we deal with the question of the stability of the climate under perturbations of the dynamical system map. Consequently, we also address the ability of state-space systems to reproduce the climate after training. In particular, Theorem \ref{thm:trajectory_average_with_state_space_system_bound} shows that when a $C^1$ sufficiently accurate proxy system is obtained, the climate as calculated from a large set of points is perturbed only slightly.

\subsubsection{Stability of empirical measures under perturbations of the dynamical system map.} We derive a result concerning the stability of the empirical measures obtained from an initial point evolved under a perturbed dynamical system.

\begin{theorem}\label{thm:empirical_measure_stable_bound_main}
    Let $(X,d)$ be a smooth compact manifold, and $\varphi \in \operatorname{Diff}^1(X)$ a $C^1$ structurally stable map. Thus there exists $\delta>0$ such that for $\hat{\varphi} \in \operatorname{Diff}^1(X)$ if $d_1(\varphi, \hat{\varphi})<\delta$, then $\varphi$ is topologically conjugate to $\hat{\varphi}$ by a homeomorphism $g \colon X \longrightarrow X$, that is $g \circ {\varphi} = \hat{\varphi} \circ g$. Let $\mu \in \mathcal{P}(X)$ be an invariant measure for $\varphi$. Then, for any $x \in \Basin _\varphi(\mu) \cap g(\Basin _\varphi(\mu)),$

    \begin{equation}
        \lim_{t \to\infty} d_{\mathcal{P}}\left(\frac{1}{T} \sum_{t=0}^{T-1} \delta_{\varphi^t(x)}, \frac{1}{T} \sum_{t=0}^{T-1} \delta_{\hat{\varphi}^t(x)} \right) = d_{\mathcal{P}}(\mu, g_*\mu).
    \end{equation}

    If $\varphi$ is {\bfi strongly structurally stable}, $d_{\mathcal{P}}(\mu, g_*\mu)$ may be made arbitrarily small by decreasing $\delta$.
\end{theorem}

\noindent {\bf Proof.} By Lemma \ref{lmm:top_conj_basin_supp}, $\Basin _\varphi(\mu) \cap g(\Basin _\varphi(\mu)) = \Basin _\varphi(\mu) \cap \Basin _{\hat{\varphi}} (\hat{\mu}),$ where $\hat{\mu} = g_* \mu.$ Thus for any element $x$ of this set,

\begin{equation}
    \lim_{t \to\infty} d_{\mathcal{P}}\left(\frac{1}{T} \sum_{t=0}^{T-1} \delta_{\varphi^t(x)}, \frac{1}{T} \sum_{t=0}^{T-1} \delta_{\hat{\varphi}^t(x)} \right) 
    = d_{\mathcal{P}}(\mu, \hat{\mu}) = d_{\mathcal{P}}(\mu, g_*\mu).
    % \leq \limsup_{t \to\infty} d_{\mathcal{P}}\left(\frac{1}{T} \sum_{t=0}^{T-1} \delta_{\varphi^t(x)}, \mu \right) + d_{\mathcal{P}} ( \mu, \hat{\mu} ) + d_{\mathcal{P}} \left( \hat{\mu}, \frac{1}{T} \sum_{t=0}^{T-1} \delta_{\hat{\varphi}^t(x)} \right) = d_{\mathcal{P}}(\mu, g_*\mu).
\end{equation}

If $\varphi$ is strongly structurally stable, continuity of the map $ g \mapsto g_* \mu$ (Lemma \ref{lmm:pushforward_cts_in_f}) means that $d_{\mathcal{P}}(\mu, g_*\mu) \longrightarrow 0$ as $\delta \longrightarrow 0. \quad \blacksquare$

\begin{remark}\label{rem:stab_emp_measure}
    \normalfont
    An important question here is how large the set $\Basin _\varphi(\mu) \cap g(\Basin _\varphi(\mu))$ is. If $\mu$ is ergodic for $\varphi$, and $g^{-1}$ is nonsingular with respect to $\mu$, then both $\Basin _\varphi (\mu)$ and $g(\Basin _\varphi (\mu))$ have full $\mu-$measure, so their intersection will also have full $\mu-$measure. On the other hand, in the case where $\mu$ is a physical measure for $\varphi$ with respect to a reference measure $\ell,$ $\Basin _\varphi(\mu)$ contains $\ell-$almost every point in an open neighborhood $U$ of $\supp (\mu)$. As noted in Remark \ref{rem:assumptions_validity}, structural stability assumptions can be used to guarantee that the open set $U \cap g(U)$ is nonempty. If $g^{-1}$ is nonsingular with respect to $\ell,$ then $\Basin_\varphi(\mu) \cap g(\Basin_\varphi(\mu))$ contains $\ell-$almost every point in this intersection.
\end{remark}

\subsubsection{Stability of empirical measures for the proxy system.} 

 We recall the notation for when state-space systems are used in point forecasts, as described in Section \ref{subsec:learning with state-space systems}. Consider any initial condition $\underline{m}_0 \in \mathcal{M}$ for the underlying dynamical system. Corresponding to this we have a trajectory $\underline{u} \in \mathcal{U}^{\mathbb{Z}}$ in the observation space given by $\underline{u}_t = \omega(\underline{m}_t) = \omega(\phi^t(\underline{m}_0))$. Our state-space system yields a sequence $\underline{x} \in \widetilde{\mathcal{X}}^{\mathbb{Z}}$. Due to approximation in choosing $\hat{h}$, we end up with the predicted sequence $\hat{\underline{x}}$ where $\hat{\underline{x}}_t = \hat{\Phi}^t(\underline{x}_0)$ for $t=1,2,3,\dots$. The predictions $(\hat{\underline{u}}_1, \hat{\underline{u}}_2, \dots)$ are given by $\hat{\underline{u}}_t = \hat{h} ( \hat{\Phi}^{t-1}(\underline{x}_0)).$
 
 \begin{theorem}\label{thm:trajectory_average_with_state_space_system_bound}
    Suppose that $(\widetilde{\mathcal{X}}, d)$ is a smooth compact manifold, and $\Phi \in \operatorname{Diff}^1(\widetilde{\mathcal{X}})$ is a $C^1$ structurally stable map. Thus there exists $\delta>0$ such that if $\hat{\Phi} \in \operatorname{Diff}^1(\widetilde{\mathcal{X}})$, and $d_1(\Phi, \hat{\Phi}) < \delta$, then $\Phi$ is topologically conjugate to $\hat{\Phi}$ by a homeomorphism $g \colon \widetilde{\mathcal{X}} \longrightarrow \widetilde{\mathcal{X}}$, that is $ g \circ \Phi = \hat{\Phi} \circ g.$ Let $ \xi \in \mathcal{P}(\widetilde{\mathcal{X}})$ be an invariant probability measure for $\Phi$. For any initial condition $\underline{m}_0\in \mathcal{M}$, if $\underline{x}_0 = \zeta(\underline{m}_0) \in \Basin_\Phi(\xi) \cap g( \Basin_\Phi(\xi)),$ then
    
    \begin{equation}
        \lim_{t \to\infty} d_{\mathcal{P}, \mathcal{U}}\left(\frac{1}{T} \sum_{t=0}^{T-1} \delta_{\underline{u}_t}, \frac{1}{T} \sum_{t=0}^{T-1} \delta_{\hat{\underline{u}}_t} \right)  =  d_{\mathcal{P}, \mathcal{U}}(h_* \xi, \hat{h}_* g_* \xi) \leq d_{\mathcal{P}, \mathcal{U}}(h_* \xi, \hat{h}_* \xi) + d_{\mathcal{P}, \mathcal{U}}(\hat{h}_* \xi, \hat{h}_* g_* \xi).
    \end{equation}

    If $\Phi$ is {\bfi strongly structurally stable}, then the bound may be made arbitrarily small as $d_1(h, \hat{h}) \longrightarrow 0.$
\end{theorem}

\noindent {\bf Proof.} Since $\underline{x}_0 = \zeta(\underline{m}_0) \in \Basin_\Phi(\xi) \cap g( \Basin_\Phi(\xi)),$ $\frac{1}{T} \sum_{t=0}^{T-1} \delta_{\Phi^t(\underline{x}_0)} \longrightarrow \xi$ and $\frac{1}{T} \sum_{t=0}^{T-1} \delta_{\hat{\Phi}^t(\underline{x}_0)}\longrightarrow \hat{\xi}$ where $\hat{\xi} = g_* \xi$, as $T \longrightarrow \infty.$ Now, $\delta_{\underline{u}_t} = h_* \delta_{\underline{x}_{t-1}}$ and likewise $\delta_{\hat{\underline{u}}_t} = \hat{h}_* \delta_{\hat{\underline{x}}_{t-1}}$. Thus by Lemma \ref{lmm:pushforward_cts_in_nu},

\begin{align}
    \lim_{t \to\infty} d_{\mathcal{P}, \mathcal{U}}\left(\frac{1}{T} \sum_{t=0}^{T-1} \delta_{\underline{u}_t}, \frac{1}{T} \sum_{t=0}^{T-1} \delta_{\hat{\underline{u}}_t} \right) 
    &= d_{\mathcal{P}, \mathcal{U}}\left( h_* \left(\frac{1}{T} \sum_{t=0}^{T-1} \delta_{\underline{x}_t} \right), \hat{h}_* \left(\frac{1}{T} \sum_{t=0}^{T-1} \delta_{\underline{\hat{x}}_t} \right) \right)\\
    % &\leq \limsup_{t \to \infty} d_{\mathcal{P}, \mathcal{U}}\left( h_* \left(\frac{1}{T} \sum_{t=0}^{T-1} \delta_{\underline{x}_t} \right), h_* \xi \right) + d_{\mathcal{P}, \mathcal{U}}(h_* \xi, \hat{h}_* \hat{\xi}) \\
    % &+ d_{\mathcal{P}, \mathcal{U}}\left( \hat{h}_* \hat{\xi}, \hat{h}_* \left( \frac{1}{T} \sum_{t=0}^{T-1} \delta_{\hat{\underline{u}}_t} \right) \right)\\
    &= d_{\mathcal{P}, \mathcal{U}} (h_* \xi, \hat{h}_* g_* \xi) \leq d_{\mathcal{P}, \mathcal{U}}(h_* \xi, \hat{h}_* \xi) + d_{\mathcal{P}, \mathcal{U}}(\hat{h}_* \xi, \hat{h}_* g_* \xi).
\end{align}

Note that $d_1(\Phi, \hat{\Phi}) \longrightarrow 0$ as $d_1(h, \hat{h}) \longrightarrow 0.$ Thus if $\Phi$ is strongly structurally stable, $d_\infty(\text{id}, g) \longrightarrow 0$ as $d_1(h, \hat{h}) \longrightarrow 0$ and so by Lemmas \ref{lmm:pushforward_cts_in_f} and \ref{lmm:pushforward_cts_in_nu}, $d_{\mathcal{P}, \mathcal{U}}(h_* \xi, \hat{h}_* \xi) + d_{\mathcal{P}, \mathcal{U}}(\hat{h}_* \xi, \hat{h}_* g_* \xi) \longrightarrow 0$ as $d_1(h, \hat{h}) \longrightarrow 0. \quad \blacksquare$

\medskip
\begin{figure}[!htb]
\centering
\begin{subfigure}{\textwidth}
    \centering
    \includegraphics[scale=0.32]{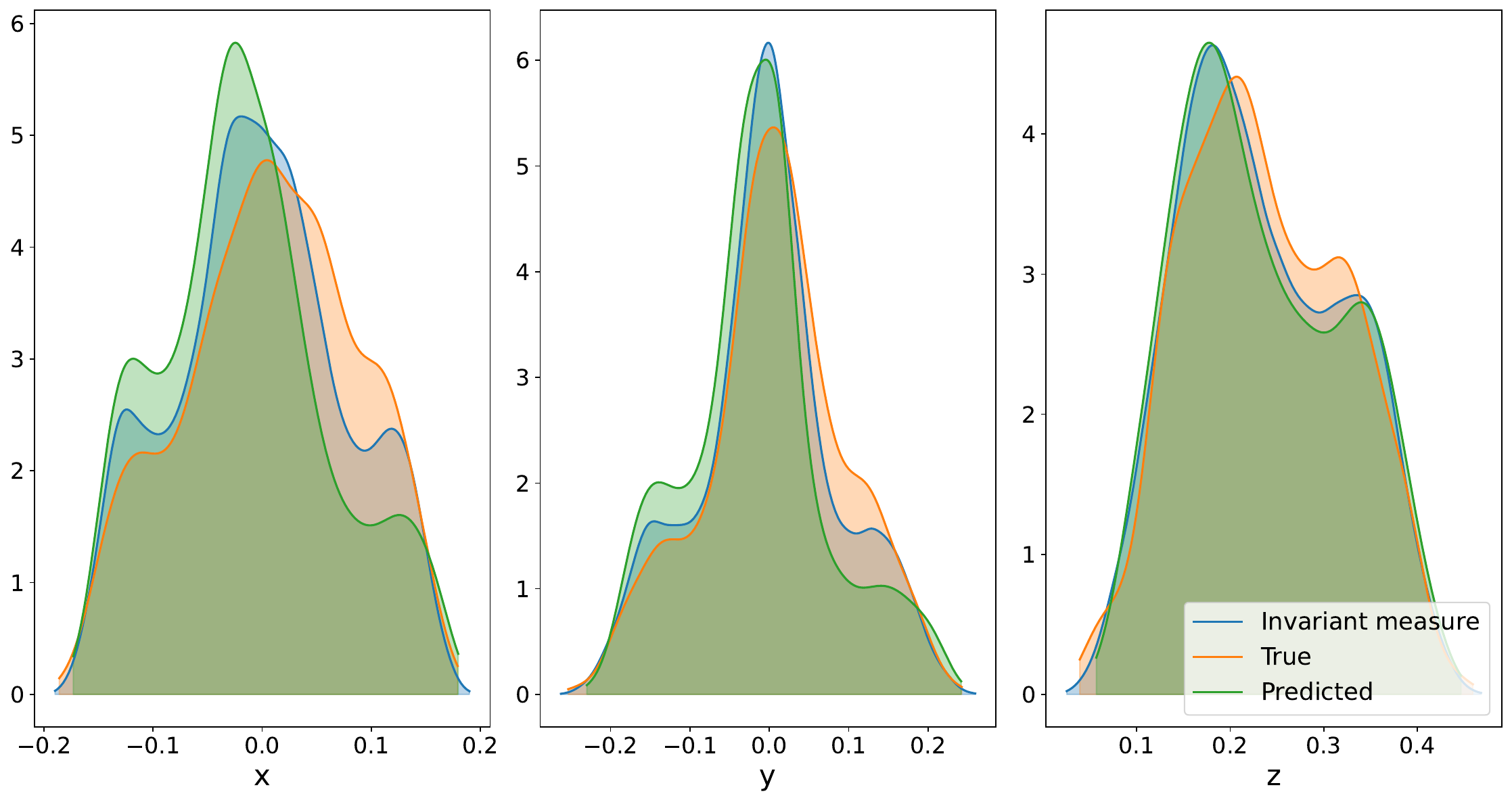}
    \caption{Empirical distributions: invariant measure of the true system (blue); empirical measure calculated up to time $\tau=80$ for an initial condition evolved under the true system (orange); empirical measure calculated up to time $\tau=80$ for an initial condition evolved under the proxy system (green).}
    \label{fig:densities_axes_traj}
\end{subfigure}

\vspace{1em}

\begin{subfigure}{\textwidth}
    \centering
    \includegraphics[scale=0.3]{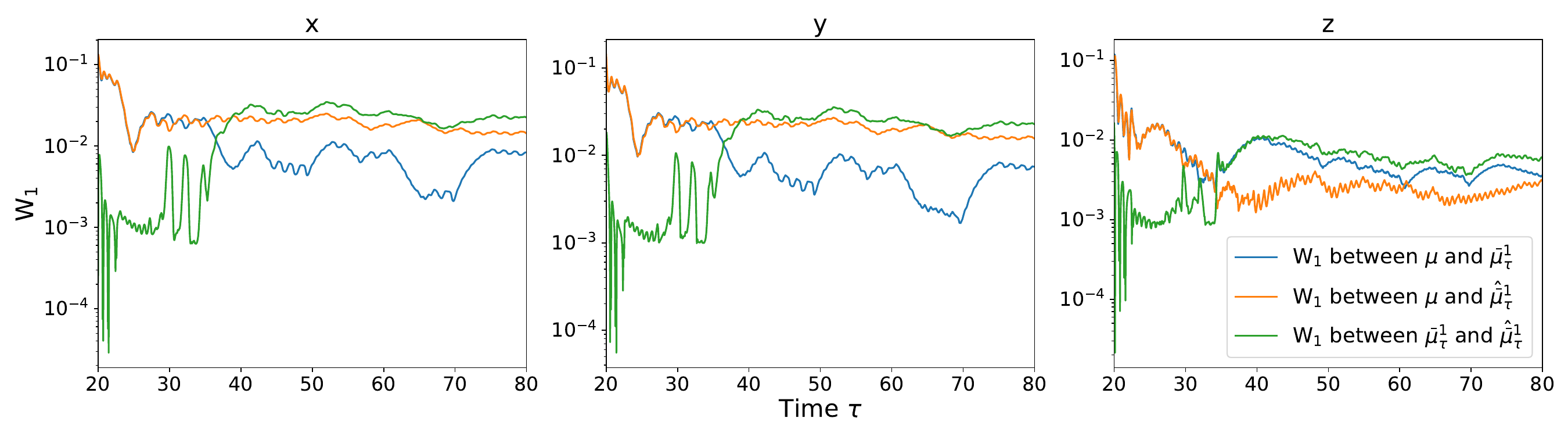}
    \caption{Convergence of empirical distributions: Wasserstein-1 distance between the invariant distribution and Ces\`{a}ro average for a trajectory of the true system (blue), between the invariant distribution and Ces\`{a}ro average for a trajectory of the proxy system (orange), between a Ces\`{a}ro average over a trajectory of the true and proxy systems (green).}
    \label{fig:Wass1_transport_single_trajectory_figure_panel}
\end{subfigure}

\caption{Convergence of empirical measures. The empirical measures produced by the true and proxy systems through a Ces\`{a}ro average approximate the true invariant distribution. The error between the true and predicted Ces\`{a}ro averages stabilizes around $\tau = 40$ (18.1 Lyapunov times), thus validating Theorem \ref{thm:trajectory_average_with_state_space_system_bound}.}
\label{fig:combined_trajectory}
\end{figure}

\medskip

\subsection{Additional numerics.}\label{subsec:additional_numerics}

\subsubsection{Numerical verification of Theorem \ref{thm:trajectory_average_with_state_space_system_bound}.}\label{sec:numerics_empirical_measure}

To validate Theorem \ref{thm:trajectory_average_with_state_space_system_bound}, we selected an initial condition $\underline{m}_{t=1000}$ in $\mu^1_{t=1000}$ and considered the evolution of the Ces\`{a}ro averages.

\begin{align}
    \overline{\mu}^1_t &= \frac{1}{t-1000} \sum_{i=1000}^{t-1001} \delta_{\underline{m}_{i}}\\
    \hat{\overline{\mu}}^1_t &= \frac{1}{t-1000} \sum_{i=1000}^{t-1001} \delta_{\hat{\underline{m}}_{i}.}
\end{align}

This was compared to the invariant distribution which was approximated by aggregating the last 100 time steps of all 1000 initial conditions in $\mu^1_{t=1000}$. Figure \ref{fig:combined_trajectory} (a) plots the final empirical distribution at $t=4000$ ($\tau = 40$). Figure \ref{fig:combined_trajectory} (b) plots the convergence of the two Ces\`{a}ro averages to one another and to the invariant distribution. Both the true and proxy Ces\`{a}ro averages approach the invariant distribution, though the true system gets closer than the proxy. Theorem \ref{thm:trajectory_average_with_state_space_system_bound} states that the discrepancy between the true and proxy empirical measures is bounded. Indeed, considering the green curve, we see that the error increases once prediction is started, but by around $\tau=40$ (18.1 Lyapunov times of prediction) it settles at a constant value. This constant value is comparable to the distance between the empirical distribution calculated from the proxy system and the invariant distribution of Lorenz.

\medskip

\subsubsection{Invariant distribution of the proxy system.}\label{sec:invariant_distribution}

\begin{figure}[!htb]
\centering
\includegraphics[scale = 0.4]{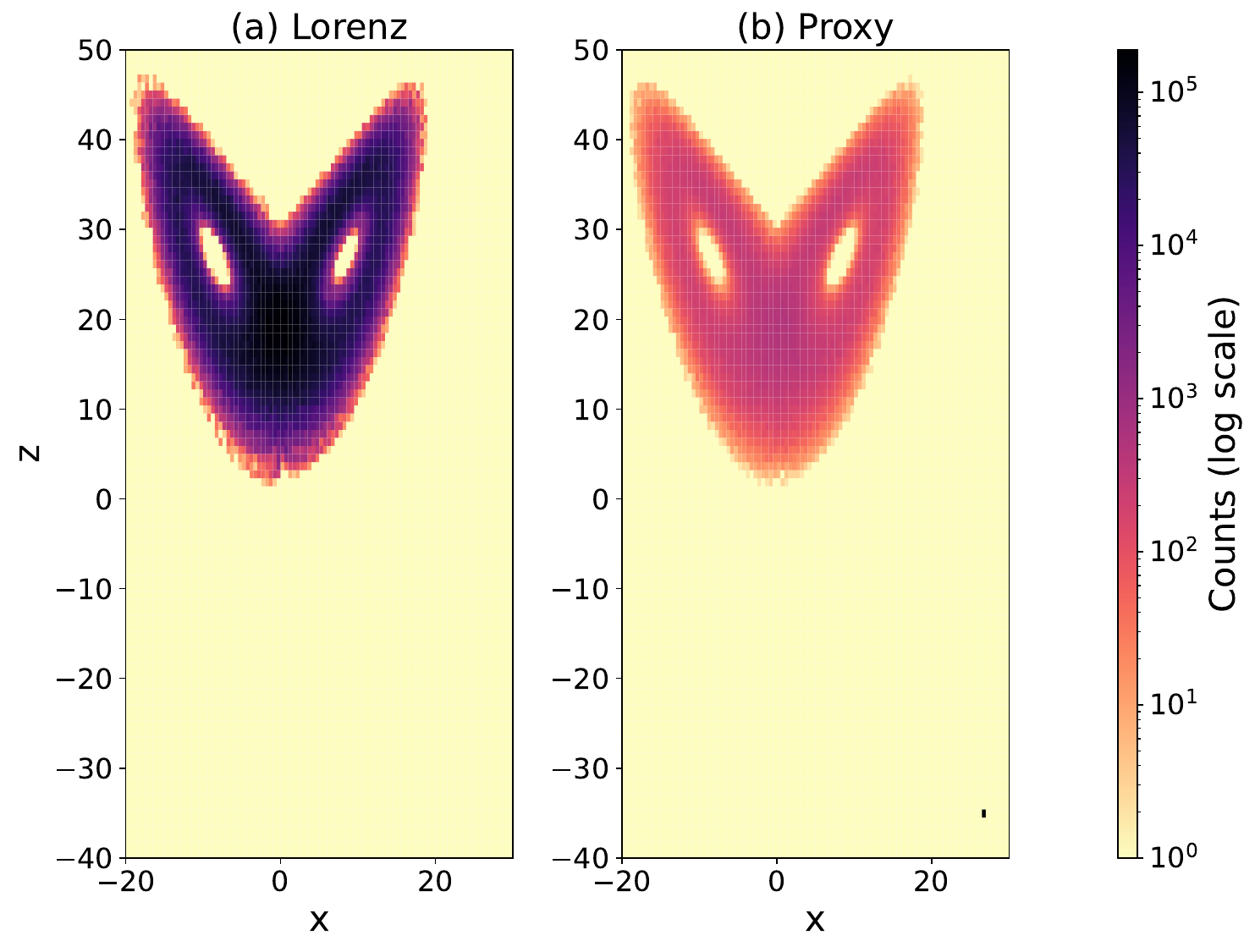}
\caption{Invariant measure of (a) the Lorenz system and (b) the proxy system. The proxy system accurately captures the invariant distribution of the Lorenz system, but also exhibits an attracting fixed point at (26.736, 18.134, -35.464).}
\label{fig:invariant_measures}
\end{figure}

To find the invariant measure of the true and proxy system, we randomly selected 1000 initial conditions from the uniform distribution $\mathcal{U}([-300,300)\times[-300,300)\times[-300,300))$ and evolved them for 1500 time steps. We discarded the first 1000 time steps and aggregated all the points from the 1000 trajectories over the last 500 time steps to approximate the invariant measure. Figure \ref{fig:invariant_measures} shows the invariant measures of the true and proxy systems. The proxy system evolves in the state space, and is then mapped back to the observation space through the readout. It captures the invariant measure on the Lorenz attractor, but also has a Dirac measure concentrated on an attracting fixed point of the system. It `hallucinates' in that it sees a spurious attracting fixed point which does not exist in the system being learnt. Other instances of the ESN that came up during the numerical work exhibited attracting limit cycles. This underscores our observation that the structural stability assumption is not strictly necessary to obtain stable density forecasts. The proxy system is clearly not conjugate to the original Lorenz system, nevertheless, as was seen in Section \ref{sec:numerics}, the proxy system yields stable density forecasts for initial conditions started close enough to the attractor. Thus at least for the Lorenz system, which satisfies linear response and stable mixing, a sufficiently well-trained proxy will give stable density predictions for distributions that start near enough to the attractor.

\bigskip

\footnotesize
\addcontentsline{toc}{section}{References}
\bibliographystyle{wmaainf}
\bibliography{GOLibrary.bib} % JMLouw: I added in a reference for Kingmans subadditive ergodic theorem
% \bibliography{/Users/JPO/Dropbox/Public/GOLibrary}
% \bibliography{/Users/Lyudmila/Dropbox/Mendeley/GOLibrary}
% \bibliography{/Volumes/Staff/Mila/Library/BibTex/GOLibrary}
\end{document}